\documentclass[twocolumn]{aastex62}
\usepackage{amsmath}
\usepackage{amssymb}
\usepackage{mathbbol}
 \usepackage{mathrsfs}
\usepackage{dsfont}
\usepackage{amsfonts}
\usepackage{color}
\usepackage{amsopn}
\usepackage{bm}
\usepackage{graphicx,longtable,times,threeparttable,multirow}
\usepackage{longtable,times,threeparttable,multirow}
\usepackage{float}

\DeclareGraphicsExtensions{.eps,.eps.gz,.epsi, .jpg, .png}

\newcommand{\teff}{\ensuremath{T_{\mathrm {eff}}\,}}
\newcommand{\logg}{\ensuremath{{\mathrm {log}\, } g\,}}
\newcommand{\feh}{\ensuremath{[{\mathrm {Fe/H}}]\,}}
\newcommand{\mk}{\ensuremath{M_{K{\mathrm s}}\,}}
\newcommand{\msun}{\ensuremath{M_\odot\,}}
\newcommand{\Hii}{\ensuremath{\mathrm {H\,\textsc{ii}}\,}}
\newcommand{\Ha}{\ensuremath{\mathrm H_\alpha\,}}
\newcommand{\ks}{\ensuremath{m_{K{\mathrm s}}\,}}
\newcommand{\aks}{\ensuremath{A_{K{\mathrm s}}\,}}

\shorttitle{Data-driven OB star Distances}
\shortauthors{Xiang et al.}

\begin{document}

\title{Data-driven spectroscopic estimates of absolute magnitude, distance and binarity \\
--- method and catalog of 16,002 O- and B-type stars from LAMOST}

\author{Mao-Sheng Xiang}
\affil{Max-Planck Institute for Astronomy, K\"onigstuhl 17, D-69117 Heidelberg, Germany}
\email{{\rm email: } {\em mxiang@mpia.de}}
\author{Hans-Walter Rix}
\affil{Max-Planck Institute for Astronomy, K\"onigstuhl 17, D-69117 Heidelberg, Germany}
\email{{\rm email: } {\em rix@mpia.de}}
\author{Yuan-Sen Ting}\thanks{Hubble fellow}
\affil{Institute for Advanced Study, Princeton, NJ 08540, USA}
\affil{Department of Astrophysical Sciences, Princeton University, Princeton, NJ 08544, USA}
\affil{Observatories of the Carnegie Institution of Washington, 813 Santa
Barbara Street, Pasadena, CA 91101, USA}
\affil{Research School of Astronomy \& Astrophysics, Australian National University, Canberra, ACT 2611, Australia}
\author{Eleonora Zari}
\affil{Max-Planck Institute for Astronomy, K\"onigstuhl 17, D-69117 Heidelberg, Germany}
\author{Kareem El-Badry}
\affil{Department of Astronomy and Theoretical Astrophysics Center, University of California Berkeley, Berkeley, CA 94720, USA}
\affil{Max-Planck Institute for Astronomy, K\"onigstuhl 17, D-69117 Heidelberg, Germany}
\author{Hai-Bo Yuan}
\affil{Department of Astronomy, Beijing Normal University, Beijing 100875, P. R. China}
\author{Wen-Yuan Cui}
\affil{Department of Physics, Hebei Normal University, Shijiazhuang 050024, People’s Republic of China}

\begin{abstract}{We present a data-driven method to estimate absolute magnitudes for O- and B-type stars from the LAMOST spectra, which we combine with \textit{Gaia} parallaxes to infer distance and binarity. The method applies a neural network model trained on stars with precise \textit{Gaia} parallax to the spectra and predicts $K_{\rm s}$-band absolute magnitudes \mk with a precision of 0.25\,mag, which corresponds to a precision of 12\% in spectroscopic distance.  For distant stars (e.g. $>5$\,kpc), the inclusion of constraints from spectroscopic \mk significantly improves the distance estimates compared to inferences from \textit{Gaia} parallax alone. Our method accommodates for emission line stars by first identifying them via PCA reconstructions and then treating them separately for the \mk estimation.  We also take into account unresolved binary/multiple stars, which we identify through deviations in the spectroscopic \mk from the geometric \mk inferred from \textit{Gaia} parallax. This method of binary identification is particularly efficient for unresolved binaries with near equal-mass components and thus provides an useful supplementary way to identify unresolved binary or multiple-star systems. We present a catalog of spectroscopic \mk, extinction, distance, flags for emission lines, and binary classification for 16,002 OB stars from LAMOST DR5. As an illustration of the method, we determine the \mk and distance to the enigmatic LB-1 system, where \cite{LiuJ2019} had argued for the presence of a black hole and incorrect parallax measurement, and we do not find evidence for errorneous {\it Gaia} parallax.} 

\end{abstract}

\section{Introduction} \label{intro}
O-type and B-type (OB) stars constitute the population of massive, young, and luminous stars in a galaxy.  They play a significant role in many aspects of astrophysics. They are important factories for element production \citep{Thielemann1985, Timmes1995, Woosley1995, Chieffi2004, Nomoto2006} and act as major sources of ionization and energetic feedback to the ISM and IGM \citep{Freyer2003, Freyer2006, Hopkins2014, Mackey2015, Struck2020}.  They often form in binaries and are candidate companions for, or precursors of, black holes and associated gravitational wave events \citep{Abbott2016a, Abbott2016b, Abbott2017, Belczynski2016, LiuJ2019}. In our Galaxy and elsewhere, they serve as signposts of star formation and diagnostics of the initial mass function \citep[e.g.][]{Lequeux1979, Humphreys1984, Reed2005, Bartko2010}. They also serve as tracers of the structure and dynamics of the Galactic disk, including features like spiral arms \citep[e.g.][]{Torra2000, Shu2016, Xu2018, Chen2019, Cheng2019, Li2019, Wang2020}.  Knowledge of the luminosities and distances of the OB stars in the Milky Way is fundamental to all such analyses.

For OB stars within $\lesssim 2$~kpc from the Sun, high precision distance estimates are obtainable using parallaxes from \textit{Gaia} \citep{Prusti2016, Brown2018, Lindegren2018}. At the largest distances that these luminous stars can be observed out to ($\gtrsim 5$~kpc), however, parallax-based distance estimates become suboptimal \citep[e.g.][Zari et al. in prep.]{Shull2019}. This calls for developing alternative ways to accurately determine the distances to such stars.  

Photometric distance estimation has a long and successful history: for stars on the lower main sequence, e.g., G and K dwarfs, unreddened broad-band colors alone serve as excellent luminosity predictors, as long as the metallicity is known to within $\lesssim 0.5$~dex \citep[e.g.][]{Ivezic2008,Juric2008}. For OB stars, on the other hand, the time spent on or near their zero-age main sequence is so short that, at a given color, their luminosities can range over an order of magnitude.  Distant young OB stars also tend to be in directions of high dust extinction with on-going star formation, further limiting the accuracy of the derived de-reddened colors. On top of that, there is also a lack of color variation beyond the Rayleigh-Jeans tail of the OB stars at  $\sim$\,4000~\AA.  This combination of factors makes photometric luminosity and distance estimates for hot stars particularly challenging. 

The spectra of OB stars contain much more information than photometric colors and can yield powerful constraints on luminosities and distances. There are two conceptually different approaches to infer spectroscopic distances.  The first is to derive the basic stellar parameters, \teff , \logg and \feh, and then, for example, employ the ``flux-weighted gravity luminosity" relationship (FGLR) \citep{Kudritzki03}, which relates the $g_f\equiv g / \teff^4$ of a star to its bolometric luminosity \citep{Kudritzki2020}.  With this relationship, the \teff\ and \logg derived from the high-resolution spectra of single stars yield luminosities and distances with precisions of $< 20\%
$  and $10\%$, respectively, for luminous hot stars.  Similarly, stellar isochrones have been used to infer stellar absolute magnitudes and distances from stellar parameters \teff, \logg, and \feh, as has been widely implemented on spectroscopic survey data sets \citep[e.g.][]{Carlin2015, Yuan2015, Wang2016, Xiang2017a, Coronado2018, Green2020, Queiroz2020}.  So far this has been mostly applied to FGK stars, but it should provide decent distance estimates akin to the FGLR method for hot luminous stars.

The second approach is to learn the luminosity and distance {\sl directly} from the data, without the detour of first deriving the stellar parameters. This can be done if numerous examples of the spectra of stars with independently known distances exist, which is the case in the age of \textit{Gaia} and extensive spectroscopic surveys. \citet{Jofre2015} inferred stellar distance with spectroscopically identified twin stars that have accurate parallax measurements, and achieved a distance precision better than 10\% from high-resolution spectra. \citet{Hogg19} demonstrated that the distances of red giant branch stars with SDSS/APOGEE \citep{Majewski2017, SDSS_DR14} can be inferred to within $<10\%$ using a data-driven model trained on \textit{Gaia} parallaxes.
\citet{Xiang2017b} deduced absolute magnitudes from the LAMOST low-resolution spectra for AFGK stars using stars in common with $Hipparcos$ \citep{Perryman1997} as the training set, and achieved a precision of better than 0.26\,mag in \mk estimates (corresponding to a distance precision of 12\%) through a verification with \textit{Gaia} parallaxes \citep{Xiang2017a}.  

For OB stars, such a data-driven approach for spectroscopic luminosity estimation comes with two additional complications. First, a non-negligible fraction of massive stars are in close binaries, often with comparable luminosities \citep{Sana2012, Sana2013}. Second, the optical spectra of OB stars, such as those from LAMOST survey, frequently show emission lines arising either from their corona, their surrounding disks or nearby \Hii regions.  Eliminating these outliers is crucial for achieving robust spectroscopic luminosity estimates.

In this work, we set up a data-driven method to estimate the spectroscopic luminosity and, by implication, the distance of individual OB stars. We employ a training set built from stars with precise \textit{Gaia} parallaxes in a neural network model to map observed spectra to $K_{\rm s}$-band absolute magnitudes \mk.  The method is designed to account for issues stemming from binarity and emission lines.  We combine spectroscopic \mk and \textit{Gaia} parallaxes to identify binary stars as those that are over-luminous compared to the single stars. We apply the method to the LAMOST low-resolution ($R\simeq1800$) spectra for a set of 16,002 OB stars from \citet{LiuZ2019}, which is by far the most extensive set of spectra for luminous, hot stars.  We prefer this data-driven approach over the FGLR, given the complications of validating the accuracy of the stellar parameters determined for OB stars from low-resolution spectra.  

Our validation of the results suggests that the combination of the spectroscopic \mk with the \textit{Gaia} DR2 parallax yields a median distance uncertainty of only 8\% for the LAMOST OB star sample. These distances are presented together with a catalog of \mk, extinction, and flags for binary and emission lines for the 16,002 LAMOST OB stars. The results allow a reassessment for the distance to the binary system LB-1, which has been recently suggested to hold a 70$M_\odot$ black hole, the most massive stellar mass black hold ever found \citep{LiuJ2019}.  

This paper is laid out as follows: Section\,\ref{method} gives an overview of the methods. Section\,\ref{pca} introduces the identification of emission lines in LAMOST OB star spectra with a PCA reconstruction method; Section\,\ref{extinction} introduces our extinction estimation; Section\,\ref{nnmodel} presents the data-driven method for \mk estimation. Section\,\ref{binaryeffect} introduces our method for binary identification. Section\,\ref{distance} describes the inference of distance using the spectroscopic \mk together with \textit{Gaia} parallaxes. Section\,\ref{lb1} discusses our estimates of absolute magnitude and distance of the LB-1 system. We summarize in Section\,\ref{conclusion}.

\section{method overview} \label{method}

\begin{figure}
\centering
\includegraphics[width=0.45\textwidth]{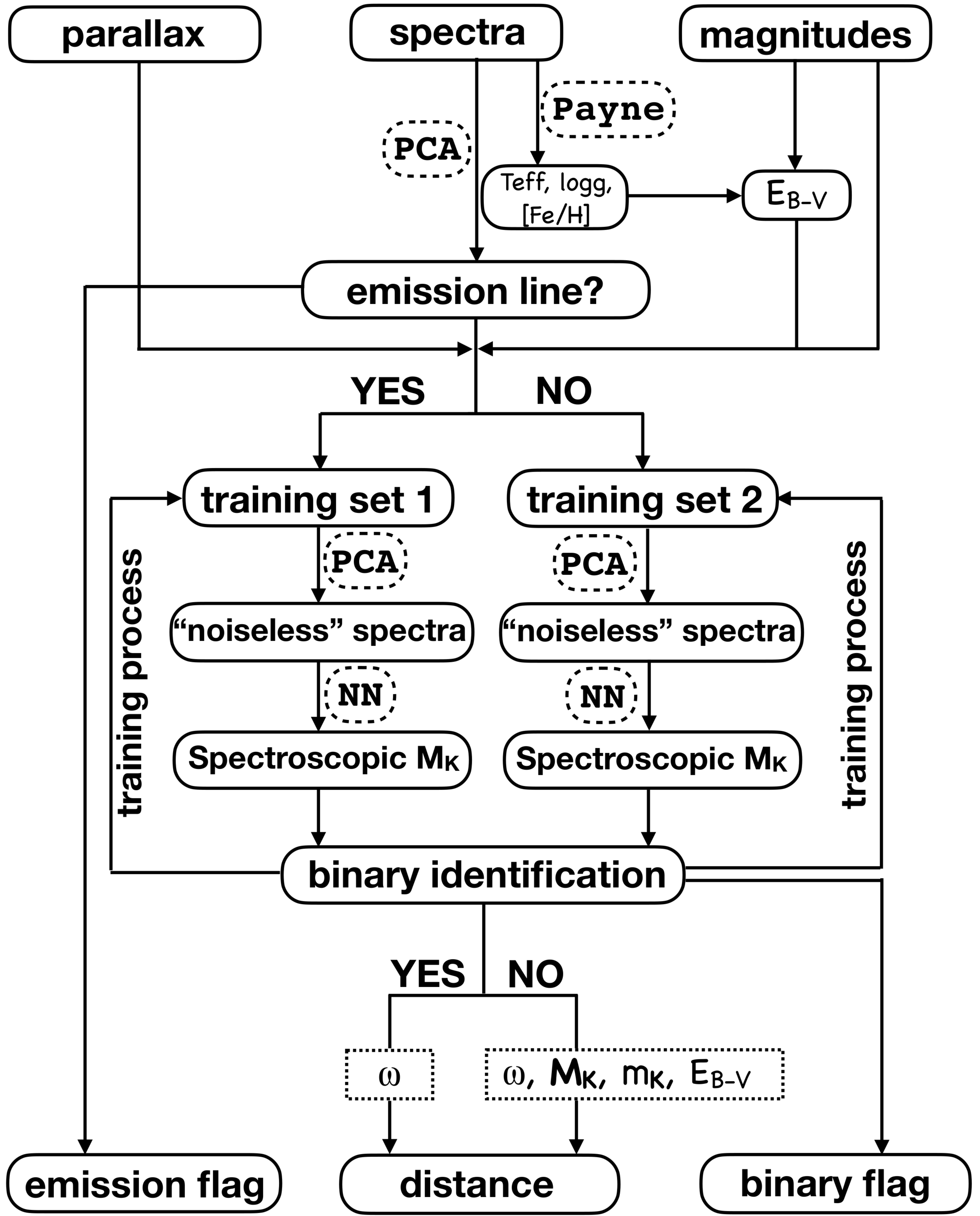}
\caption{Schematic illustration of our data-driven approach for deriving absolute magnitudes and distances for OB stars, including the identification of stars with emission lines and stars in binary or multiple star systems. Astrometric parallaxes, spectra, and apparent magnitudes are adopted as inputs.}
\label{fig:Fig1}
\end{figure}
We derive absolute magnitudes \mk from survey spectra with a data-driven neural network model trained on a subset of selected stars with precise \textit{Gaia} parallaxes (Section\,\ref{nnmodel}). In light of the non-negligible impact from emission lines on our data-driven models, we identify the spectra containing emission lines using a PCA reconstruction method (Section\,\ref{pca}). For these stars, we derive \mk using a separate neural network model, with the emission-line wavelength regions masked. Considering the neural network model is sensitive to spectral noise, we adopt the PCA-reconstructed spectra, for both emission and non-emission stars, as an approximate of the `noiseless' spectra for \mk estimation. With the spectroscopic \mk estimates, we infer distances in combination with the apparent magnitudes and \textit{Gaia} parallaxes (Section\,\ref{distance}).  Binary and multiple star systems are discarded iteratively from the training sets used for \mk estimation.  To identify these systems, which should be over-luminous compared to their single-star counterparts, we measure the deviation between the spectro-photometric parallax and the \textit{Gaia} astrometric parallax (Section\,\ref{binaryeffect}).  To correct for the extinction of individual stars, we use intrinsic colors estimated from synthetic photometry (Section\,\ref{extinction}).  A schematic description of our method is shown in Fig.\,\ref{fig:Fig1}. 

Our approach leverages the information contained in spectra, astrometric parallax and photometric magnitudes, and can be applied to a wide range of stellar types. For the present study, however,  we restrict our analysis to LAMOST spectra of OB stars. We adopt the OB star catalog of \citet{LiuZ2019}, which contains a total of 16,032 OB stars from the fifth data release (DR5)\footnote{http://dr5.lamost.org} of the LAMOST Galactic surveys \citep{Deng2012, Zhao2012, Liu2014}. These OB stars are identified with line indices, and further inspected by eye, leading to a high purity of the sample \citep{LiuZ2019}. We also adopt the astrometric parallax measurements from \textit{Gaia} DR2 \citep{Brown2018}, following \citet{Leung2019} to correct for the zero-point offset as a function of G-band magnitude.  For the photometric input, we use the 2MASS $K_{\rm s}$-band measurements \citep{Skrutskie2006}.

\begin{figure*}
\centering
\includegraphics[width=160mm]{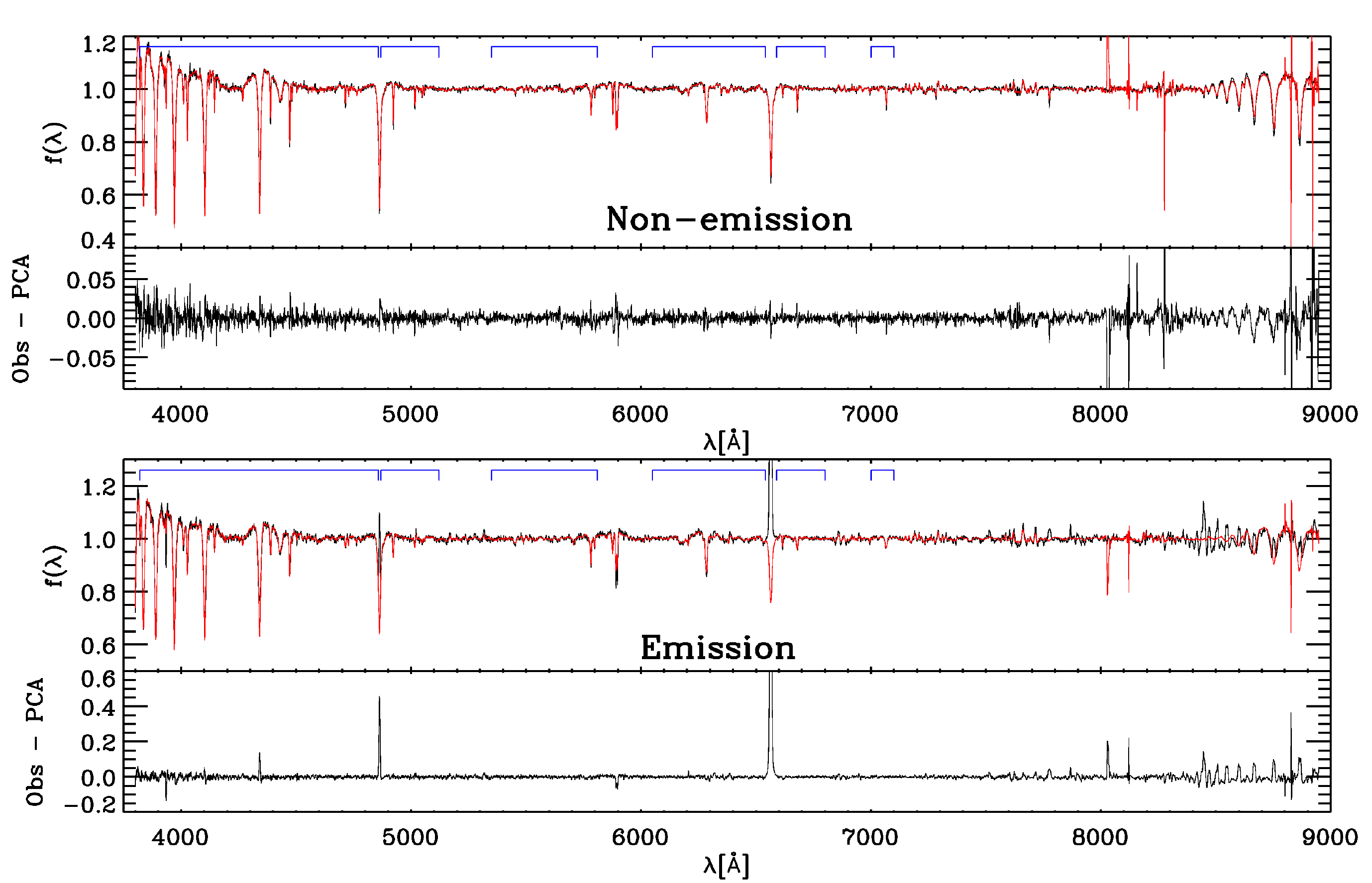}
\caption{
Two example spectra reconstructed with a PCA method for a non-emission (top) and an emission (bottom) star. In each panel, the black is the LAMOST spectrum, while the red is the PCA-reconstructed spectrum. Marked in blue are the clean wavelength windows with which we construct the principal component coefficients for the PCA reconstruction. Note that,  throughout this work, we normalize the spectra with a pseudo-continuum derived by smoothing the spectra with a Gaussian kernel of 50{\AA} in width, as such some pixels in the continuum-normalized spectra can exceed unity.}
\label{fig:Fig2}
\end{figure*}

\begin{figure}
\centering
\includegraphics[width=85mm]{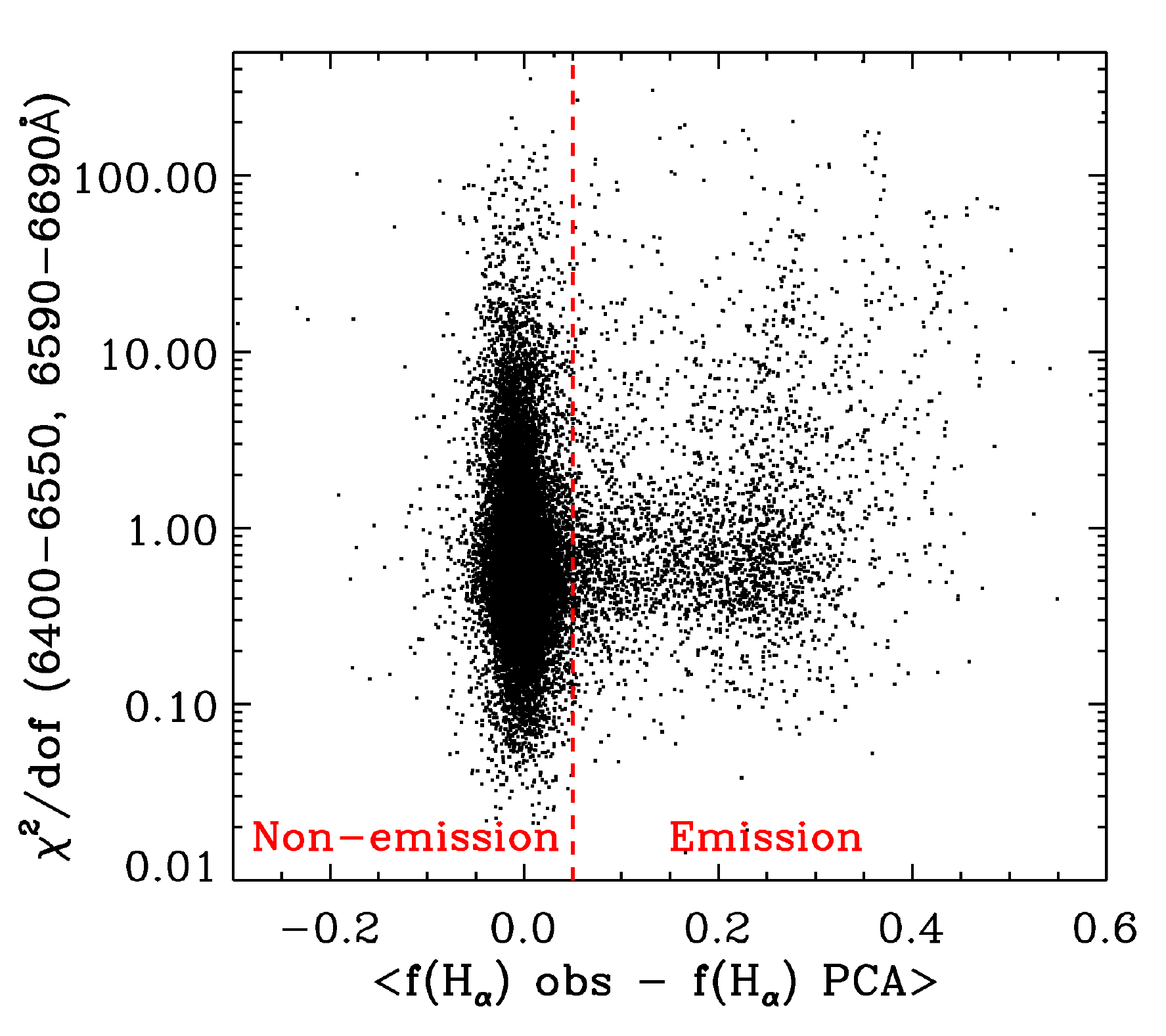}
\caption{Observed \Ha flux excess for emission spectra with respect to the PCA reconstruction. The horizontal axis shows the differences in mean \Ha fluxes in $\lambda$6571--6557{\AA} between the observed LAMOST spectra and the reconstructed PCA spectra. The vertical axis shows the  reduced $\chi^2$ across the wavelength range that encapsulates the \Ha features ($\lambda$6400--6550{\AA}, $\lambda$6590--6690{\AA}). Spectra with emission lines exhibit large difference in \Ha line between the LAMOST and PCA-reconstructed spectra, and are clearly separated from the majority of stars. The vertical dashed line delineates our criterion to select OB spectra with emission lines.}
\label{fig:Fig3}
\end{figure}

\section{Identifying emission lines with PCA reconstruction} \label{pca}
The spectra for a considerable fraction of OB stars can exhibit emission lines, either from their associated H\,{\sc ii} regions or from surrounding gas disks.  Since this emission does not necessarily correlate with the properties of the star, or its \mk, these emission lines can confound our estimation of \mk. We therefore opt to remove all objects with emission lines in the spectra from our main training set and devise a separate strategy to estimate \mk for these objects, using only the part of the spectrum free of emission lines.  Almost all of the spectra with emission lines are found to exhibit \Ha emission. Some of them also exhibit other Hydrogen lines in the Balmer series and Paschen series, as well as some metal lines. Nonetheless, in most cases, the \Ha emission line is the most prominent. As such, we adopt \Ha as a key diagnosis to identify OB stellar spectra with emission lines.

To automatically identify spectra with emission lines, we adopt a  principal component analysis (PCA) reconstruction method as described below. In a nut shell, we first define a set of clean wavelength windows that suffer minimal impact from emission lines with which we will determine the coefficient of the principal component of the spectrum. We then reconstruct the full spectrum using the coefficients as well as the full spectrum eigenbases derived from a sample of non-emission stars. We identify the stars with emission lines through the \Ha difference between the observed spectra and the PCA reconstruction counterparts. This process is implemented iteratively to obtain a sample without emission lines. In practice, we found that only one iteration is sufficient since further iterations will not make significant change to the results. 

Given a set of $M$ spectra $\bm{x}_i$ (with $i = 1, ..., M$) each containing $N$ pixels, PCA projects the spectra in spectral space to an eigenspace, where the eigenvalues represent the amount of variance held by each orthonormal eigenbasis. The eigenbases are obtained through diagonalizing the covariance matrix,
\begin{equation}
    {\bm C} = {\bm X}{\bm X}^{\rm T}, 
\end{equation}
\begin{equation}
{\bm C}{\bm \xi} = \lambda{\bm \xi},
\end{equation}
where ${\bm X}$ is the $N\times M$-dimensional array of spectra and $\lambda$ is the eigenvalue associated with the eigenbasis (eigenvector) ${\bm \xi}$. 
Entries in ${\bm X}$ are standardized by subtracting, from each pixel, the mean flux of the training set and then normalized by the standard deviation. The eigenvalues and eigenvectors are computed using the $trired.pro$ and $triql.pro$ scripts in {\sc IDL}. 

In practice, we first consider only pixels in the clean windows (as shown in Fig.~\ref{fig:Fig2}) and use those pixels to construct the matrix ${\bm X}$ of the training spectra. We adopt the matrix ${\bm X}$ to determine the principal components (eigenvectors/eigenbases) $\{{\bm \xi}\}$ in this restricted space. For any given training spectrum, ${\bm x}$, we calculate the principal component coefficients by projecting ${\bm x}$ onto each principal component ${\bm \xi}$. Since the eigenvectors are by definition normalized, the projection is simply the dot product of the two vectors, $
p = {\bm x} \cdot {\bm \xi}$. The collection of all principal coefficients ${\bm p}$ for all training spectra constitute a $M \times K$ matrix, ${\bm P}$, where $K$ is the number of principal components, and $M$ is the number of training spectra. Here we choose only the top $K=100$ principal components in order to denoise and omit irrelevant information in the spectra.

With matrix ${\bm P}$ in place, we then reconstruct the full spectra as follow. Let $\bar{\bm X}$ to be the corresponding full spectra matrix for the training spectra, we search for an array ${\bm B}$ such that ${\bm P} {\bm B} = \bar{\bm X}^T $. In other other words, we approximate the eigenbases for the full spectra with which $\bar{\bm X}$ shares the same principal component coefficients in the full spectral space as those from ${\bm X}$ in the restricted space. Practically, the matrix ${\bm B}$ is solved by inverting ${\bm P}$ with the Gram-Schmidt orthogonalization method. Let ${\bm P'}$ be principal coefficients of the test spectra ${\bm X'}$ determined by projecting ${\bm X'}$ onto the eigenvectors $\{{\bm \xi}\}$, we can then reconstruct the full test spectra via ${\bm X''}^T = {\bm P'} {\bm B}$. Here ${\bm X''}$ is the PCA-reconstructed spectra of the test spectra ${\bm X'}$.

Fig.\,\ref{fig:Fig2} shows that the full stellar spectra can be well-reconstructed using the PCA method. Objects with emission lines are identified using residuals between the PCA-reconstructions and the LAMOST spectra. Fig.\,\ref{fig:Fig3} shows the mean \Ha flux residual determined with our technique versus the reduced $\chi^2$ measured around the \Ha line.  A clear branch of stars with flux excess at the position of the \Ha line is present.  We deem a star has \Ha emission if the flux excess exceeds the red vertical dashed line as delineated in Fig.~\ref{fig:Fig3}.  Such a criterion identifies 2074 of the 16,002 unique LAMOST OB stars (13\%) as emission-line stars\footnote{We note that there could be multiple visits for the same star in the LAMOST database. For those stars, we adopt the results from the visit with the highest S/N.}. Fig.\,\ref{fig:Fig3} also shows a small fraction of stars with large $\chi^2$. These stars are found to have erroneous spectra due to multiple reasons, such as data artifacts or wrong wavelength calibration. Fig.\,\ref{fig:FigA1} in the Appendix shows a few examples of such erroneous spectra.

\section{extinction} \label{extinction}

As OB stars are mostly located in the Galactic disk with on-going star formation, they can suffer from serious interstellar extinction. Accurate correction for extinction is thus necessary to obtain accurate geometric \mk. Throughout this study, we refer {\it geometric} \mk to be the ``apparent'', distance-corrected, \mk inferred using {\it Gaia} parallax and 2MASS $K_{\rm s}$ apparent magnitudes. We note that extinction is an issue even we work with the infrared $K_{\rm s}$ band. A reddening $E_{B-V}$ of 1\,mag, which is not uncommon for distant OB stars, can cause a $\sim0.3$\,mag extinction in the $K_{\rm s}$ band \citep[e.g.][]{Yuan2013, Wang2019}, leading to a distance bias of 13\%. 

There are a variety of possible ways that we could obtain extinction for our OB star sample:  e.g., through direct use of an existing 3D reddening map \citep[e.g.][]{Green2019, Chen2019b}, the application of the Rayleigh-Jeans Color Excess method with infrared colors \citep[RJCE;][]{Majewski2011}, or deducing from the intrinsic colors of OB stars either empirically \citep[e.g.][]{Deng2020} or theoretically from synthetic models.  We adopt the latter in this study. 

Deriving the intrinsic color requires a robust estimation of the stellar parameters. To achieve that, we leverage the stellar parameters derived for our OB star sample via the implementation of the spectral fitting codes {\sc the Payne} \citep{Ting2019} in the hot star regime (Xiang et al. 2020, {\it in prep.}). Adopting \teff, \logg, and \feh  derived in this companion work, we estimate the intrinsic colors of the stars with the MIST isochrones \citep{Choi2016}. We have tested that using the PARSEC isochrones \citep{Bressan2012} or the empirical \teff--color relation of \citet{Deng2020} only incur a difference of $<0.03$\,mag for the $E_{B-V}$ estimate, which is negligible for the purposes of this study.

We estimate  $E_{B-V}$  through the observed color excesses in the 2MASS $J$ and $H$ magnitudes \citep{Skrutskie2006}, Gaia DR2 BP/RP \citep{Evans2018}, and the $g,r,i$ magnitudes from the XSTPS-GAC survey \citep[][]{Zhang2014, Liu2014}. For bright ($r<13$\,mag) stars that the XSTPS-GAC photometry saturates, we adopt the APASS $B,V,g,r,i$ photometric magnitdues \citep{Henden2012, Munari2014} instead.  To estimate $E_{B-V}$ (and subsequently $A_{Ks}$), we opt to avoid using $K_{\rm s}$-band photometry as it can be contaminated by emission from surrounding gas and dust disk. With all these $N$ photometry bands, we construct $N-1$ colors using only the photometry of adjacent bands. When compared to the intrinsic color, each observed color gives an estimate of $E_{B-V}$. We take the average of these $E_{B-V}$ estimates weighted by uncertainties of the colors, which are assumed to be the quadratic sum of the uncertainties of the two photometric bands. On top of that, the uncertainties in $E_{B-V}$ estimates are further derived through the propagation of uncertainties in photometric colors. Note that we have assigned a minimal uncertainty of 0.014\,mag in all the colors, including the Gaia $BP-RP$.  

To convert color excesses into $E_{B-V}$, we adopt extinction coefficients determined by convolving the public Kurucz model spectra \citep{Castelli2003, Kurucz2005} with the \citet{Fitzpatrick1999} extinction curve, assuming $R_V=3.1$. In principle, the extinction coefficient can vary according to the stellar parameters \citep[e.g.][]{Green2020}.  However, considering we only focus on the hot stars, for simplicity, we derive the extinction coefficients using a fixed Kurucz spectrum with $\teff=12,000$\,K, $\logg=4.5$, and $\feh=0$. 

\begin{figure}
\centering
\includegraphics[width=85mm]{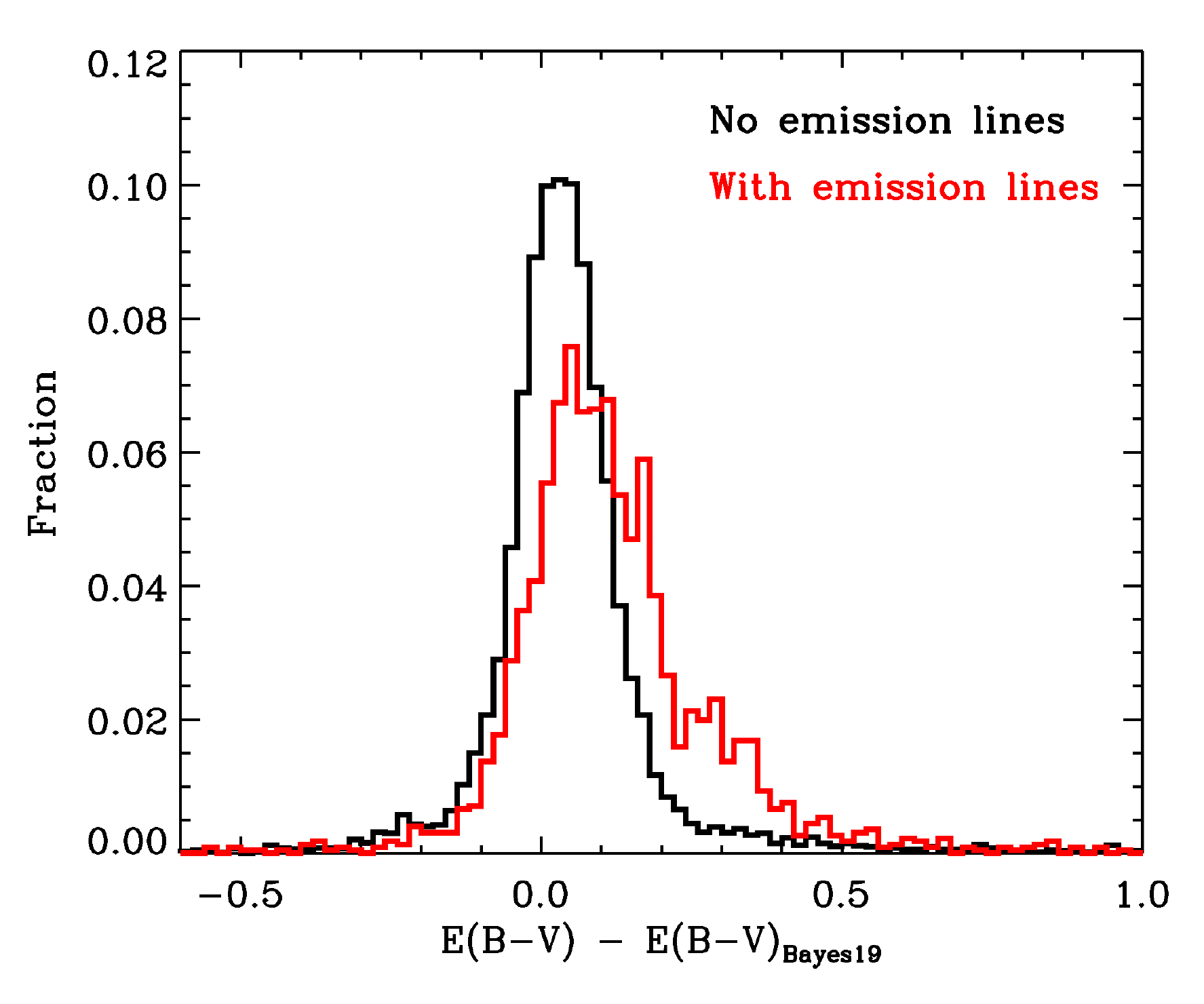}
\caption{Comparison of the $E_{B-V}$ derived in this work and the $E_{B-V}$ from the 3D map of \citet{Green2019}. We only consider stars with reliable \textit{Gaia} parallax measurements ($\varpi/\sigma_{\varpi}>5$). For stars without emission lines, the mean and dispersion of the $E_{B-V}$ difference are 0.04\,mag and 0.09\,mag, respectively, whereas for stars with emission lines in the spectra we find a shifted mean of 0.11\,mag and a standard deviation of 0.13\,mag. The larger $E_{B-V}$  for stars with emission lines indicates that these stars are more extincted by the associated H\,{\sc ii} regions or by their surrounding gas and dust.}
\label{fig:Fig4}
\end{figure}

Fig.\,\ref{fig:Fig4} shows a comparison of the derived $E_{B-V}$ with $E_{B-V}$ interpolated from the 3D map of \citet{Green2019}, using the Gaia distance from \citet{Bailer-Jones2018}. The differences for the majority of stars are smaller than 0.1\,mag, which corresponds to a negligible extinction difference of $\lesssim\,$0.03\,mag in $K_{\rm s}$-band. There are a small fraction of stars with large $E_{B-V}$ differences. Many of them turn out to be stars with emission lines. The large discrepancy could be caused by the fact that these stars are distant objects that are outside the feasible range of the \citet{Bailer-Jones2018} distance and/or the \citet{Green2019} reddening map. Alternatively, especially for stars with emission lines, they could suffer from additional extinction from the dense \Hii regions or surrounding disks. Due to the possibility of contamination in the $K_{\rm s}$ band flux by the surrounding environment, we caution about distance estimates from stars with emission line flag in this study.

\section{Data-driven \mk estimation} \label{nnmodel}
The absolute magnitude (luminosity) is an intrinsic astrophysical property of a star that is derivable from a stellar spectrum. It is related to the stellar parameters via the Stefan-Boltzmann equation 
\begin{equation}
  L = 4\pi R^2T_{\rm eff}^4,
\end{equation}
as well as the gravity equation
\begin{equation}
  g = \frac{GM}{R^2},
\end{equation}
where \teff, $R$, $M$, and $g$ are respectively the effective temperature, radius, mass and surface gravity of the star. Equations 3 and 4 together yield
\begin{equation}
  \log\frac{L}{L_\odot} = \log\frac{M}{M_\odot} + 4\log\frac{T_{\rm eff}}{T_{\rm eff, \odot}} - \log\frac{g}{g_\odot}
\end{equation}
Recall that \teff and \logg are basic stellar parameters that are derivable from stellar spectra. Furthermore, the stellar mass itself also implicitly depends on $T_{\rm eff}$, $\log g$, abundance [X/H] and rotation velocities, all of which are stellar properties that are readily measurable from stellar spectra. Consequently, one would expect that there exists an empirical relation which connects stellar spectra to the absolute magnitude of the stars. In this study, we will model this relation with a neural network.

\subsection{The neural-network model}
We consider a feed-forward multi-layer perceptron neural network model that maps the LAMOST spectra to the absolute magnitude \mk of the stars. Adopting the Einstein sum notation, our neural network contains two-layers and can be succinctly written as
\begin{equation}
  \mk = w \cdot \sigma\left(\tilde{w}_i\sigma\left(w_{{\lambda}i}f_\lambda + b_i\right)+\tilde{b}\right),
\end{equation}
where $\sigma$ is the Sigmoid activation function; $w$ and $b$ are weights and biases of the network to be optimized; the index $i$ denotes the neurons; and $\lambda$ denotes the wavelength pixels. We adopt 100 neurons for both layers. The training process is carried out with the $Pytorch$ package in {\em Python}. To reduce over-fitting in the training process, we also employ the dropout method \citep{srivastava14a} with a dropout parameter of 0.2. 

Considering the impact of emission lines, we set up two neural network models. One neural network is constructed for stars without emission lines, using the whole wavelength range except for $\lambda$5720--6050\AA, $\lambda$6270--6380\AA, $\lambda$6800--6990\AA, $\lambda$7100-7320\AA, $\lambda$7520--7740\AA, and $\lambda$8050--8350\AA; these wavelength windows contain prominent absorption bands of the earth atmospheres and strong {\rm Na}\,{\sc i D} absorption lines from interstellar medium. The other neural network is for stars with emission lines, using only wavelength windows that are devoid of strong emission lines as demonstrated in Fig.~\ref{fig:Fig2}. Finally, considering the neural network model is sensitive to spectral noise, we attempted to denoise spectra through the PCA reconstruction, using only the first 100 PCs. For non-emission stars, the PCA-reconstructed spectra are reconstructed using the full wavelength range; while for emission stars, the PCA-reconstructed spectra are reconstructed using only the clean wavelength windows. 

\subsection{The training and test set} \label{trainingset}
We define two training sets that correspond to the two neural network models set up above, for stars with or without emission lines.  We also define a test set that we use to verify the $\mk$ estimation.

The training and test sets adopt stars with good parallax measurements ($\varpi/\sigma_{\varpi} > 10$).  To derive a more robust empirical relation, we require the training stars to have a spectral S/N (per pixel) higher than 50. Stars that meet these requirements are divided into two groups: 4/5 of them are adopted as the training set to train the neural network model, while the remaining 1/5 constitute the test set, in combination with stars with good parallaxes but lower spectral S/N. Roughly 200 stars that are not in the \citet{LiuZ2019} sample but have $M_{K{\rm s}}<-1.5$\,mag are also added to the training set. The inclusion of these stars is to enlarge the sample size at the brighter end where there is limited number of training stars. Although not in the original LAMOST OB stars catalog, all these additional training stars have $T_{\rm eff}>7000$\,K according to the LAMOST DR5 stellar parameter catalog of \citet{Xiang2019}. Their spectra are further manually inspected to ensure they are early-type stars. About half of them are OB stars that exhibit He absorption lines, while the others are likely late-B or early A-type stars. In total, we obtain 6861 stars for the training set for stars without emission lines and 7161 stars for training set for stars with emission lines.  

Since the geometric ``apparent'' \mk for binaries are biased, the training process is iterated to exclude the binaries. Binaries are singled out based on significant deviation between the predicted \mk and the geometric \mk  (see Section\,\ref{binaryeffect} for details).  In practice, two  iterations are implemented, as we find negligible change in the resultant \mk estimates after these iterations.  

\begin{figure*}
\centering
\includegraphics[width=160mm]{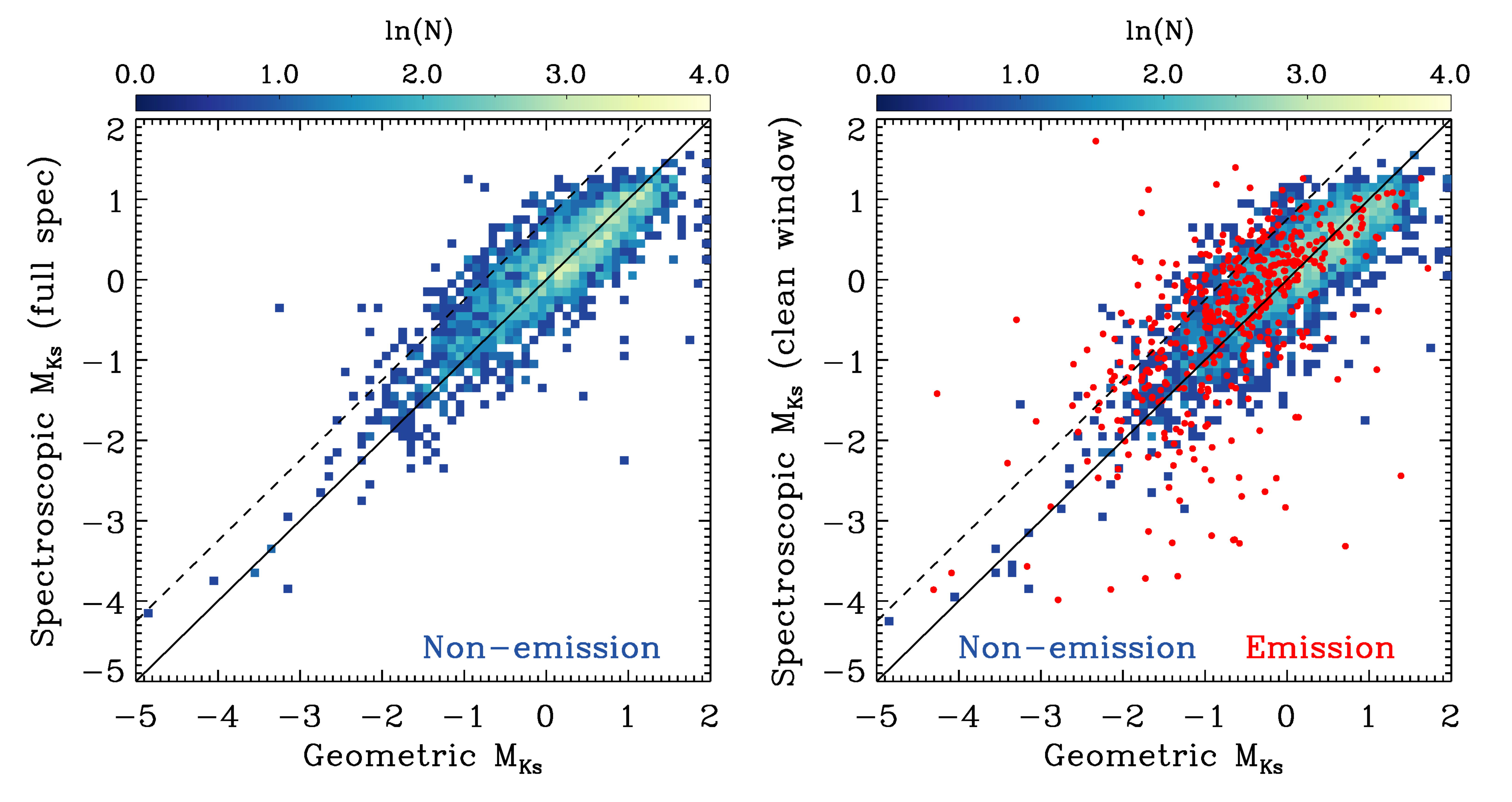}
\caption{Comparison between the geometric \mk inferred from \textit{Gaia} parallaxes and the spectroscopic \mk derived from LAMOST spectra for test stars. We only show results for test stars that have precise \textit{Gaia} parallax measurements ($\varpi/\sigma_{\varpi} > 10$). The left panel demonstrates the spectroscopic \mk derived from the full spectrum for stars without emission lines. The right panel illustrates the spectroscopic \mk derived using only wavelength pixels that are devoid of emission lines for both non-emission stars (blue/green background) and emission line stars (red dots). In both panels, the solid line delineates the 1:1 line. The dashed line shows an offset of 0.75\,mag from the 1:1 line, the offset one would expect from equal mass binaries. Stars that are close to the dashed line have geometric\mk brighter than the spectroscopic \mk, signaling the possibility of binaries or multiple systems.}
\label{fig:Fig5}
\end{figure*}

Fig.\,\ref{fig:Fig5} shows the comparison of the resultant spectroscopic \mk estimates with the geometric \mk for the test stars with ${\rm S/N}>20$. The Figure shows good overall consistency between the two sets of \mk estimates across the range roughly from $-4$\,mag to 1.5\,mag.  Below $\mk\sim0$\,mag, more stars have fainter spectroscopic \mk than the geometric \mk. This is due to a contribution to the geometric \mk from binaries at these magnitudes. Recall that, the geometric \mk is derived from the apparent magnitudes, where both stars contribute.

We note that there are some subdwarfs and white dwarfs in the LAMOST OB star sample. They typically have geometric \mk fainter than 2\,mag and are not shown in the figure. For these stars, our spectroscopic \mk estimates, which are trained primarily on normal OB star spectra, can be problematic. The spectroscopic \mk for stars fainter than 2\,mag should therefore be used with caution.

\begin{figure}
\centering
\includegraphics[width=85mm]{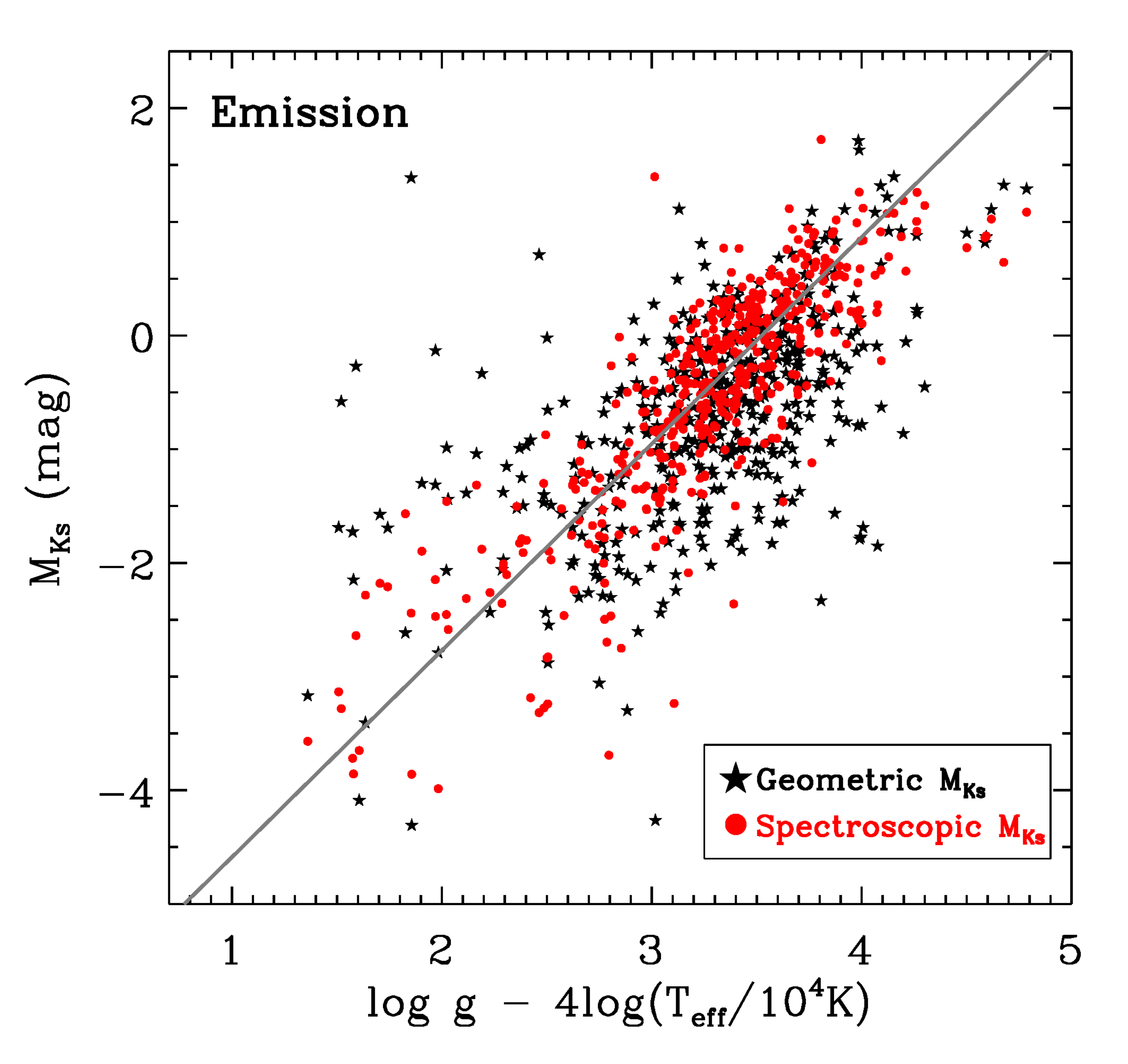}
\caption{Temperature-weighted gravity versus \mk for test stars with emission lines. The temperature-weighted gravity of a star is an indicator of its bolometric luminosity.  Spectroscopic \mk measurements are shown with dot symbols while geometric \mk are shown with star symbols. The solid line indicates the best-fit linear relation from the sample of non-emission stars.}
\label{fig:Fig6}
\end{figure}

Fig.\,\ref{fig:Fig5} also demonstrates that the scatter between the spectroscopic \mk and the geometric \mk is larger for stars with emission lines than for the non-emission stars. Particularly, there are a number of emission line stars that exhibit spectroscopic \mk much brighter than their geometric \mk. Peculiarly, we found that both spectroscopic \mk and geometric \mk of these stars appear to be robust. On the one hand, as illustrated in Fig.\,\ref{fig:Fig6}, the spectroscopic \mk for these stars is consistent with the luminosity predicted by the temperature-weighted gravity \citep{Kudritzki2020}. On the other hand, for about half of these stars, their \textit{Gaia} re-normalized unit weight error (RUWE) \footnote{A detail explanation on the RUWE can be found in the public DPAC document from Lindegren L., titled "Re-normalising the astrometric chi-square in \textit{Gaia} DR2", via \url{https://www.cosmos.esa.int/web/gaia/public-dpac-documents} (at the bottom of the page)} values are around 1.0, suggesting that at least half of these outliers have decent astrometry measurements. Investigating the nature of these stars is beyond the scope of this study, but one possibility is that they might be stripped stars as a consequence of binary evolution. As a result of the stripping, they are in reality fainter (as probed by the geometric \mk) while exhibiting similar spectra to a main-sequence or subgiant star, and hence a brighter spectroscopic \mk (see also Section\,\ref{lb1}).

\begin{figure}
\centering
\includegraphics[width=85mm]{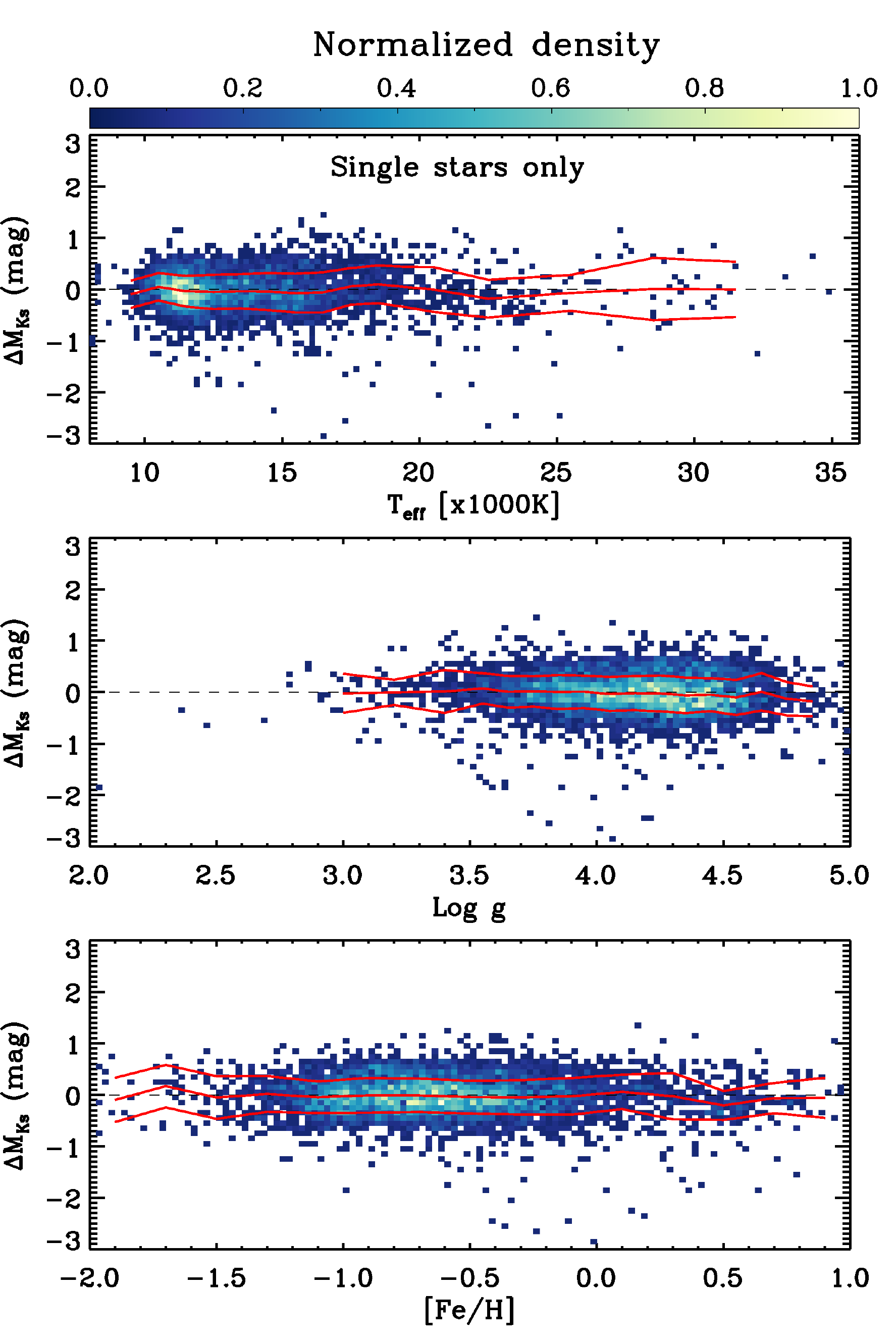}
\caption{Differences between spectroscopic \mk and geometric \mk as a function of stellar parameters. Only results for single stars are shown. The solid lines delineate the median and standard deviation as a function of stellar parameters.}
\label{fig:Fig7}
\end{figure}

Fig.\,\ref{fig:Fig7} illustrates the difference between the spectroscopic \mk and the geometric \mk for the test stars as a function of stellar parameters \teff, \logg and \feh. The stellar parameters were derived from LAMOST spectra in a parallel work (Xiang et al. {\it in prep.}). Only results from single stars without emission lines are shown. This figure demonstrates that our spectroscopic \mk estimates do not exhibit bias with respect to stellar parameters.

\begin{figure*}
\centering
\includegraphics[width=160mm]{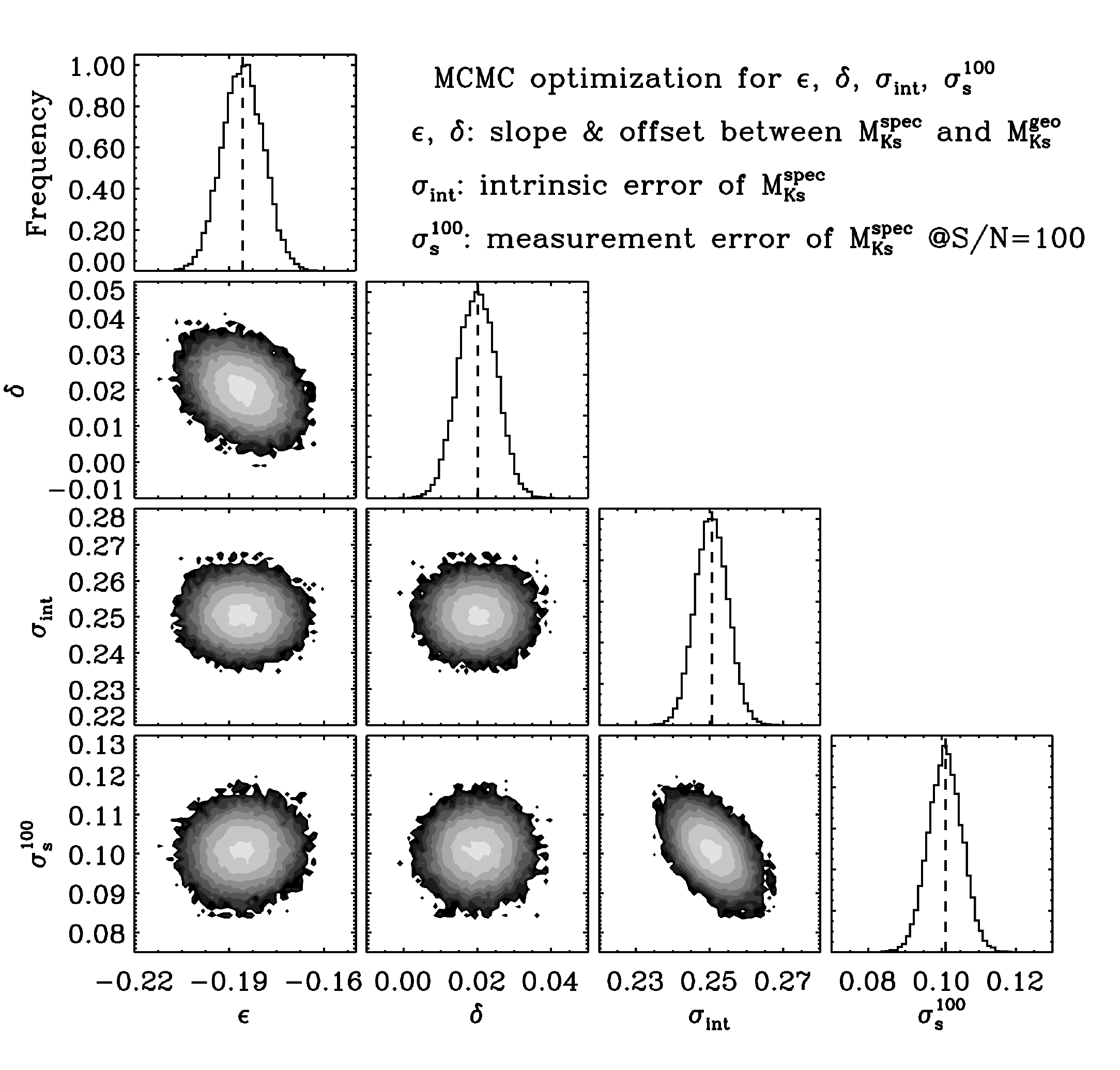}
\caption{MCMC fitting of the intrinsic uncertainty and measurement uncertainty in our \mk estimates. Binaries were excluded iteratively. The intrinsic uncertainty $\sigma_{\rm int}$ indicates the epistemic uncertainty in \mk estimates, whereas the measurement uncertainty $\sigma_s$ quantifies the aleatoric uncertainty induced by spectral noise. The MCMC results show an intrinsic uncertainty of $\sigma_{\rm int} =\,$ 0.25\,mag and a typical measurement uncertainty of $\sigma_s = \,$0.10\,mag at S/N$\,=\,$100. } 
\label{fig:Fig8}
\end{figure*}

\begin{figure}
\centering
\includegraphics[width=85mm]{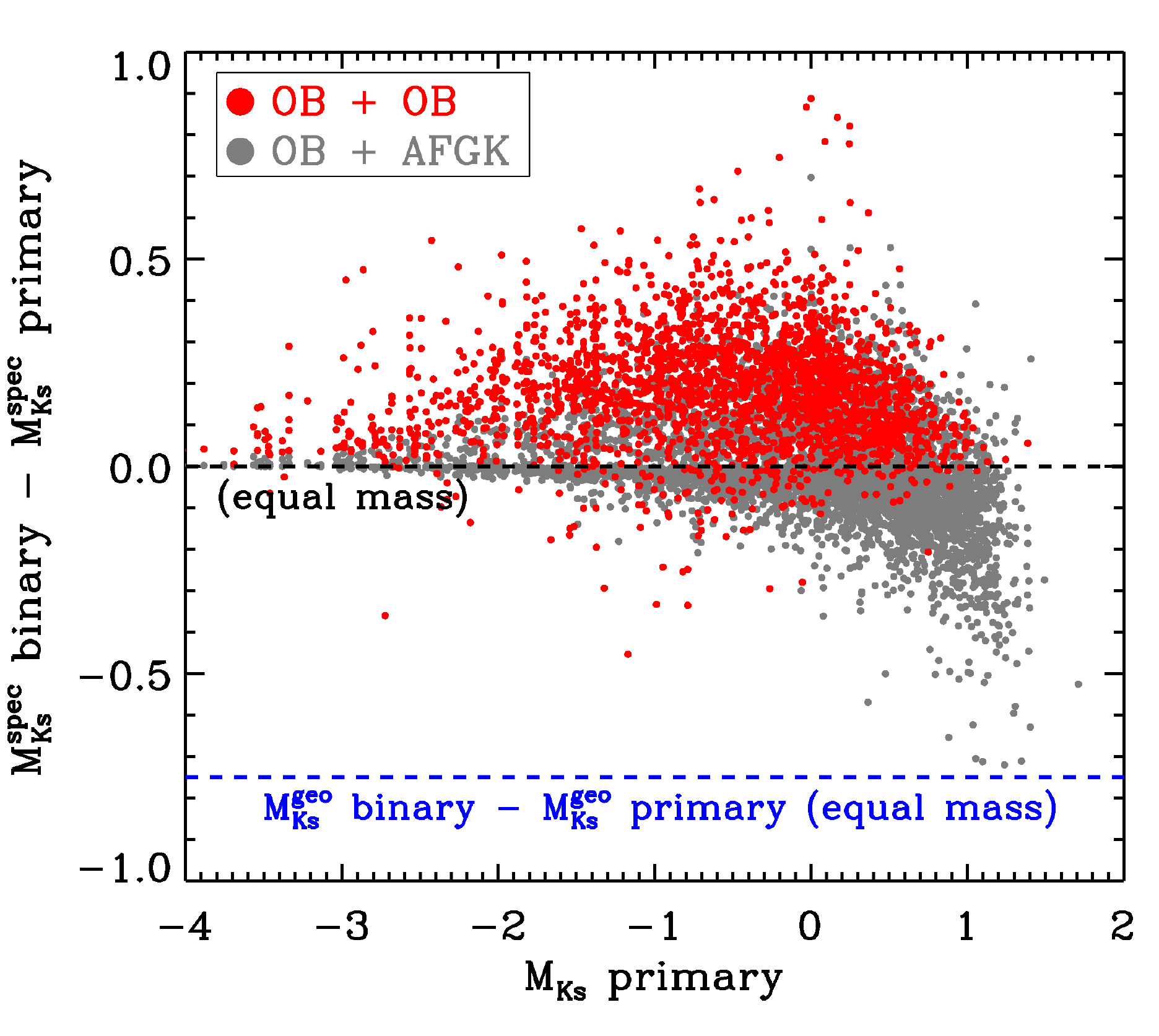}
\caption{
Differences between spectroscopic \mk estimates derived from mock composite binary spectra versus that for the primaries of the composites. Red symbols show OB+OB binary systems and the grey symbols are binary systems composed of an OB star and a companion with another (AFGK) spectral type. In most cases, the composite spectra cause the inferred spectroscopic \mk to be slightly fainter than the primary, further facilitating the identification of binaries through the difference between geometric \mk and spectroscopic \mk. For example, the geometric \mk for equal-mass binaries are 0.75\,mag brighter than their primaries (dashed line in blue), due to the contribution from the secondary. We note that for some binaries composed of a late-B type primary with $\mk\gtrsim0.5$\,mag and an AFGK type secondary, the spectroscopic \mk could be brighter than that of the primary. As such, the binary identification in this regime is less efficient (see text for details).}
\label{fig:Fig9}
\end{figure}

\subsection{Measurement uncertainty and intrinsic uncertainty} \label{intuncertainty}

In this section, we evaluate the quality of our spectroscopic \mk estimates. The nature of the uncertainty can be aleatoric or epistemic. For the former, the uncertainty in \mk is caused by uncertainties in the spectra. The latter can be caused by multiple sources. For example, the spectra simply do not contain the full information of the luminosity of the stars. On top of that, our data-driven models might also be suboptimal in extracting such information. In the following, we will quantify both the aleatoric and epistemic uncertainties using the test set. 

To make a complete accounting of the uncertainties in our \mk estimates requires careful characterization of both the contribution from uncertainties in the geometric \mk and the additional scatter raised as a result of unresolved binaries. In order to minimize the impact of binary stars, we calculate the measurement uncertainty and intrinsic uncertainty with an iterative approach. In each iteration, we identify and discard all likely binaries that have more than $2\sigma$ difference between their spectroscopic \mk and geometric \mk. 

The measurement uncertainty and intrinsic uncertainty are estimated in a Bayesian framework. In the following, we will denote the ground truth geometric absolute magnitude of each star as $M^g$ and the ground truth spectroscopic absolute magnitude as $M^s$. We assume that there is a linear relation between the expected $mean$ spectroscopic absolute magnitude $\bar{M}^s$ and $M^g$, with a slope close to 1, with a correction term $\varepsilon$,  and an intercept of $\delta$, 
\begin{equation}
\bar{M}^s = (1+\varepsilon)M^g + \delta.
\end{equation}
We further assume that, for a given $\bar{M}^s$, due to epistemic uncertainty, the $M^s$ of the individual stars are distributed as a Gaussian with an intrinsic uncertainty of $\sigma_{\rm int}$, i.e.
\begin{equation}
P(M^s|\bar{M}^s,\sigma_{\rm int}) = \mathcal{G}(\bar{M}^s, \sigma_{\rm int}). 
\end{equation}
The estimated absolute magnitudes $M_o^s$ are themselves assumed to distribute around a given $M^s$ as a Gaussian distribution with width set by the aleatoric measurement uncertainty $\sigma_s$, 
\begin{equation}
P(M_o^s|M^s,\sigma_s) = \mathcal{G}(M^s, \sigma_s). 
\end{equation}
For simplicity, we assume that the aleatoric measurement uncertainty depends only on, and scale linearly with, the spectral S/N. In particular, we have
\begin{equation}
\sigma_s = \sigma_s^{100}/[({\rm S/N})/100],
\end{equation}
where $\sigma_s^{100}$ is the uncertainty for spectra with ${\rm S/N}=100$.

Finally, for the geometric absolute magnitude, we assume a flat prior on the $M^g$. We can deduce that
\begin{equation}
P(M^g|m_0,  \delta_{m_0}, \varpi, \delta \varpi) = 
\mathcal{G}(m_0+5\log\varpi-10 - M^g, \sigma_{\varpi,m_0}),
\end{equation}
\begin{equation}
\sigma_{\varpi,m_0}\equiv\sqrt{\bigg(\frac{5}{\ln{10}}\frac{\delta\varpi}{\varpi}\bigg)^2+\delta{m_0}^2},
\end{equation}
where $\varpi$ and $\delta\varpi$ are the \textit{Gaia} parallax and its uncertainty in unit of {\em mas}. $m_0$ and $\delta{m_0}$ are the dereddened apparent magnitude and its uncertainty, respectively. The latter is computed for each star as the quadratic sum of the uncertainties in the photometric magnitude and the extinction estimate. 

Combining all these ingredients, we arrived at the final posterior probability distribution for the parameters {$\varepsilon$, $\delta$,  $\sigma_{\rm int}$, $\sigma_s^{100}$},
\begin{equation}
\begin{aligned}
&P(\varepsilon, \delta, \sigma_{\rm int}, \sigma_s^{100}\vert\{{\mathbf M}_{o,i}^s, (\mathrm{S/N})_i, {\mathbf m}_{0,i}, {\mathbf \varpi}_i, \mathbf{\delta}_{i,m_0}, \delta \varpi_i \}) = \\
&\left(\prod_{i=1}^{N}P({\mathbf M}_{o,i}^s, ({\mathrm {S/N}})_i, {\mathbf m}_{0,i}, {\mathbf \varpi}_i, \vert \varepsilon, \delta, \sigma_{\rm int}, \sigma_s^{100}, \mathbf{\delta}_{i,m_0}, \delta \varpi_i)\right)\cdot\\
&P_{prior}(\varepsilon, \delta, \sigma_{\rm int}, \sigma_s^{100}),
\end{aligned}
\end{equation}
where $N$ is the number of stars, and the likelihood reads 
\begin{equation}
\begin{aligned}
&P({\mathbf M}_{o,i}^s, ({\mathrm {S/N}})_i, {\mathbf m}_{0,i}, {\mathbf \varpi}_i, \vert \varepsilon, \delta, \sigma_{\rm int}, \sigma_s^{100}, \mathbf{\delta}_{i,m_0}, \delta \varpi_i)=\\
&\mathcal{G}\Bigg({\mathbf M}_{o,i}^s-[(1+\varepsilon)({\mathbf m}_{0,i}+5\log{\mathbf \varpi}_i-10)+\delta],\\
&\sqrt{\sigma_{\rm int}^2+\left(\frac{\sigma_s^{100}}{\rm ({\mathbf {S/N}})_i/100}\right)^2+\left(\frac{5}{\ln{10}}\frac{\delta{\mathbf \varpi}_i}{{\mathbf \varpi}_i}\right)^2 + \delta_{i,m_0}^2} \Bigg)
\end{aligned}
\end{equation}
In this work, we adopt a flat prior for all the parameters and sample the posterior with Markov Chain Monte Carlo (MCMC). 

Fig.\,\ref{fig:Fig8} displays the results of the MCMC fitting to non-emission single stars. The mean of posterior suggests an aleatoric measurement uncertainty of $\sigma_s^{100}=\,$0.10\,mag for spectra with S/N = 100 and an epistemic intrinsic uncertainty $\sigma_{\rm int}=\,$0.25\,mag.  We find a moderate slope correction of  $\varepsilon = -0.18$ which is likely due to the lingering presence of binaries, considering that our $2\sigma$ cut can leave a considerable number of binaries with \mk excess smaller than $2\sigma$ ($\lesssim0.5$\,mag) in the sample. The presence of lingering binaries also implies that the intrinsic uncertainty from the MCMC posterior is a conservative estimate as binaries can contribute to part of the scatter.

\section{binary identification} \label{binaryeffect}
Binaries are ubiquitous and play important roles in astrophysics \citep{Abt1983, Duchene2013, Moe2017}. In star clusters, where all member stars share the same distance and the same age, binary stars in the lower main sequence are recognizable as they are brighter than all other single stars that distribute along a well-described locus in the color-magnitude diagram \citep[e.g.][]{Hurley1998, Kouwenhoven2005, Li2013}. The identification of binary stars in the field is more complicated due to the mixture of multiple stellar populations.  There are a number of tailored approaches that have been implemented for binary identification in the field, such as interferometry \citep[e.g.][]{Raghavan2010}, eclipsing transits \citep[e.g.][]{Qian2017, Zhang2017, LiuN2018, Yang2020}, color displacements \citep[e.g.][]{Pourbaix2004, Yuan2015b}, astrometric noise excess \citep[e.g.][]{Kervella2019, Penoyre2020, Belokurov2020}, common phase space motion  \citep[e.g.][]{Andrews2017, Oh2017, Coronado2018b, El-Badry2018c, Hollands2018}, radial velocity variations \citep[e.g.][]{Matijevic2011, Gao2014, Gao2017, Price-Whelan2017, Badenes2018, Tian2018, Tian2020}, spectroscopic binaries with double lines \citep[e.g.][]{Fernandez2017, Merle2017, Traven2017, Skinner2018, Traven2020} and full spectral fitting \citep{El-Badry2018b, Traven2020} 

The various methods listed above have been extensively employed to identity and characterize binaries from large surveys. \citet{Belokurov2020} demonstrated that \textit{Gaia} RUWE is an efficient tool for identifying unresolved, short-period binaries with low-to-intermediate mass ratios. Short-period binaries have also been characterized through their double-line spectra from high-resolution spectroscopic surveys \citep[e.g.][]{Fernandez2017, Merle2017, Traven2020}, or through radial velocity variations with both high- and low-resolution surveys \citep[e.g.][]{Gao2017, Badenes2018, Tian2018}. Unresolved binaries with longer period, which exhibit single lines in their spectra, are generally harder to identify, but not impossible. Based on full spectral fitting technique, \citet{El-Badry2018b} have characterized thousands of main-sequence binaries from the APOGEE spectra, many of them long-period binaries.

Nonetheless, for long-period binaries, the method presented in \citet{El-Badry2018b} is only mostly effective for systems with intermediate mass ratios ($0.4\lesssim q\lesssim0.85$; where $q=m_2/m_1$). It remains a challenge to identify unresolved single-line binaries with higher mass ratios ($q\gtrsim0.85$). In this regime, identifying binaries via the binary sequence in the HR diagram for cool stars with $\teff\lesssim5200$\,K is possible \citep[e.g.][]{Babusiaux2018, Coronado2018, LiuC2019}. While this method is not as applicable to the hotter ($\gtrsim5200$\,K) stars or giants due to the larger intrinsic variation of luminosity (at a given \teff).

Here we present a method of binary identification that tackle this challenging regime (long-period binaries with hot stars), leveraging differences between spectroscopic \mk and geometric \mk, or analogously, differences between the spectro-photometric parallaxes deduced from the spectroscopic \mk and the parallaxes determined with \textit{Gaia}. The method is particularly efficient for identifying single-line binaries with large mass ratios, e.g., binaries with equal-mass components, thus serves as a complementary among the aforementioned approaches. A brief introduction on the philosophy of the method has been laid out and applied to LAMOST AFGK stars in \citet{Xiang2019}. Here we present a more detailed description based on the application to LAMOST OB stars. 

The basic idea is that, since the observed apparent magnitude of an unresolved binary/multiple star system is brighter than any individual star in the system\footnote{In general, the light centroids of binaries exhibit only little variation, resulting in only minor systematics in the astrometry, especially for binaries with similar mass companions.}, we should expect the geometric \mk for binary systems to be brighter than the spectroscopic \mk. This is possible because, while geometric \mk reflects the contributions from both stars faithfully, the spectroscopic \mk mostly reflect the dominant star in the system. To demonstrate the latter, we build an empirical library of mock binary spectra using the LAMOST spectra for single stars.  The mock test allows us to compare, and measure the differences between the spectroscopic \mk derived from composites and that from the individual components.  

\begin{figure}
\centering
\includegraphics[width=85mm]{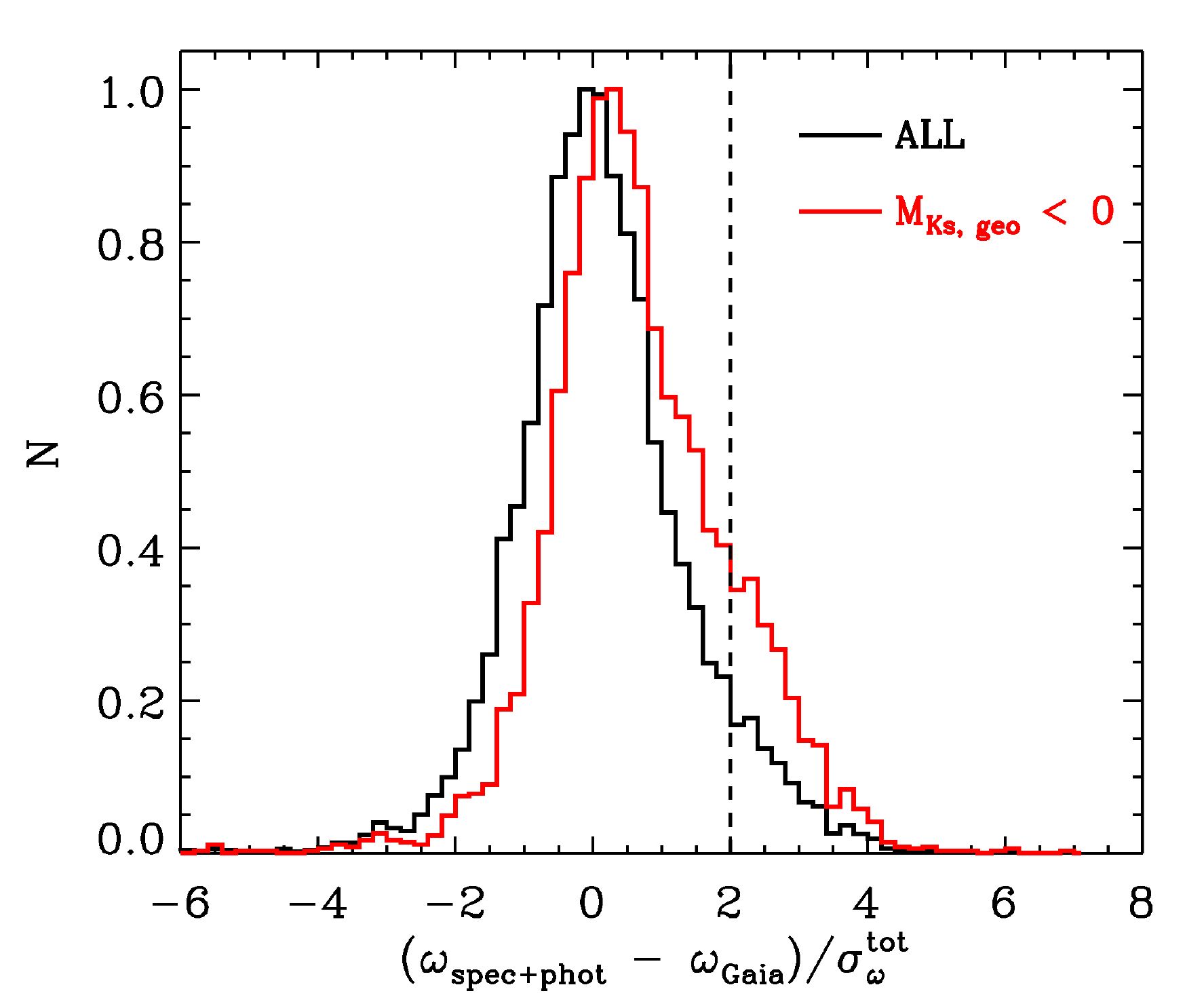}
\caption{The normalized differences between the inferred spectro-photometric parallax and the \textit{Gaia} parallax for individual LAMOST OB stars. Only the results from stars with robust \textit{Gaia} parallaxes ($\varpi/\sigma_{\varpi}>10$) and decent spectral quality (${\rm S/N}>50$) are shown. The histogram in black shows results from stars of all geometric \mk while the red one shows only results for stars with geometric $\mk<0$. Our method is more effective for finding binaries for the latter. The vertical dashed line delineates the criterion adopted for binary identification.} 
\label{fig:Fig10}
\end{figure}

\begin{figure*}
\centering
\includegraphics[width=160mm]{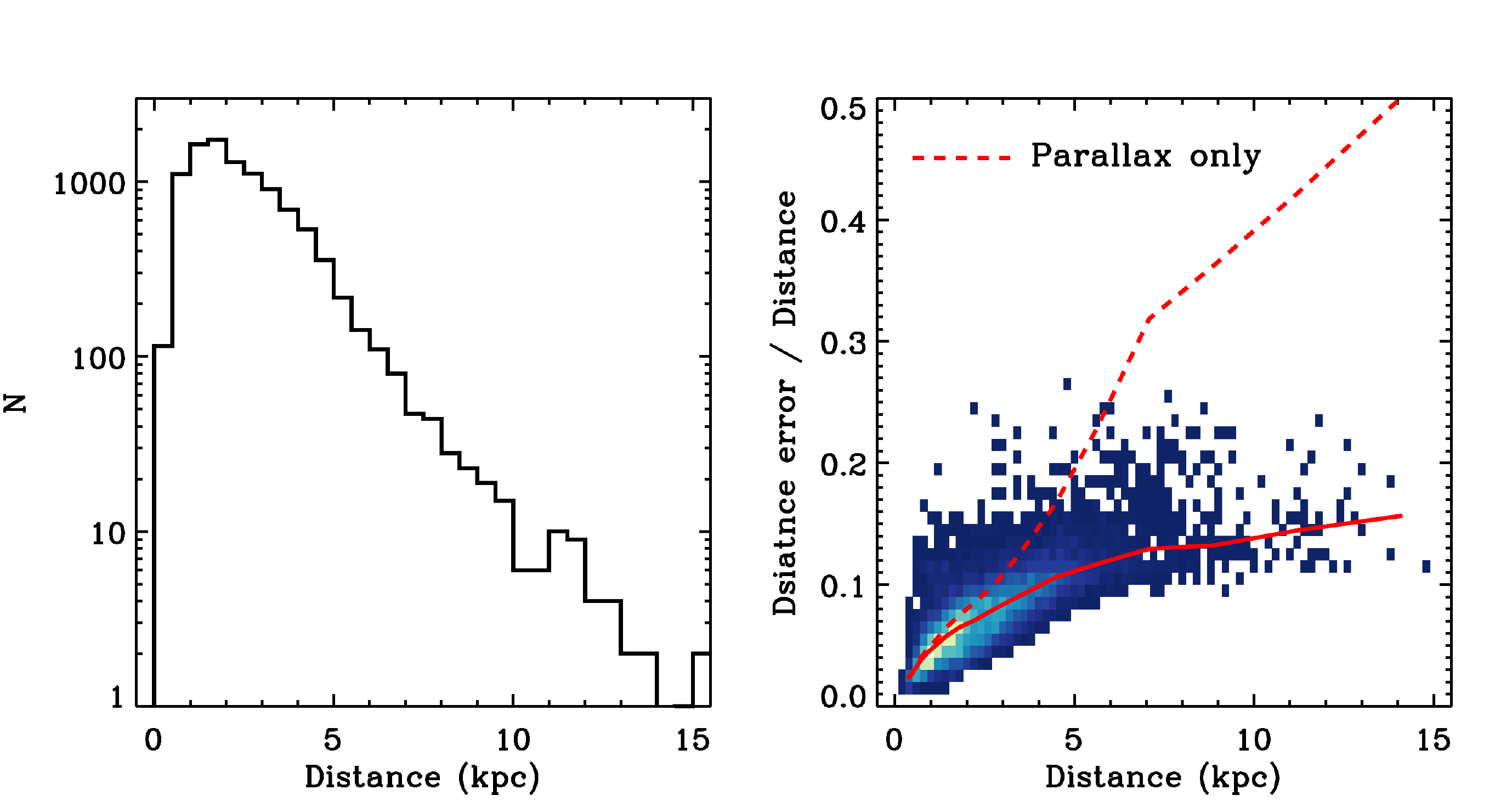}
\caption{Left: Distribution of Heliocentric distance of the LAMOST OB star sample. Right: Relative distance  uncertainty ($\sigma_d/d=\ln{10}\times\sigma_{\log{d}}$) as a function of distance. Only results for single stars are shown. The solid line delineates the median value of the relative distance uncertainty at various distances. The dashed line delineates the relative distance uncertainty in the case that only the \textit{Gaia} parallax is adopted to infer the distance. The improvement to distance estimates using the spectroscopic \mk is clearly visible for distant stars. }
\label{fig:Fig11}
\end{figure*}

To generate the composite spectra, we scale the fluxes of the LAMOST spectra to the same distance. We restrict our spectral library to stars with robust \textit{Gaia} parallax ($\varpi/\sigma_{\varpi}>10$) and spectral ${\rm S/N}>100$, to insure high data quality. Accurate spectral flux calibration is necessary to generate realistic composite spectra. For this purpose, we adopt the LAMOST spectra deduced with the flux calibration method of \citet{Xiang2015}. The calibrated spectral SEDs have a relative precision of $\sim$\,10\% in the wavelength range of $\lambda$4000--9000\AA. We ensure that, for spectra in our library, the scaled fluxes are consistent with the photometry in all individual  passbands, including the \textit{Gaia} G-band, the XSTPS-GAC and APASS $g$, $r$ and $i-$band. Each of the spectra are further de-reddened using the reddening estimates in Section\,\ref{extinction}, assuming the extinction curve from \citet{Fitzpatrick1999}. We assemble a set of binaries by taking an OB star for the primary star and a star of any (O/B/A/F/G/K) spectral type for the secondary. Fig.\,\ref{fig:FigA2} in the Appendix shows a few examples of the composite spectra.  

Fig.\,\ref{fig:Fig9} shows the difference between the \mk derived from the composite binary spectra by simply treating them as single-star spectra and the \mk derived from the spectra of the primaries in these composites. In most cases, the spectroscopic \mk for the binary is comparable to that of the single, primary component.  As expected, the spectroscopic \mk for equal-mass binary systems are identical to those of the primary component; the normalized spectra of the two component stars are identical. A similar result also applies for binary systems with small mass ratios. In this case, the secondary contributes minimally to the spectrum. Our mock test suggests that their spectroscopic \mk are fainter than that of the primary (by $\sim0.2$\,mag on average). Note that this is consistent with the findings of \citet{El-Badry2018}, which suggest that, for AFGK stars in binaries, the binary spectrum yields a larger \logg than the single star. In any case, this effect facilitates the identification of binaries, as the difference between the geometric \mk and spectroscopic \mk would be even larger. In short, our experiment concludes that, unlike geometric \mk, spectroscopic \mk mostly reflects the contribution from the dominant stars, regardless of the mass ratio of the binary systems.

Nonetheless, we note that for some binary systems composed of an OB-type primary star with $\mk\gtrsim0.5$\,mag and an A/F/G/K-type secondary star, our spectroscopic \mk estimates from the composite spectra could be brighter than the primary by more than $0.2$\,mag. This is particularly common for systems with a late B-type primary. For these systems, the Balmer lines of the composite are shallower than the primary (Fig.\,\ref{fig:FigA2}). The neural network model predicts a brighter \mk because the model is trained on single OB stars, for which the strength of Balmer lines decreases with increasing temperature, and hence a brighter \mk. 

To identify binary stars, we compute the spectro-photometric parallax (in {\rm mas})
\begin{equation}
\varpi_s = 10^{-0.2(m_{K_{\rm s}}-\mk-10-A_{K_{\rm s}})}, 
\end{equation}
and derive the S/N of the parallax excess of $\varpi_s$ with respect to the \textit{Gaia} astrometric parallax $\varpi$, 
\begin{equation}
S/N_{\Delta\varpi} = \frac{\varpi_s - \varpi}{\sqrt{\delta\varpi_s^2 + \delta\varpi^2}}, 
\end{equation}
where $\delta\varpi_s$ and $\delta\varpi$ are the measurement uncertainty of the spectro-photometric parallax and the \textit{Gaia} parallax, respectively. 
The $\delta\varpi_s$ is defined via
\begin{equation}
\frac{\delta\varpi_s}{\varpi_s} = 0.2\ln10\sqrt{\sigma_s^2+\sigma_{\rm int}^2+\sigma_{m_{Ks}}^2+\sigma_{A_{Ks}}^2}, 
\end{equation}
where $\sigma_s$ and $\sigma_{\rm int}^2$ are the (S/N-dependent) measurement uncertainty and the intrinsic uncertainty, derived in Section\,\ref{intuncertainty}; $\sigma_{m_{Ks}}$ is the photometric uncertainty for the 2MASS $K_{\rm s}$ magnitude, and $\sigma_{A_{Ks}}$ the uncertainty of the extinction estimate. We adopt a 2$\sigma$ criterion, i.e., we assign a star to be a binary if ${\rm S/N}_{\Delta\varpi}>2$. In total, 1597 of the 16,002 LAMOST OB stars in our sample (10.0\%) are identified as binary stars, 13,257 (82.8\%) are marked as single stars, and 1148 (7.2\%) stars are unclassified due to a lack of either a \textit{Gaia} parallax or 2MASS photometric magnitudes.

Fig.\,\ref{fig:Fig10} shows the differences between spectro-photometric parallax and \textit{Gaia} parallax for individual LAMOST OB stars. We show results from stars with robust \textit{Gaia} parallaxes ($\varpi/\sigma_{\varpi}>10$) and decent spectral quality (${\rm S/N}>50$).  Especially among stars with $\mk<0$\,mag, there is a clear positive tail, contributed by binary/multiple star systems. These stars are over-luminous compared to their single stars counterpart. 

Most of the binary stars selected with our method have a small RUWE value ($\sim$1). This partly reflects that our method is efficient for identifying binaries with large mass ratios, especially equal-mass binaries, for which the Gaia RUWE value is small due to negligible wobbles of the light centroids. Viewed in this way, our method complements approaches which identify binaries through large astrometric wobbles quantified by the Gaia RUWE value \citep[e.g.][]{Belokurov2020}.

Nonetheless, a few caveats apply. As discussed above, many binaries with late B-type primaries (with $\mk\gtrsim0.5$\,mag) can be missed. This is also illustrated in Fig.~\ref{fig:Fig10}, where the positive tail diminishes as we consider the full sample.  The quality of the \textit{Gaia} parallax is another limit to the effectiveness of this method. Since our method relies on the comparison between the spectro-photometric parallax and the {\it Gaia} astrometric parallax, the results are less robust in recognizing distant binary systems that have larger Gaia parallax uncertainty.  

\begin{figure}
\centering
\includegraphics[width=85mm]{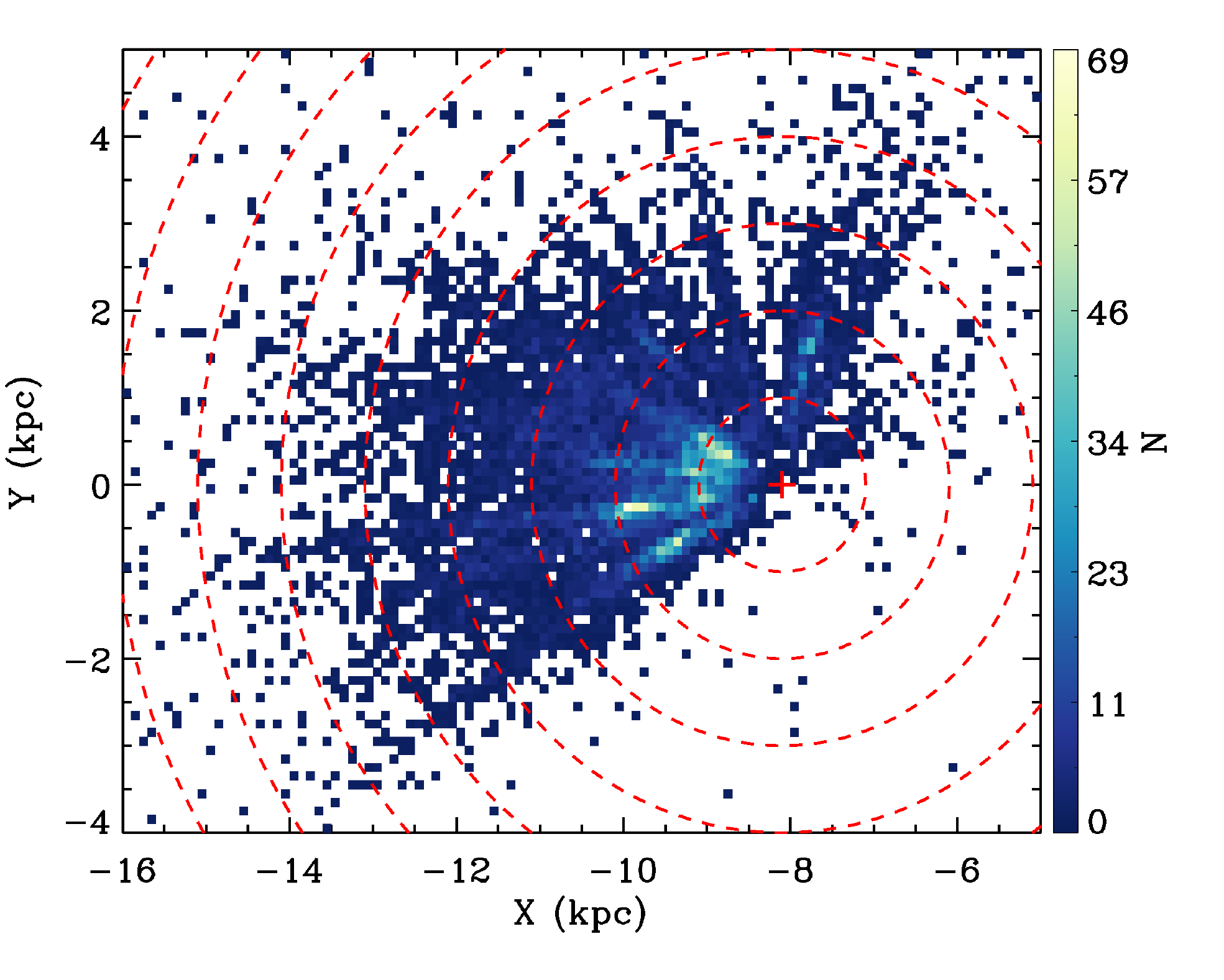}
\caption{Spatial distribution of the LAMOST OB star sample in the disk $X$-$Y$ plane in Galactic Cartesian coordinates. The plus symbol designates the position of the Sun ($X=-8.1$\,kpc, $Y=0$\,kpc). The dashed rings delineate constant distances from the Sun in step of 1\,kpc.}
\label{fig:Fig12}
\end{figure}

\section{distance} \label{distance}
The distance to the star can be estimated by combining its \textit{Gaia} parallax with the spectro-photometric distance derived from the spectroscopic \mk. We estimate the distance using Bayesian scheme presented below. 

In terms of \mk, \ks, extinction $A_{K_{\rm s}}$, and \textit{Gaia} parallax $\varpi$, the probability distribution function of distance $d$ is
\begin{equation}
P(d|\varpi, \mk, \ks, \aks) = P(d|\varpi)P(d|\mk, \ks, \aks),
\end{equation}
where
\begin{equation}
P(d|\varpi) = P(\varpi|d)P(d),
\end{equation}
\noindent
and 
\begin{equation}
\begin{aligned}
&P(d|\mk, \ks, \aks) \\ 
&=  P(\mk|d, \ks, \aks)P(\ks,\aks|d)P(d). 
\end{aligned}
\end{equation}
\noindent
The likelihood function can be written as
\begin{equation}
P(\varpi|d) = \mathcal{G}\left(1/d, \delta\varpi\right), 
\end{equation}
\begin{equation}
\begin{aligned}
&P(\mk|d, \ks, \aks) \\
&= \mathcal{G}\left(\mk-(\ks - 5\log~d + 5 - \aks), \sigma_{\rm tot}\right),
\end{aligned}
\end{equation}
\begin{equation}
 \sigma_{\rm tot} = \sqrt{\delta \mk^2+\delta \ks^2+\delta \aks^2}.
\end{equation}
The extinction \aks is derived by
\begin{equation}
\aks = R_{K\rm s} \times E_{B-V},
\end{equation}
and $R_{K\rm s}$ is the extinction coefficient in the 2MASS $K_{\rm s}$  passband.  We adopt a flat prior $P(d)$, and $P(m,A\vert d)$. Note that, for binary systems, we adopt the \textit{Gaia} parallax alone for distance estimation, as their spectro-photometric distances might be biased.

We sample the posterior distribution function (PDF) for individual stars with a fine distance step of 0.1\,pc, and adopt the mode of the PDF as the distance estimate and the 16 and 84 percentile as the $1\sigma$ estimates. We also sample the PDF in logarithmic distance scale. The PDF in logarithmic distance are close to Gaussian, and we thus adopt the PDF-weighted mean and standard deviation as the estimates of the logarithmic distance and its uncertainty, respectively. 
The left panel of Fig.\,\ref{fig:Fig11} shows the distribution of distances to our sample of LAMOST OB stars.  While the majority of stars are located within 3\,kpc from the Sun, a number of them could lie beyond 10\,kpc.  Fig.\ref{fig:Fig11} illustrates the relative distance uncertainty as a function of distance. The median distance uncertainty of the sample is 8\%, and the distance uncertainty only increases moderately with distance; the distance uncertainty is about $\sim$14\% at 15 kpc. The Figure also shows the distance uncertainty when only the \textit{Gaia} parallax is adopted. It illustrates that the inclusion of the spectroscopic \mk outperforms Gaia distance estimates for stars further than 2\,kpc from the Sun, and the improvement becomes critical for stars more distant than about 5\,kpc. Although not shown, we have also checked the distance of \citet{Bailer-Jones2018} for our sample stars, and found good consistency for stars with $d\lesssim5$\,kpc, a regime where their distance estimates are robust and are not dominated by the priors imposed in their studies.

Fig.\,\ref{fig:Fig12} shows the LAMOST OB star sample in the $X$-$Y$ plane in Galactic Cartesian coordinates ($X$, $Y$, $Z$). The Sun is assumed to be located at position $X=-8.1$\,kpc, $Y=0$ and $Z=0$. The figure highlights the wide spatial range $-16<X-5$\,kpc, $-4<Y<5$\,kpc covered by the sample. A small number of stars outside these ranges are not shown here. The data exhibits over-densities at approximately the distance of  Persues arm (e.g. at $X=-10$\,kpc and $Y\simeq-0.3$\,kpc), which is about 2\,kpc from the Sun \citep[e.g.][]{Xu2006}.

Finally, our estimates of \mk, extinction, and distance and flags for binary and emission lines for the 16,002 LAMOST OB stars are made publicly available\footnote{The catalog will be published along with the article. It can also be accessed via a temporary link 
https://keeper.mpdl.mpg.de/f/56d86145cfb0417eb8a8/?dl=1}. Table\,\ref{table:table1} presents a summary of the catalog.  
 
\begin{figure}
\centering
\includegraphics[width=85mm]{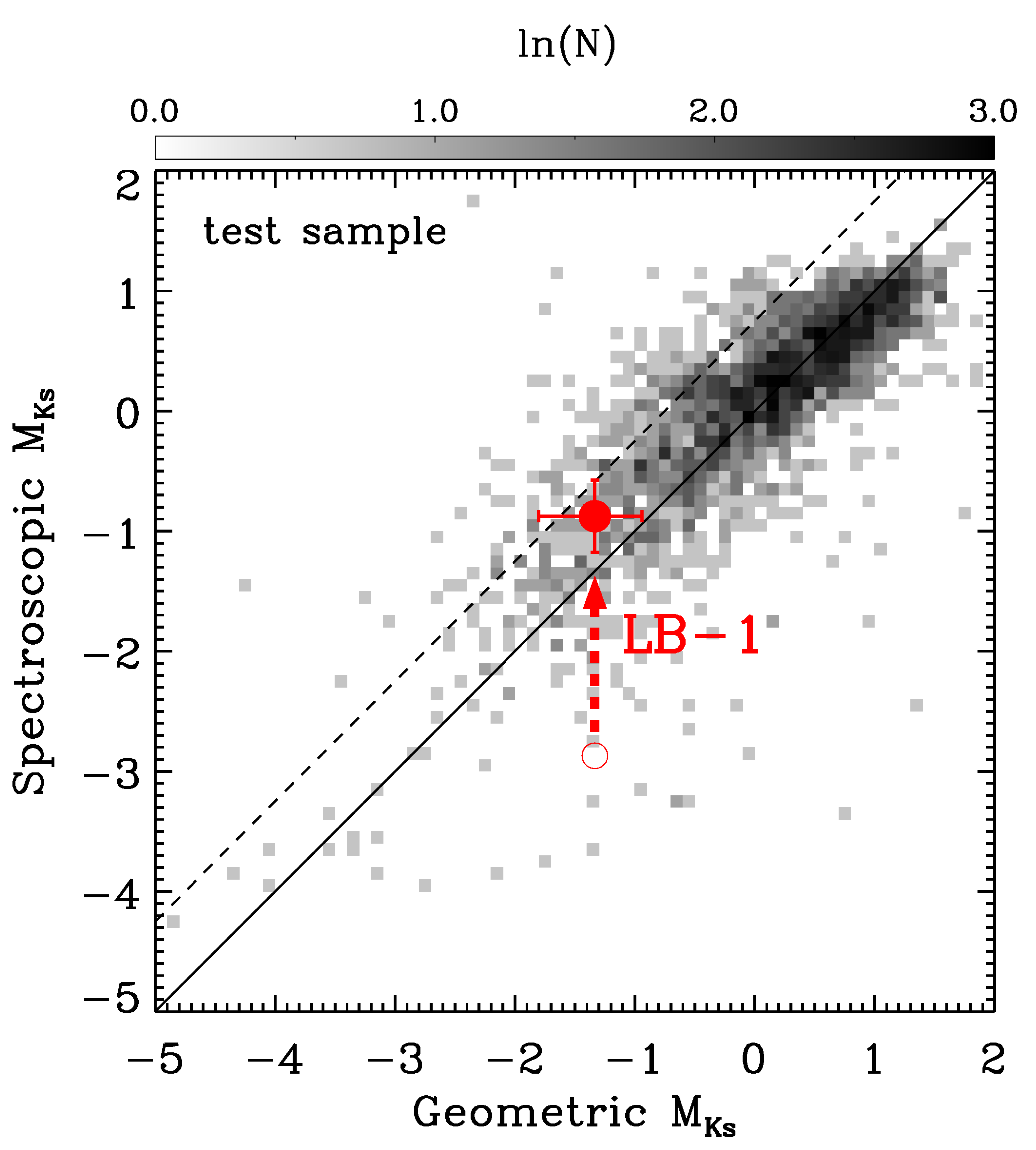}
\caption{Comparison of spectroscopic \mk with geometric \mk for a test star sample with precise Gaia parallax ($\widetilde{\omega}/\sigma_{\widetilde{\omega}} > 10$). The inferred spectroscopic \mk and geometric \mk of the LB-1 system are highlighted in red solid symbol. The open circle ($-2.9$\,mag) shows the \mk inferred from distance in \citet{LiuJ2019}, which is at odd with our spectroscopic \mk. The solid line delineates the 1:1 line, while the dashed line delineates an offset of 0.75\,mag to the 1:1 line.}
\label{fig:Fig13}
\end{figure}

\begin{figure*}
\centering
\includegraphics[width=180mm]{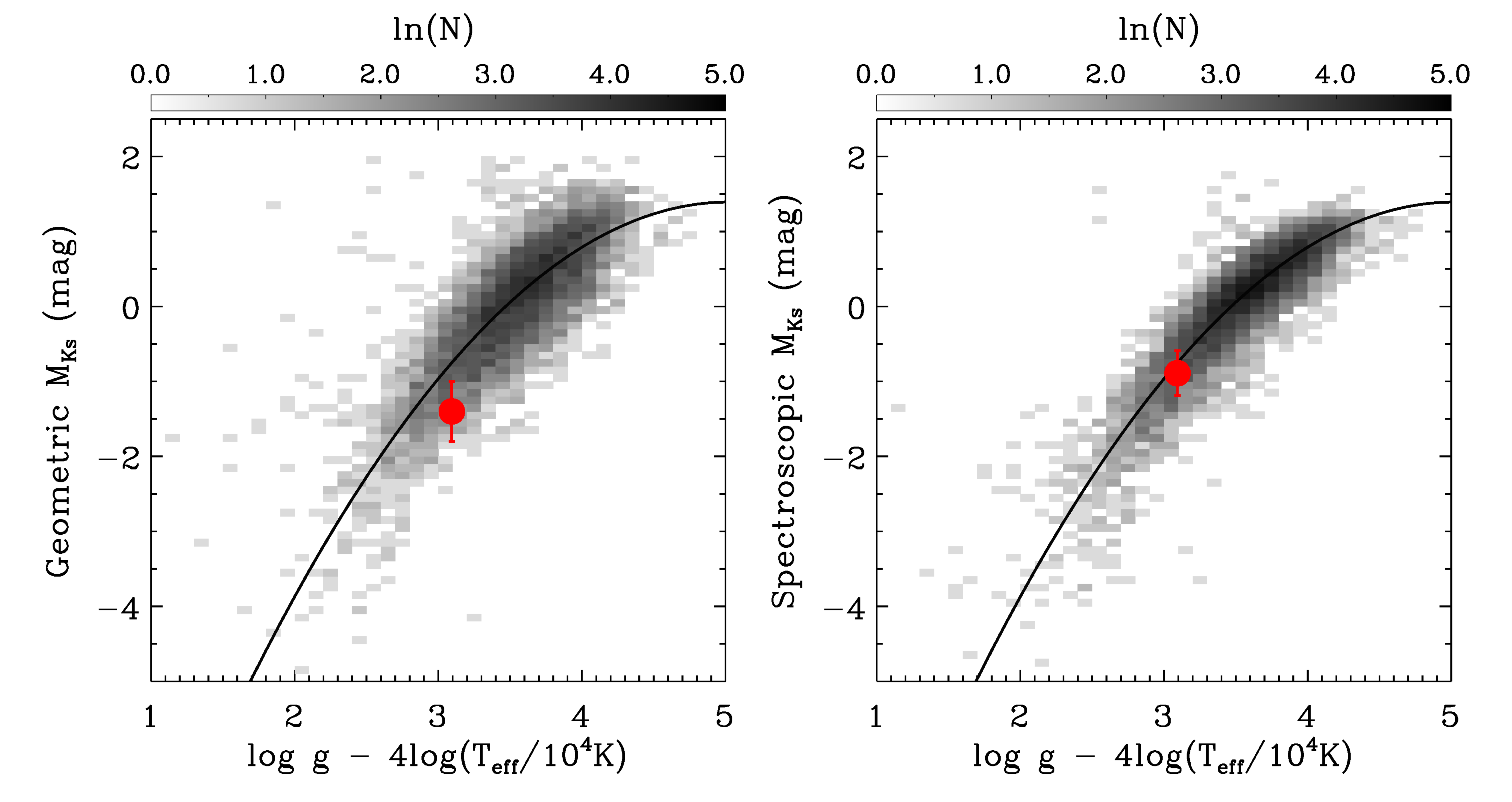}
\caption{The temperature-weighted gravity versus the geometric \mk (the left panel) and the spectroscopic \mk (the right panel). The temperature-weighted gravity is adopted as a ``grouth truth'' luminosity indicator. Only stars with precise Gaia parallax  ($\widetilde{\omega}/\sigma_{\widetilde{\omega}} > 10$) are shown. The red dot with error bar highlights the LB-1 B-star companion. The solid line is a second-order polynomial fit to the {\em geometric} \mk as a function of the temperature-weighted gravity. The spectroscopic \mk estimate for LB-1 is perfectly consistent with the temperature-weighted gravity. The geometric \mk is slightly brighter than prediction from the gravity, but the deviation is within $2\sigma$. 
}
\label{fig:Fig14}
\end{figure*}

\section{The Distance to the LB-1 B-Star System} \label{lb1}
LB-1 (LS V +22 25; RA=92$^\circ$.95450, Dec=22$^\circ$.82575) is a binary system discovered by \citet{LiuJ2019}. It was purported to consist of a B-type star that exhibits periodic radial velocity variations with an amplitude of around 50\,km/s and a period of 78.9\,day; the system also exhibits broad emission Hydrogen lines that show periodic radial velocity variations of $\sim10$\,km/s \citep{LiuJ2019, LiuJ2020}. \citet{LiuJ2019} interpreted LB-1 as a B-star orbiting a $68^{+11}_{-13}$\,$M_\odot$ black hole as the unseen primary companion, which, if true, would be the most massive stellar mass black hole ever been found. Since its discovery, there have been various controversies about the origin of this system \citep{Abdul-Masih2020, El-Badry2020a, El-Badry2020b, Eldridge2020, Irrgang2020, LiuJ2020, Rivinius2020, Shenar2020, SimonDaz2020, Yungelson2020}. Alternative explanations have been proposed. In particular, spectral disentangling \citep{Shenar2020} has all but demonstrated that LB-1 is an SB2 binary with two luminous components: it shows a Be star, with its broad absorption lines and a surrounding emission-line disk, and a luminous hot star with low log $g$,
presumably recently stripped to its current mass of $\sim1\msun$.  This has led \citet{Shenar2020} and \citet{El-Badry2020a} to conclude that the low mass and high luminosity of the stripped star leads to the high radial velocity variations in the combined spectrum. There is no need, and presumably no room, for a black hole. 

\begin{table*}
\caption{Descriptions for the distance catalog for 16,002 OB stars in LAMOST DR5. \tablenotemark{1}}
\label{table:table1}
\begin{tabular}{ll}
\hline
 Field &    Description  \\
\hline
 specid &   LAMOST spectra ID in the format of ``date-planid-spid-fiberid" \\
 fitsname & Name of the LAMOST spectral .FITS file \\
 ra &   Right ascension from the LAMOST DR5 catalog (J2000; deg)\\
 dec &  Declination from the LAMOST DR5 catalog (J2000; deg)  \\
 uniqflag & Flag to indicate repeat visits; uqflag = 1 means unique star, uqflag = 2, 3, ..., $n$ indicates the $n$th repeat visit \\
 &  For stars with repeat visits, the uniqflag is sorted by the spectral S/N, with uqflag = 1 having the highest S/N \\
 star\_id  & A unique ID for each unique star based on its RA and Dec, in the format of ``Sdddmmss$\pm$ddmmss" \\
 snr\_g &  Spectral signal-to-noise ratio per pixel in SDSS g-band \\
 rv  & Radial velocity from LAMOST (km/s) \\
 rv\_err &  Uncertainty in radial velocity (km/s) \\
 $\mk$  & $\mk$ estimated from LAMOST spectra \\
 $\mk$\_err &  Uncertainty in $\mk$ \\
 $\mk$\_geo  & Geometric $\mk$ inferred from \textit{Gaia} parallaxes and 2MASS apparent magnitudes\\
 $\mk$\_geo\_err &  Uncertainty in $\mk$\_geo \\
 dis & Distance at the mode of the distance probability density function (PDF) \\ 
 dis\_low & Distance at 16 percentile of the cumulative probability distribution function \\
 dis\_high & Distance at 84 percentile of the cumulative probability distribution function \\
 logdis & PDF-weighted mean logarithmic distance  \\
 logdis\_err & Uncertainty in logdis \\
 ebv & Reddening estimated in this work \\
 ebv\_err & Uncertainty in ebv \\
 snr\_dparallax & Excess in spectro-photometric parallax with respect to the \textit{Gaia} astrometric parallax \\
 binary\_flag & Flag of binarity; 1 = binary (${\rm snr\_dparallax} \geq 2$), 0 = single (${\rm snr\_dparallax} < 2$), $-9$ = unknown \\
 em\_flag & Flag of emission lines; 1 = with emission lines, 0 = no emission lines \\
 gaia\_id & \textit{Gaia} DR2 Source ID \\
 parallax & \textit{Gaia} DR2 parallax (mas) \\
 parallax\_error & Uncertainty in gaia\_parallax (mas) \\
 parallax\_offset & Offset of \textit{Gaia} parallax according to the offset -- G magnitude relation of \citet{Leung2019} \\
 pmra & \textit{Gaia} DR2 proper motion in RA direction \\
 pmra\_error & Uncertainty in pmra \\
 pmdec & \textit{Gaia} DR2 proper motion in Dec direction \\
 pmdec\_error & Uncertainty in pmdec \\
 ruwe & \textit{Gaia} DR2 RUWE \\
 J & 2MASS J-band magnitude \\
 J\_err & Uncertainty in J \\
 H & 2MASS H-band magnitude \\
 H\_err & Uncertainty in H \\
 $K_{\rm s}$ & 2MASS $K_{\rm s}$-band magnitude \\
 $K_{\rm s}$\_err & Uncertainty in $K_{\rm s}$ \\
 X/Y/Z & 3D position in the Galactic Cartesian coordinates (kpc)\\
 \hline
\end{tabular}
\tablenotetext{1}{Due to multiple visits of common stars, the catalog contains 27,784 entries for 16,002 unique stars. This is slightly different from the original catalog of \citet{LiuZ2019}, which contains 22,901 spectra entries for 16,032 stars. We have increased the number of spectra by including all repeat visits in the LAMOST DR5 database. All the spectra can be found on the LAMOST DR5 website.}
\end{table*} 

Disentangling various explanations for LB-1 is clearly beyond the scope of this paper. Nonetheless, \citet{LiuJ2019} in their initial analysis also came to the conclusion that the parallax-based distance to LB-1 had to be incorrect. This is what we can test here: our catalog contains \mk and distance measurements from 12 individual LAMOST spectra for the LB-1 system, all with ${\rm S/N}>300$. From these, we obtain a \mk of $-0.89\pm0.30$\,mag, which implies a distance of $1.87^{+0.24}_{-0.15}$\,kpc in the case that the luminous light is dominated a single B star. As shown in Fig.\,\ref{fig:Fig13}, our spectroscopic \mk estimate is consistent with the geometric \mk from {\it Gaia} parallax and is in tension with \citet{LiuJ2019}.

As an independent check, we adopt stellar parameters for LB-1 derived from the LAMOST spectra by fitting the Kurucz 1D-LTE ATLAS12 model spectra \citep{Kurucz1970, Kurucz1993} with {\sc The Payne} (Xiang et al., {\it in prep.}) and evaluate the temperature-weighted gravity \citep{Kudritzki2020} as a luminosity indicator.  Fig.\,\ref{fig:Fig14} illustrates that the LB-1 system stellar parameters (assuming a single star fit) is in line with the relation between the temperature-weighted gravity and the \mk for the majority of stars, leading credence to our spectroscopic \mk estimate. In short, our exploration cannot confirm that the Gaia parallax for LB-1 is errorneous. Future high-quality data sets, such as the coming Gaia DR3, may help to better understand these results.

\section{summary} \label{conclusion}
In this study, we have presented a data-driven approach for deriving $K_{\rm s}$-band absolute magnitudes \mk for OB stars from low-resolution ($R\simeq1800$) LAMOST spectra.  Our method uses a neural network model trained on a set of stars with good parallaxes from \textit{Gaia} DR2. Applying to a test data set, we find that the neural network is capable of delivering \mk with 0.25\,mag precision from the LAMOST OB star spectra. We have also applied the method separately to stars with emission lines in their spectra. The emission line spectra are identified through comparing the observed spectra and the PCA reconstruction of the spectra. 

We verify that the \mk estimated from the composite spectrum of a binary system is comparable to, or slightly fainter than, the \mk of the primary star.  This is in contrast to the geometric \mk calculated from \textit{Gaia} parallaxes, as both components of the binary contribute to the geometric \mk. We propose a new method of binary identification, leveraging differences between the spectroscopic \mk and the geometric \mk. The method is particularly effective for identifying equal-mass binaries or multiple-star systems, since the geometric \mk of these systems are much brighter than their primaries. Our method is generic and can be applied to any combined astrometric and spectroscopic data beyond this study. 

With the spectroscopic \mk determinations, we derive accurate distances to 16,002 OB stars from the LAMOST sample of \citet{LiuZ2019}.  The median distance uncertainty for our sample stars is 8\%, and the distance uncertainty for the most distant  stars at more than 10\,kpc away is about 14\%. We present a value-added catalogue of OB stars for future studies of the structure and dynamics of the Galactic disk. Besides absolute magnitudes and distances, the catalog presents also emission-line flags and binary flags for the LAMOST OB stars, significantly expanding the number of known emission-line objects and binaries for massive stars, especially those with mass ratios close to unity. Our method yields a spectral \mk of $0.89\pm0.30$\,mag from the  LAMOST spectra of LB-1, which corresponds to distance of $1.87^{+0.24}_{-0.15}$\,kpc in the case that the light is dominated by a single B star.

\vspace{7mm} \noindent {\bf Acknowledgments}{
H.-W. Rix acknowledges funding by the Deutsche Forschungsgemeinschaft (DFG, German Research Foundation) -- Project-ID 138713538 -- SFB 881 (``The Milky Way System'', subproject A03). M.-S. Xiang is grateful for DR. Bodem for the successful dental surgery and the attentive care from him during recovery. YST is grateful to be supported by the NASA Hubble Fellowship grant HST-HF2-51425.001 awarded by the Space Telescope Science Institute.}

This work has made use of data acquired through the Guoshoujing Telescope. Guoshoujing Telescope (the Large Sky Area Multi-Object Fiber Spectroscopic Telescope; LAMOST) is a National Major Scientific Project built by the Chinese Academy of Sciences. Funding for the project has been provided by the National Development and Reform Commission. LAMOST is operated and managed by the National Astronomical Observatories, Chinese Academy of Sciences.

This work has also made use of data from the European Space Agency (ESA) mission \textit{Gaia}, processed by the \textit{Gaia} Data Processing and Analysis Consortium (DPAC). Funding for the DPAC has been provided by national institutions, in particular the institutions participating in the \textit{Gaia} Multilateral Agreement.

\bibliography{reference.bib}

\begin{thebibliography}{}
\expandafter\ifx\csname natexlab\endcsname\relax\def\natexlab#1{#1}\fi
\providecommand{\url}[1]{\href{#1}{#1}}

\bibitem[{{Abbott} {et~al.}(2016{\natexlab{a}}){Abbott}, {Abbott}, {Abbott},
  {Abernathy}, {Acernese}, {Ackley}, {Adams}, {Adams}, {Addesso}, {Adhikari},
  {Adya}, {Affeldt}, {Agathos}, {Agatsuma}, {Aggarwal}, {Aguiar}, {Aiello},
  {Ain}, {Ajith}, {Allen}, {Allocca}, {Altin}, {Anderson}, {Anderson}, {Arai},
  {Arain}, {Araya}, {Arceneaux}, {Areeda}, {Arnaud}, {Arun}, {Ascenzi},
  {Ashton}, {Ast}, {Aston}, {Astone}, {Aufmuth}, {Aulbert}, {Babak}, {Bacon},
  {Bader}, {Baker}, {Baldaccini}, {Ballardin}, {Ballmer}, {Barayoga},
  {Barclay}, {Barish}, {Barker}, {Barone}, {Barr}, {Barsotti}, {Barsuglia},
  {Barta}, {Bartlett}, {Barton}, {Bartos}, {Bassiri}, {Basti}, {Batch},
  {Baune}, {Bavigadda}, {Bazzan}, {Behnke}, {Bejger}, {Belczynski}, {Bell},
  {Bell}, {Berger}, {Bergman}, {Bergmann}, {Berry}, {Bersanetti}, {Bertolini},
  {Betzwieser}, {Bhagwat}, {Bhandare}, {Bilenko}, {Billingsley}, {Birch},
  {Birney}, {Birnholtz}, {Biscans}, {Bisht}, {Bitossi}, {Biwer}, {Bizouard},
  {Blackburn}, {Blair}, {Blair}, {Blair}, {Bloemen}, {Bock}, {Bodiya}, {Boer},
  {Bogaert}, {Bogan}, {Bohe}, {Bojtos}, {Bond}, {Bondu}, {Bonnand}, {Boom},
  {Bork}, {Boschi}, {Bose}, {Bouffanais}, {Bozzi}, {Bradaschia}, {Brady},
  {Braginsky}, {Branchesi}, {Brau}, {Briant}, {Brillet}, {Brinkmann},
  {Brisson}, {Brockill}, {Brooks}, {Brown}, {Brown}, {Brown}, {Buchanan},
  {Buikema}, {Bulik}, {Bulten}, {Buonanno}, {Buskulic}, {Buy}, {Byer},
  {Cabero}, {Cadonati}, {Cagnoli}, {Cahillane}, {Bustillo}, {Callister},
  {Calloni}, {Camp}, {Cannon}, {Cao}, {Capano}, {Capocasa}, {Carbognani},
  {Caride}, {Casanueva Diaz}, {Casentini}, {Caudill}, {Cavagli{\`a}},
  {Cavalier}, {Cavalieri}, {Cella}, {Cepeda}, {Baiardi}, {Cerretani},
  {Cesarini}, {Chakraborty}, {Chalermsongsak}, {Chamberlin}, {Chan}, {Chao},
  {Charlton}, {Chassand e-Mottin}, {Chen}, {Chen}, {Cheng}, {Chincarini},
  {Chiummo}, {Cho}, {Cho}, {Chow}, {Christensen}, {Chu}, {Chua}, {Chung},
  {Ciani}, {Clara}, {Clark}, {Cleva}, {Coccia}, {Cohadon}, {Colla}, {Collette},
  {Cominsky}, {Constancio}, {Conte}, {Conti}, {Cook}, {Corbitt}, {Cornish},
  {Corsi}, {Cortese}, {Costa}, {Coughlin}, {Coughlin}, {Coulon}, {Countryman},
  {Couvares}, {Cowan}, {Coward}, {Cowart}, {Coyne}, {Coyne}, {Craig},
  {Creighton}, {Creighton}, {Cripe}, {Crowder}, {Cruise}, {Cumming},
  {Cunningham}, {Cuoco}, {Dal Canton}, {Danilishin}, {D'Antonio}, {Danzmann},
  {Darman}, {Da Silva Costa}, {Dattilo}, {Dave}, {Daveloza}, {Davier},
  {Davies}, {Daw}, {Day}, {De}, {DeBra}, {Debreczeni}, {Degallaix}, {De
  Laurentis}, {Del{\'e}glise}, {Del Pozzo}, {Denker}, {Dent}, {Dereli},
  {Dergachev}, {DeRosa}, {De Rosa}, {DeSalvo}, {Dhurandhar}, {D{\'\i}az}, {Di
  Fiore}, {Di Giovanni}, {Di Lieto}, {Di Pace}, {Di Palma}, {Di Virgilio},
  {Dojcinoski}, {Dolique}, {Donovan}, {Dooley}, {Doravari}, {Douglas},
  {Downes}, {Drago}, {Drever}, {Driggers}, {Du}, {Ducrot}, {Dwyer}, {Edo},
  {Edwards}, {Effler}, {Eggenstein}, {Ehrens}, {Eichholz}, {Eikenberry},
  {Engels}, {Essick}, {Etzel}, {Evans}, {Evans}, {Everett}, {Factourovich},
  {Fafone}, {Fair}, {Fairhurst}, {Fan}, {Fang}, {Farinon}, {Farr}, {Farr},
  {Favata}, {Fays}, {Fehrmann}, {Fejer}, {Feldbaum}, {Ferrante}, {Ferreira},
  {Ferrini}, {Fidecaro}, {Finn}, {Fiori}, {Fiorucci}, {Fisher}, {Flaminio},
  {Fletcher}, {Fong}, {Fournier}, {Franco}, {Frasca}, {Frasconi}, {Frede},
  {Frei}, {Freise}, {Frey}, {Frey}, {Fricke}, {Fritschel}, {Frolov}, {Fulda},
  {Fyffe}, {Gabbard}, {Gair}, {Gammaitoni}, {Gaonkar}, {Garufi}, {Gatto},
  {Gaur}, {Gehrels}, {Gemme}, {Gendre}, {Genin}, {Gennai}, {George}, {Gergely},
  {Germain}, {Ghosh}, {Ghosh}, {Ghosh}, {Giaime}, {Giardina}, {Giazotto},
  {Gill}, {Glaefke}, {Gleason}, {Goetz}, {Goetz}, {Gondan}, {Gonz{\'a}lez},
  {Castro}, {Gopakumar}, {Gordon}, {Gorodetsky}, {Gossan}, {Gosselin},
  {Gouaty}, {Graef}, {Graff}, {Granata}, {Grant}, {Gras}, {Gray}, {Greco},
  {Green}, {Greenhalgh}, {Groot}, {Grote}, {Grunewald}, {Guidi}, {Guo},
  {Gupta}, {Gupta}, {Gushwa}, {Gustafson}, {Gustafson}, {Hacker}, {Hall},
  {Hall}, {Hammond}, {Haney}, {Hanke}, {Hanks}, {Hanna}, {Hannam}, {Hanson},
  {Hardwick}, {Harms}, {Harry}, {Harry}, {Hart}, {Hartman}, {Haster},
  {Haughian}, {Healy}, {Heefner}, {Heidmann}, {Heintze}, {Heinzel}, {Heitmann},
  {Hello}, {Hemming}, {Hendry}, {Heng}, {Hennig}, {Heptonstall}, {Heurs},
  {Hild}, {Hoak}, {Hodge}, {Hofman}, {Hollitt}, {Holt}, {Holz}, {Hopkins},
  {Hosken}, {Hough}, {Houston}, {Howell}, {Hu}, {Huang}, {Huerta}, {Huet},
  {Hughey}, {Husa}, {Huttner}, {Huynh-Dinh}, {Idrisy}, {Indik}, {Ingram},
  {Inta}, {Isa}, {Isac}, {Isi}, {Islas}, {Isogai}, {Iyer}, {Izumi}, {Jacobson},
  {Jacqmin}, {Jang}, {Jani}, {Jaranowski}, {Jawahar}, {Jim{\'e}nez-Forteza},
  {Johnson}, {Johnson-McDaniel}, {Jones}, {Jones}, {Jonker}, {Ju}, {Haris},
  {Kalaghatgi}, {Kalogera}, {Kandhasamy}, {Kang}, {Kanner}, {Karki},
  {Kasprzack}, {Katsavounidis}, {Katzman}, {Kaufer}, {Kaur}, {Kawabe},
  {Kawazoe}, {K{\'e}f{\'e}lian}, {Kehl}, {Keitel}, {Kelley}, {Kells},
  {Kennedy}, {Keppel}, {Key}, {Khalaidovski}, {Khalili}, {Khan}, {Khan},
  {Khan}, {Khazanov}, {Kijbunchoo}, {Kim}, {Kim}, {Kim}, {Kim}, {Kim}, {Kim},
  {King}, {King}, {Kinzel}, {Kissel}, {Kleybolte}, {Klimenko}, {Koehlenbeck},
  {Kokeyama}, {Koley}, {Kondrashov}, {Kontos}, {Koranda}, {Korobko}, {Korth},
  {Kowalska}, {Kozak}, {Kringel}, {Krishnan}, {Kr{\'o}lak}, {Krueger}, {Kuehn},
  {Kumar}, {Kumar}, {Kuo}, {Kutynia}, {Kwee}, {Lackey}, {Landry}, {Lange},
  {Lantz}, {Lasky}, {Lazzarini}, {Lazzaro}, {Leaci}, {Leavey}, {Lebigot},
  {Lee}, {Lee}, {Lee}, {Lee}, {Lenon}, {Leonardi}, {Leong}, {Leroy},
  {Letendre}, {Levin}, {Levine}, {Li}, {Libson}, {Littenberg}, {Lockerbie},
  {Logue}, {Lombardi}, {London}, {Lord}, {Lorenzini}, {Loriette}, {Lormand},
  {Losurdo}, {Lough}, {Lousto}, {Lovelace}, {L{\"u}ck}, {Lundgren}, {Luo},
  {Lynch}, {Ma}, {MacDonald}, {Machenschalk}, {MacInnis}, {Macleod},
  {Maga{\~n}a-Sandoval}, {Magee}, {Mageswaran}, {Majorana}, {Maksimovic},
  {Malvezzi}, {Man}, {Mandel}, {Mandic}, {Mangano}, {Mansell}, {Manske},
  {Mantovani}, {Marchesoni}, {Marion}, {M{\'a}rka}, {M{\'a}rka}, {Markosyan},
  {Maros}, {Martelli}, {Martellini}, {Martin}, {Martin}, {Martynov}, {Marx},
  {Mason}, {Masserot}, {Massinger}, {Masso-Reid}, {Matichard}, {Matone},
  {Mavalvala}, {Mazumder}, {Mazzolo}, {McCarthy}, {McClelland}, {McCormick},
  {McGuire}, {McIntyre}, {McIver}, {McManus}, {McWilliams}, {Meacher},
  {Meadors}, {Meidam}, {Melatos}, {Mendell}, {Mendoza-Gandara}, {Mercer},
  {Merilh}, {Merzougui}, {Meshkov}, {Messenger}, {Messick}, {Meyers},
  {Mezzani}, {Miao}, {Michel}, {Middleton}, {Mikhailov}, {Milano}, {Miller},
  {Millhouse}, {Minenkov}, {Ming}, {Mirshekari}, {Mishra}, {Mitra},
  {Mitrofanov}, {Mitselmakher}, {Mittleman}, {Moggi}, {Mohan}, {Mohapatra},
  {Montani}, {Moore}, {Moore}, {Moraru}, {Moreno}, {Morriss}, {Mossavi},
  {Mours}, {Mow-Lowry}, {Mueller}, {Mueller}, {Muir}, {Mukherjee}, {Mukherjee},
  {Mukherjee}, {Mukund}, {Mullavey}, {Munch}, {Murphy}, {Murray}, {Mytidis},
  {Nardecchia}, {Naticchioni}, {Nayak}, {Necula}, {Nedkova}, {Nelemans},
  {Neri}, {Neunzert}, {Newton}, {Nguyen}, {Nielsen}, {Nissanke}, {Nitz},
  {Nocera}, {Nolting}, {Normandin}, {Nuttall}, {Oberling}, {Ochsner}, {O'Dell},
  {Oelker}, {Ogin}, {Oh}, {Oh}, {Ohme}, {Oliver}, {Oppermann}, {Oram},
  {O'Reilly}, {O'Shaughnessy}, {Ott}, {Ottaway}, {Ottens}, {Overmier}, {Owen},
  {Pai}, {Pai}, {Palamos}, {Palashov}, {Palomba}, {Pal-Singh}, {Pan}, {Pan},
  {Pankow}, {Pannarale}, {Pant}, {Paoletti}, {Paoli}, {Papa}, {Paris},
  {Parker}, {Pascucci}, {Pasqualetti}, {Passaquieti}, {Passuello},
  {Patricelli}, {Patrick}, {Pearlstone}, {Pedraza}, {Pedurand }, {Pekowsky},
  {Pele}, {Penn}, {Perreca}, {Pfeiffer}, {Phelps}, {Piccinni}, {Pichot},
  {Pickenpack}, {Piergiovanni}, {Pierro}, {Pillant}, {Pinard}, {Pinto},
  {Pitkin}, {Poeld}, {Poggiani}, {Popolizio}, {Post}, {Powell}, {Prasad},
  {Predoi}, {Premachandra}, {Prestegard}, {Price}, {Prijatelj}, {Principe},
  {Privitera}, {Prix}, {Prodi}, {Prokhorov}, {Puncken}, {Punturo}, {Puppo},
  {P{\"u}rrer}, {Qi}, {Qin}, {Quetschke}, {Quintero}, {Quitzow-James}, {Raab},
  {Rabeling}, {Radkins}, {Raffai}, {Raja}, {Rakhmanov}, {Ramet}, {Rapagnani},
  {Raymond}, {Razzano}, {Re}, {Read}, {Reed}, {Regimbau}, {Rei}, {Reid},
  {Reitze}, {Rew}, {Reyes}, {Ricci}, {Riles}, {Robertson}, {Robie}, {Robinet},
  {Rocchi}, {Rolland}, {Rollins}, {Roma}, {Romano}, {Romano}, {Romanov},
  {Romie}, {Rosi{\'n}ska}, {Rowan}, {R{\"u}diger}, {Ruggi}, {Ryan}, {Sachdev},
  {Sadecki}, {Sadeghian}, {Salconi}, {Saleem}, {Salemi}, {Samajdar}, {Sammut},
  {Sampson}, {Sanchez}, {Sandberg}, {Sandeen}, {Sand ers}, {Sanders},
  {Sassolas}, {Sathyaprakash}, {Saulson}, {Sauter}, {Savage}, {Sawadsky},
  {Schale}, {Schilling}, {Schmidt}, {Schmidt}, {Schnabel}, {Schofield},
  {Sch{\"o}nbeck}, {Schreiber}, {Schuette}, {Schutz}, {Scott}, {Scott},
  {Sellers}, {Sengupta}, {Sentenac}, {Sequino}, {Sergeev}, {Serna},
  {Setyawati}, {Sevigny}, {Shaddock}, {Shaffer}, {Shah}, {Shahriar}, {Shaltev},
  {Shao}, {Shapiro}, {Shawhan}, {Sheperd}, {Shoemaker}, {Shoemaker}, {Siellez},
  {Siemens}, {Sigg}, {Silva}, {Simakov}, {Singer}, {Singer}, {Singh}, {Singh},
  {Singhal}, {Sintes}, {Slagmolen}, {Smith}, {Smith}, {Smith}, {Smith}, {Son},
  {Sorazu}, {Sorrentino}, {Souradeep}, {Srivastava}, {Staley}, {Steinke},
  {Steinlechner}, {Steinlechner}, {Steinmeyer}, {Stephens}, {Stevenson},
  {Stone}, {Strain}, {Straniero}, {Stratta}, {Strauss}, {Strigin}, {Sturani},
  {Stuver}, {Summerscales}, {Sun}, {Sutton}, {Swinkels}, {Szczepa{\'n}czyk},
  {Tacca}, {Talukder}, {Tanner}, {T{\'a}pai}, {Tarabrin}, {Taracchini},
  {Taylor}, {Theeg}, {Thirugnanasambandam}, {Thomas}, {Thomas}, {Thomas},
  {Thorne}, {Thorne}, {Thrane}, {Tiwari}, {Tiwari}, {Tokmakov}, {Tomlinson},
  {Tonelli}, {Torres}, {Torrie}, {T{\"o}yr{\"a}}, {Travasso}, {Traylor},
  {Trifir{\`o}}, {Tringali}, {Trozzo}, {Tse}, {Turconi}, {Tuyenbayev},
  {Ugolini}, {Unnikrishnan}, {Urban}, {Usman}, {Vahlbruch}, {Vajente},
  {Valdes}, {Vallisneri}, {van Bakel}, {van Beuzekom}, {van den Brand}, {Van
  Den Broeck}, {Vand er-Hyde}, {van der Schaaf}, {van Heijningen}, {van
  Veggel}, {Vardaro}, {Vass}, {Vas{\'u}th}, {Vaulin}, {Vecchio}, {Vedovato},
  {Veitch}, {Veitch}, {Venkateswara}, {Verkindt}, {Vetrano}, {Vicer{\'e}},
  {Vinciguerra}, {Vine}, {Vinet}, {Vitale}, {Vo}, {Vocca}, {Vorvick}, {Voss},
  {Vousden}, {Vyatchanin}, {Wade}, {Wade}, {Wade}, {Waldman}, {Walker},
  {Wallace}, {Walsh}, {Wang}, {Wang}, {Wang}, {Wang}, {Wang}, {Ward}, {Ward},
  {Warner}, {Was}, {Weaver}, {Wei}, {Weinert}, {Weinstein}, {Weiss}, {Welborn},
  {Wen}, {We{\ss}els}, {Westphal}, {Wette}, {Whelan}, {Whitcomb}, {White},
  {Whiting}, {Wiesner}, {Wilkinson}, {Willems}, {Williams}, {Williams},
  {Williamson}, {Willis}, {Willke}, {Wimmer}, {Winkelmann}, {Winkler}, {Wipf},
  {Wiseman}, {Wittel}, {Woan}, {Worden}, {Wright}, {Wu}, {Yablon}, {Yakushin},
  {Yam}, {Yamamoto}, {Yancey}, {Yap}, {Yu}, {Yvert}, {Zadro{\.Z}ny},
  {Zangrando}, {Zanolin}, {Zendri}, {Zevin}, {Zhang}, {Zhang}, {Zhang},
  {Zhang}, {Zhao}, {Zhou}, {Zhou}, {Zhu}, {Zucker}, {Zuraw}, {Zweizig}, {LIGO
  Scientific Collaboration}, \& {Virgo Collaboration}}]{Abbott2016a}
{Abbott}, B.~P., {Abbott}, R., {Abbott}, T.~D., {et~al.} 2016{\natexlab{a}},
  \prl, 116, 061102

\bibitem[{{Abbott} {et~al.}(2016{\natexlab{b}}){Abbott}, {Abbott}, {Abbott},
  {Abernathy}, {Acernese}, {Ackley}, {Adams}, {Adams}, {Addesso}, {Adhikari},
  {Adya}, {Affeldt}, {Agathos}, {Agatsuma}, {Aggarwal}, {Aguiar}, {Aiello},
  {Ain}, {Ajith}, {Allen}, {Allocca}, {Altin}, {Anderson}, {Anderson}, {Arai},
  {Araya}, {Arceneaux}, {Areeda}, {Arnaud}, {Arun}, {Ascenzi}, {Ashton}, {Ast},
  {Aston}, {Astone}, {Aufmuth}, {Aulbert}, {Babak}, {Bacon}, {Bader}, {Baker},
  {Baldaccini}, {Ballardin}, {Ballmer}, {Barayoga}, {Barclay}, {Barish},
  {Barker}, {Barone}, {Barr}, {Barsotti}, {Barsuglia}, {Barta}, {Bartlett},
  {Bartos}, {Bassiri}, {Basti}, {Batch}, {Baune}, {Bavigadda}, {Bazzan},
  {Bejger}, {Bell}, {Berger}, {Bergmann}, {Berry}, {Bersanetti}, {Bertolini},
  {Betzwieser}, {Bhagwat}, {Bhandare}, {Bilenko}, {Billingsley}, {Birch},
  {Birney}, {Birnholtz}, {Biscans}, {Bisht}, {Bitossi}, {Biwer}, {Bizouard},
  {Blackburn}, {Blair}, {Blair}, {Blair}, {Bloemen}, {Bock}, {Boer}, {Bogaert},
  {Bogan}, {Bohe}, {Bond}, {Bondu}, {Bonnand}, {Boom}, {Bork}, {Boschi},
  {Bose}, {Bouffanais}, {Bozzi}, {Bradaschia}, {Brady}, {Braginsky},
  {Branchesi}, {Brau}, {Briant}, {Brillet}, {Brinkmann}, {Brisson}, {Brockill},
  {Broida}, {Brooks}, {Brown}, {Brown}, {Brown}, {Brunett}, {Buchanan},
  {Buikema}, {Bulik}, {Bulten}, {Buonanno}, {Buskulic}, {Buy}, {Byer},
  {Cabero}, {Cadonati}, {Cagnoli}, {Cahillane}, {Calder{\'o}n Bustillo},
  {Callister}, {Calloni}, {Camp}, {Cannon}, {Cao}, {Capano}, {Capocasa},
  {Carbognani}, {Caride}, {Casanueva Diaz}, {Casentini}, {Caudill},
  {Cavagli{\`a}}, {Cavalier}, {Cavalieri}, {Cella}, {Cepeda}, {Cerboni
  Baiardi}, {Cerretani}, {Cesarini}, {Chamberlin}, {Chan}, {Chao}, {Charlton},
  {Chassande-Mottin}, {Cheeseboro}, {Chen}, {Chen}, {Cheng}, {Chincarini},
  {Chiummo}, {Cho}, {Cho}, {Chow}, {Christensen}, {Chu}, {Chua}, {Chung},
  {Ciani}, {Clara}, {Clark}, {Cleva}, {Coccia}, {Cohadon}, {Colla}, {Collette},
  {Cominsky}, {Constancio}, {Conte}, {Conti}, {Cook}, {Corbitt}, {Cornish},
  {Corsi}, {Cortese}, {Costa}, {Coughlin}, {Coughlin}, {Coulon}, {Countryman},
  {Couvares}, {Cowan}, {Coward}, {Cowart}, {Coyne}, {Coyne}, {Craig},
  {Creighton}, {Cripe}, {Crowder}, {Cumming}, {Cunningham}, {Cuoco}, {Dal
  Canton}, {Danilishin}, {D'Antonio}, {Danzmann}, {Darman}, {Dasgupta}, {Da
  Silva Costa}, {Dattilo}, {Dave}, {Davier}, {Davies}, {Daw}, {Day}, {De},
  {DeBra}, {Debreczeni}, {Degallaix}, {De Laurentis}, {Del{\'e}glise}, {Del
  Pozzo}, {Denker}, {Dent}, {Dergachev}, {De Rosa}, {DeRosa}, {DeSalvo},
  {Devine}, {Dhurand har}, {D{\'\i}az}, {Di Fiore}, {Di Giovanni}, {Di
  Girolamo}, {Di Lieto}, {Di Pace}, {Di Palma}, {Di Virgilio}, {Dolique},
  {Donovan}, {Dooley}, {Doravari}, {Douglas}, {Downes}, {Drago}, {Drever},
  {Driggers}, {Ducrot}, {Dwyer}, {Edo}, {Edwards}, {Effler}, {Eggenstein},
  {Ehrens}, {Eichholz}, {Eikenberry}, {Engels}, {Essick}, {Etzel}, {Evans},
  {Evans}, {Everett}, {Factourovich}, {Fafone}, {Fair}, {Fairhurst}, {Fan},
  {Fang}, {Farinon}, {Farr}, {Farr}, {Favata}, {Fays}, {Fehrmann}, {Fejer},
  {Fenyvesi}, {Ferrante}, {Ferreira}, {Ferrini}, {Fidecaro}, {Fiori},
  {Fiorucci}, {Fisher}, {Flaminio}, {Fletcher}, {Fong}, {Fournier}, {Frasca},
  {Frasconi}, {Frei}, {Freise}, {Frey}, {Frey}, {Fritschel}, {Frolov}, {Fulda},
  {Fyffe}, {Gabbard}, {Gair}, {Gammaitoni}, {Gaonkar}, {Garufi}, {Gaur},
  {Gehrels}, {Gemme}, {Geng}, {Genin}, {Gennai}, {George}, {Gergely},
  {Germain}, {Ghosh}, {Ghosh}, {Ghosh}, {Giaime}, {Giardina}, {Giazotto},
  {Gill}, {Glaefke}, {Goetz}, {Goetz}, {Gondan}, {Gonz{\'a}lez}, {Gonzalez
  Castro}, {Gopakumar}, {Gordon}, {Gorodetsky}, {Gossan}, {Gosselin}, {Gouaty},
  {Grado}, {Graef}, {Graff}, {Granata}, {Grant}, {Gras}, {Gray}, {Greco},
  {Green}, {Groot}, {Grote}, {Grunewald}, {Guidi}, {Guo}, {Gupta}, {Gupta},
  {Gushwa}, {Gustafson}, {Gustafson}, {Hacker}, {Hall}, {Hall}, {Hamilton},
  {Hammond}, {Haney}, {Hanke}, {Hanks}, {Hanna}, {Hannam}, {Hanson},
  {Hardwick}, {Harms}, {Harry}, {Harry}, {Hart}, {Hartman}, {Haster},
  {Haughian}, {Healy}, {Heidmann}, {Heintze}, {Heitmann}, {Hello}, {Hemming},
  {Hendry}, {Heng}, {Hennig}, {Henry}, {Heptonstall}, {Heurs}, {Hild}, {Hoak},
  {Hofman}, {Holt}, {Holz}, {Hopkins}, {Hough}, {Houston}, {Howell}, {Hu},
  {Huang}, {Huerta}, {Huet}, {Hughey}, {Husa}, {Huttner}, {Huynh-Dinh},
  {Indik}, {Ingram}, {Inta}, {Isa}, {Isac}, {Isi}, {Isogai}, {Iyer}, {Izumi},
  {Jacqmin}, {Jang}, {Jani}, {Jaranowski}, {Jawahar}, {Jian},
  {Jim{\'e}nez-Forteza}, {Johnson}, {Johnson-McDaniel}, {Jones}, {Jones},
  {Jonker}, {Ju}, {K}, {Kalaghatgi}, {Kalogera}, {Kandhasamy}, {Kang},
  {Kanner}, {Kapadia}, {Karki}, {Karvinen}, {Kasprzack}, {Katsavounidis},
  {Katzman}, {Kaufer}, {Kaur}, {Kawabe}, {K{\'e}f{\'e}lian}, {Kehl}, {Keitel},
  {Kelley}, {Kells}, {Kennedy}, {Key}, {Khalili}, {Khan}, {Khan}, {Khan},
  {Khazanov}, {Kijbunchoo}, {Kim}, {Kim}, {Kim}, {Kim}, {Kim}, {Kim}, {Kim},
  {Kimbrell}, {King}, {King}, {Kissel}, {Klein}, {Kleybolte}, {Klimenko},
  {Koehlenbeck}, {Koley}, {Kondrashov}, {Kontos}, {Korobko}, {Korth},
  {Kowalska}, {Kozak}, {Kringel}, {Krishnan}, {Kr{\'o}lak}, {Krueger}, {Kuehn},
  {Kumar}, {Kumar}, {Kuo}, {Kutynia}, {Lackey}, {Land ry}, {Lange}, {Lantz},
  {Lasky}, {Laxen}, {Lazzarini}, {Lazzaro}, {Leaci}, {Leavey}, {Lebigot},
  {Lee}, {Lee}, {Lee}, {Lee}, {Lenon}, {Leonardi}, {Leong}, {Leroy},
  {Letendre}, {Levin}, {Lewis}, {Li}, {Libson}, {Littenberg}, {Lockerbie},
  {Lombardi}, {London}, {Lord}, {Lorenzini}, {Loriette}, {Lormand}, {Losurdo},
  {Lough}, {Lousto}, {L{\"u}ck}, {Lundgren}, {Lynch}, {Ma}, {Machenschalk},
  {MacInnis}, {Macleod}, {Maga{\~n}a-Sandoval}, {Maga{\~n}a Zertuche}, {Magee},
  {Majorana}, {Maksimovic}, {Malvezzi}, {Man}, {Mandel}, {Mandic}, {Mangano},
  {Mansell}, {Manske}, {Mantovani}, {Marchesoni}, {Marion}, {M{\'a}rka},
  {M{\'a}rka}, {Markosyan}, {Maros}, {Martelli}, {Martellini}, {Martin},
  {Martynov}, {Marx}, {Mason}, {Masserot}, {Massinger}, {Masso-Reid},
  {Mastrogiovanni}, {Matichard}, {Matone}, {Mavalvala}, {Mazumder}, {McCarthy},
  {McClelland}, {McCormick}, {McGuire}, {McIntyre}, {McIver}, {McManus},
  {McRae}, {McWilliams}, {Meacher}, {Meadors}, {Meidam}, {Melatos}, {Mendell},
  {Mercer}, {Merilh}, {Merzougui}, {Meshkov}, {Messenger}, {Messick},
  {Metzdorff}, {Meyers}, {Mezzani}, {Miao}, {Michel}, {Middleton}, {Mikhailov},
  {Milano}, {Miller}, {Miller}, {Miller}, {Miller}, {Millhouse}, {Minenkov},
  {Ming}, {Mirshekari}, {Mishra}, {Mitra}, {Mitrofanov}, {Mitselmakher},
  {Mittleman}, {Moggi}, {Mohan}, {Mohapatra}, {Montani}, {Moore}, {Moore},
  {Moraru}, {Moreno}, {Morriss}, {Mossavi}, {Mours}, {Mow-Lowry}, {Mueller},
  {Muir}, {Mukherjee}, {Mukherjee}, {Mukherjee}, {Mukund}, {Mullavey}, {Munch},
  {Murphy}, {Murray}, {Mytidis}, {Nardecchia}, {Naticchioni}, {Nayak},
  {Nedkova}, {Nelemans}, {Nelson}, {Neri}, {Neunzert}, {Newton}, {Nguyen},
  {Nielsen}, {Nissanke}, {Nitz}, {Nocera}, {Nolting}, {Normandin}, {Nuttall},
  {Oberling}, {Ochsner}, {O'Dell}, {Oelker}, {Ogin}, {Oh}, {Oh}, {Ohme},
  {Oliver}, {Oppermann}, {Oram}, {O'Reilly}, {O'Shaughnessy}, {Ottaway},
  {Overmier}, {Owen}, {Pai}, {Pai}, {Palamos}, {Palashov}, {Palomba},
  {Pal-Singh}, {Pan}, {Pankow}, {Pannarale}, {Pant}, {Paoletti}, {Paoli},
  {Papa}, {Paris}, {Parker}, {Pascucci}, {Pasqualetti}, {Passaquieti},
  {Passuello}, {Patricelli}, {Patrick}, {Pearlstone}, {Pedraza}, {Pedurand},
  {Pekowsky}, {Pele}, {Penn}, {Perreca}, {Perri}, {Pfeiffer}, {Phelps},
  {Piccinni}, {Pichot}, {Piergiovanni}, {Pierro}, {Pillant}, {Pinard}, {Pinto},
  {Pitkin}, {Poe}, {Poggiani}, {Popolizio}, {Post}, {Powell}, {Prasad},
  {Predoi}, {Prestegard}, {Price}, {Prijatelj}, {Principe}, {Privitera},
  {Prix}, {Prodi}, {Prokhorov}, {Puncken}, {Punturo}, {Puppo}, {P{\"u}rrer},
  {Qi}, {Qin}, {Qiu}, {Quetschke}, {Quintero}, {Quitzow-James}, {Raab},
  {Rabeling}, {Radkins}, {Raffai}, {Raja}, {Rajan}, {Rakhmanov}, {Rapagnani},
  {Raymond}, {Razzano}, {Re}, {Read}, {Reed}, {Regimbau}, {Rei}, {Reid},
  {Reitze}, {Rew}, {Reyes}, {Ricci}, {Riles}, {Rizzo}, {Robertson}, {Robie},
  {Robinet}, {Rocchi}, {Rolland}, {Rollins}, {Roma}, {Romano}, {Romano},
  {Romanov}, {Romie}, {Rosi{\'n}ska}, {Rowan}, {R{\"u}diger}, {Ruggi}, {Ryan},
  {Sachdev}, {Sadecki}, {Sadeghian}, {Sakellariadou}, {Salconi}, {Saleem},
  {Salemi}, {Samajdar}, {Sammut}, {Sanchez}, {Sandberg}, {Sandeen}, {Sand ers},
  {Sassolas}, {Sathyaprakash}, {Saulson}, {Sauter}, {Savage}, {Sawadsky},
  {Schale}, {Schilling}, {Schmidt}, {Schmidt}, {Schnabel}, {Schofield},
  {Sch{\"o}nbeck}, {Schreiber}, {Schuette}, {Schutz}, {Scott}, {Scott},
  {Sellers}, {Sengupta}, {Sentenac}, {Sequino}, {Sergeev}, {Setyawati},
  {Shaddock}, {Shaffer}, {Shahriar}, {Shaltev}, {Shapiro}, {Shawhan},
  {Sheperd}, {Shoemaker}, {Shoemaker}, {Siellez}, {Siemens}, {Sieniawska},
  {Sigg}, {Silva}, {Singer}, {Singer}, {Singh}, {Singh}, {Singhal}, {Sintes},
  {Slagmolen}, {Smith}, {Smith}, {Smith}, {Son}, {Sorazu}, {Sorrentino},
  {Souradeep}, {Srivastava}, {Staley}, {Steinke}, {Steinlechner},
  {Steinlechner}, {Steinmeyer}, {Stephens}, {Stevenson}, {Stone}, {Strain},
  {Straniero}, {Stratta}, {Strauss}, {Strigin}, {Sturani}, {Stuver},
  {Summerscales}, {Sun}, {Sunil}, {Sutton}, {Swinkels}, {Szczepa{\'n}czyk},
  {Tacca}, {Talukder}, {Tanner}, {T{\'a}pai}, {Tarabrin}, {Taracchini},
  {Taylor}, {Theeg}, {Thirugnanasamband am}, {Thomas}, {Thomas}, {Thomas},
  {Thorne}, {Thrane}, {Tiwari}, {Tiwari}, {Tokmakov}, {Toland}, {Tomlinson},
  {Tonelli}, {Tornasi}, {Torres}, {Torrie}, {T{\"o}yr{\"a}}, {Travasso},
  {Traylor}, {Trifir{\`o}}, {Tringali}, {Trozzo}, {Tse}, {Turconi},
  {Tuyenbayev}, {Ugolini}, {Unnikrishnan}, {Urban}, {Usman}, {Vahlbruch},
  {Vajente}, {Valdes}, {Vallisneri}, {van Bakel}, {van Beuzekom}, {van den
  Brand}, {Van Den Broeck}, {Vand er-Hyde}, {van der Schaaf}, {van Heijningen},
  {van Veggel}, {Vardaro}, {Vass}, {Vas{\'u}th}, {Vaulin}, {Vecchio},
  {Vedovato}, {Veitch}, {Veitch}, {Venkateswara}, {Verkindt}, {Vetrano},
  {Vicer{\'e}}, {Vinciguerra}, {Vine}, {Vinet}, {Vitale}, {Vo}, {Vocca},
  {Vorvick}, {Voss}, {Vousden}, {Vyatchanin}, {Wade}, {Wade}, {Wade}, {Walker},
  {Wallace}, {Walsh}, {Wang}, {Wang}, {Wang}, {Wang}, {Wang}, {Ward}, {Warner},
  {Was}, {Weaver}, {Wei}, {Weinert}, {Weinstein}, {Weiss}, {Wen}, {We{\ss}els},
  {Westphal}, {Wette}, {Whelan}, {Whiting}, {Williams}, {Williamson}, {Willis},
  {Willke}, {Wimmer}, {Winkler}, {Wipf}, {Wittel}, {Woan}, {Woehler}, {Worden},
  {Wright}, {Wu}, {Wu}, {Yablon}, {Yam}, {Yamamoto}, {Yancey}, {Yu}, {Yvert},
  {Zadro{\.z}ny}, {Zangrando}, {Zanolin}, {Zendri}, {Zevin}, {Zhang}, {Zhang},
  {Zhang}, {Zhao}, {Zhou}, {Zhou}, {Zhu}, {Zucker}, {Zuraw}, {Zweizig},
  {Boyle}, {Hemberger}, {Kidder}, {Lovelace}, {Ossokine}, {Scheel}, {Szilagyi},
  {Teukolsky}, {LIGO Scientific Collaboration}, \& {VIRGO
  Collaboration}}]{Abbott2016b}
---. 2016{\natexlab{b}}, \prl, 116, 241103

\bibitem[{{Abbott} {et~al.}(2017){Abbott}, {Abbott}, {Abbott}, {Acernese},
  {Ackley}, {Adams}, {Adams}, {Addesso}, {Adhikari}, {Adya}, {Affeldt},
  {Afrough}, {Agarwal}, {Agathos}, {Agatsuma}, {Aggarwal}, {Aguiar}, {Aiello},
  {Ain}, {Ajith}, {Allen}, {Allen}, {Allocca}, {Altin}, {Amato}, {Ananyeva},
  {Anderson}, {Anderson}, {Angelova}, {Antier}, {Appert}, {Arai}, {Araya},
  {Areeda}, {Arnaud}, {Arun}, {Ascenzi}, {Ashton}, {Ast}, {Aston}, {Astone},
  {Atallah}, {Aufmuth}, {Aulbert}, {AultONeal}, {Austin}, {Avila-Alvarez},
  {LIGO Scientific Collaboration}, \& {Virgo Collaboration}}]{Abbott2017}
---. 2017, \prl, 119, 141101

\bibitem[{{Abdul-Masih} {et~al.}(2020){Abdul-Masih}, {Banyard}, {Bodensteiner},
  {Bordier}, {Bowman}, {Dsilva}, {Fabry}, {Hawcroft}, {Mahy}, {Marchant},
  {Raskin}, {Reggiani}, {Shenar}, {Tkachenko}, {Van Winckel}, {Vermeylen}, \&
  {Sana}}]{Abdul-Masih2020}
{Abdul-Masih}, M., {Banyard}, G., {Bodensteiner}, J., {et~al.} 2020, \nat, 580,
  E11

\bibitem[{{Abolfathi} {et~al.}(2018){Abolfathi}, {Aguado}, {Aguilar}, {Allende
  Prieto}, {Almeida}, {Ananna}, {Anders}, {Anderson}, {Andrews}, {Anguiano},
  {Arag{\'o}n-Salamanca}, {Argudo-Fern{\'a}ndez}, {Armengaud}, {Ata},
  {Aubourg}, {Avila-Reese}, {Badenes}, {Bailey}, {Balland}, {Barger},
  {Barrera-Ballesteros}, {Bartosz}, {Bastien}, {Bates}, {Baumgarten},
  {Bautista}, {Beaton}, {Beers}, {Belfiore}, {Bender}, {Bernardi}, {Bershady},
  {Beutler}, {Bird}, {Bizyaev}, {Blanc}, {Blanton}, {Blomqvist}, {Bolton},
  {Boquien}, {Borissova}, {Bovy}, {Bradna Diaz}, {Brandt}, {Brinkmann},
  {Brownstein}, {Bundy}, {Burgasser}, {Burtin}, {Busca}, {Ca{\~n}as},
  {Cano-D{\'\i}az}, {Cappellari}, {Carrera}, {Casey}, {Cervantes Sodi}, {Chen},
  {Cherinka}, {Chiappini}, {Choi}, {Chojnowski}, {Chuang}, {Chung}, {Clerc},
  {Cohen}, {Comerford}, {Comparat}, {Correa do Nascimento}, {da Costa},
  {Cousinou}, {Covey}, {Crane}, {Cruz-Gonzalez}, {Cunha}, {da Silva Ilha},
  {Damke}, {Darling}, {Davidson}, {Dawson}, {de Icaza Lizaola}, {de la
  Macorra}, {de la Torre}, {De Lee}, {de Sainte Agathe}, {Deconto Machado},
  {Dell'Agli}, {Delubac}, {Diamond-Stanic}, {Donor}, {Downes}, {Drory}, {du Mas
  des Bourboux}, {Duckworth}, {Dwelly}, {Dyer}, {Ebelke}, {Davis Eigenbrot},
  {Eisenstein}, {Elsworth}, {Emsellem}, {Eracleous}, {Erfanianfar},
  {Escoffier}, {Fan}, {Fern{\'a}ndez Alvar}, {Fernandez-Trincado}, {Fernand o
  Cirolini}, {Feuillet}, {Finoguenov}, {Fleming}, {Font-Ribera}, {Freischlad},
  {Frinchaboy}, {Fu}, {G{\'o}mez Maqueo Chew}, {Galbany}, {Garc{\'\i}a
  P{\'e}rez}, {Garcia-Dias}, {Garc{\'\i}a-Hern{\'a}ndez}, {Garma Oehmichen},
  {Gaulme}, {Gelfand }, {Gil-Mar{\'\i}n}, {Gillespie}, {Goddard}, {Gonz{\'a}lez
  Hern{\'a}ndez}, {Gonzalez-Perez}, {Grabowski}, {Green}, {Grier}, {Gueguen},
  {Guo}, {Guy}, {Hagen}, {Hall}, {Harding}, {Hasselquist}, {Hawley}, {Hayes},
  {Hearty}, {Hekker}, {Hernand ez}, {Hernandez Toledo}, {Hogg},
  {Holley-Bockelmann}, {Holtzman}, {Hou}, {Hsieh}, {Hunt}, {Hutchinson},
  {Hwang}, {Jimenez Angel}, {Johnson}, {Jones}, {J{\"o}nsson}, {Jullo}, {Khan},
  {Kinemuchi}, {Kirkby}, {Kirkpatrick}, {Kitaura}, {Knapp}, {Kneib},
  {Kollmeier}, {Lacerna}, {Lane}, {Lang}, {Law}, {Le Goff}, {Lee}, {Li}, {Li},
  {Lian}, {Liang}, {Lima}, {Lin}, {Long}, {Lucatello}, {Lundgren}, {Mackereth},
  {MacLeod}, {Mahadevan}, {Maia}, {Majewski}, {Manchado}, {Maraston},
  {Mariappan}, {Marques-Chaves}, {Masseron}, {Masters}, {McDermid}, {McGreer},
  {Melendez}, {Meneses-Goytia}, {Merloni}, {Merrifield}, {Meszaros}, {Meza},
  {Minchev}, {Minniti}, {Mueller}, {Muller-Sanchez}, {Muna}, {Mu{\~n}oz},
  {Myers}, {Nair}, {Nand ra}, {Ness}, {Newman}, {Nichol}, {Nidever},
  {Nitschelm}, {Noterdaeme}, {O'Connell}, {Oelkers}, {Oravetz}, {Oravetz},
  {Ort{\'\i}z}, {Osorio}, {Pace}, {Padilla}, {Palanque-Delabrouille},
  {Palicio}, {Pan}, {Pan}, {Parikh}, {P{\^a}ris}, {Park}, {Peirani},
  {Pellejero-Ibanez}, {Penny}, {Percival}, {Perez-Fournon}, {Petitjean},
  {Pieri}, {Pinsonneault}, {Pisani}, {Prada}, {Prakash}, {Queiroz}, {Raddick},
  {Raichoor}, {Barboza Rembold}, {Richstein}, {Riffel}, {Riffel}, {Rix},
  {Robin}, {Rodr{\'\i}guez Torres}, {Rom{\'a}n-Z{\'u}{\~n}iga}, {Ross},
  {Rossi}, {Ruan}, {Ruggeri}, {Ruiz}, {Salvato}, {S{\'a}nchez}, {S{\'a}nchez},
  {Sanchez Almeida}, {S{\'a}nchez-Gallego}, {Santana Rojas}, {Santiago},
  {Schiavon}, {Schimoia}, {Schlafly}, {Schlegel}, {Schneider}, {Schuster},
  {Schwope}, {Seo}, {Serenelli}, {Shen}, {Shen}, {Shetrone}, {Shull}, {Silva
  Aguirre}, {Simon}, {Skrutskie}, {Slosar}, {Smethurst}, {Smith}, {Sobeck},
  {Somers}, {Souter}, {Souto}, {Spindler}, {Stark}, {Stassun}, {Steinmetz},
  {Stello}, {Storchi-Bergmann}, {Streblyanska}, {Stringfellow}, {Su{\'a}rez},
  {Sun}, {Szigeti}, {Taghizadeh-Popp}, {Talbot}, {Tang}, {Tao}, {Tayar},
  {Tembe}, {Teske}, {Thakar}, {Thomas}, {Tissera}, {Tojeiro}, {Tremonti},
  {Troup}, {Urry}, {Valenzuela}, {van den Bosch}, {Vargas-Gonz{\'a}lez},
  {Vargas-Maga{\~n}a}, {Vazquez}, {Villanova}, {Vogt}, {Wake}, {Wang},
  {Weaver}, {Weijmans}, {Weinberg}, {Westfall}, {Whelan}, {Wilcots}, {Wild},
  {Williams}, {Wilson}, {Wood-Vasey}, {Wylezalek}, {Xiao}, {Yan}, {Yang},
  {Ybarra}, {Y{\`e}che}, {Zakamska}, {Zamora}, {Zarrouk}, {Zasowski}, {Zhang},
  {Zhao}, {Zhao}, {Zheng}, {Zheng}, {Zhou}, {Zhu}, {Zinn}, \&
  {Zou}}]{SDSS_DR14}
{Abolfathi}, B., {Aguado}, D.~S., {Aguilar}, G., {et~al.} 2018, \apjs, 235, 42

\bibitem[{{Abt}(1983)}]{Abt1983}
{Abt}, H.~A. 1983, \araa, 21, 343

\bibitem[{{Andrews} {et~al.}(2017){Andrews}, {Chanam{\'e}}, \&
  {Ag{\"u}eros}}]{Andrews2017}
{Andrews}, J.~J., {Chanam{\'e}}, J., \& {Ag{\"u}eros}, M.~A. 2017, \mnras, 472,
  675

\bibitem[{{Badenes} {et~al.}(2018){Badenes}, {Mazzola}, {Thompson}, {Covey},
  {Freeman}, {Walker}, {Moe}, {Troup}, {Nidever}, {Allende Prieto}, {Andrews},
  {Barb{\'a}}, {Beers}, {Bovy}, {Carlberg}, {De Lee}, {Johnson}, {Lewis},
  {Majewski}, {Pinsonneault}, {Sobeck}, {Stassun}, {Stringfellow}, \&
  {Zasowski}}]{Badenes2018}
{Badenes}, C., {Mazzola}, C., {Thompson}, T.~A., {et~al.} 2018, \apj, 854, 147

\bibitem[{{Bailer-Jones} {et~al.}(2018){Bailer-Jones}, {Rybizki}, {Fouesneau},
  {Mantelet}, \& {Andrae}}]{Bailer-Jones2018}
{Bailer-Jones}, C.~A.~L., {Rybizki}, J., {Fouesneau}, M., {Mantelet}, G., \&
  {Andrae}, R. 2018, \aj, 156, 58

\bibitem[{{Bartko} {et~al.}(2010){Bartko}, {Martins}, {Trippe}, {Fritz},
  {Genzel}, {Ott}, {Eisenhauer}, {Gillessen}, {Paumard}, {Alexander},
  {Dodds-Eden}, {Gerhard}, {Levin}, {Mascetti}, {Nayakshin}, {Perets},
  {Perrin}, {Pfuhl}, {Reid}, {Rouan}, {Zilka}, \& {Sternberg}}]{Bartko2010}
{Bartko}, H., {Martins}, F., {Trippe}, S., {et~al.} 2010, \apj, 708, 834

\bibitem[{{Belczynski} {et~al.}(2016){Belczynski}, {Holz}, {Bulik}, \&
  {O'Shaughnessy}}]{Belczynski2016}
{Belczynski}, K., {Holz}, D.~E., {Bulik}, T., \& {O'Shaughnessy}, R. 2016,
  \nat, 534, 512

\bibitem[{{Belokurov} {et~al.}(2020){Belokurov}, {Penoyre}, {Oh}, {Iorio},
  {Hodgkin}, {Evans}, {Everall}, {Koposov}, {Tout}, {Izzard}, {Clarke}, \&
  {Brown}}]{Belokurov2020}
{Belokurov}, V., {Penoyre}, Z., {Oh}, S., {et~al.} 2020, \mnras, 496, 1922

\bibitem[{{Bressan} {et~al.}(2012){Bressan}, {Marigo}, {Girardi}, {Salasnich},
  {Dal Cero}, {Rubele}, \& {Nanni}}]{Bressan2012}
{Bressan}, A., {Marigo}, P., {Girardi}, L., {et~al.} 2012, \mnras, 427, 127

\bibitem[{{Carlin} {et~al.}(2015){Carlin}, {Liu}, {Newberg}, {Beers}, {Chen},
  {Deng}, {Guhathakurta}, {Hou}, {Hou}, {L{\'e}pine}, {Li}, {Luo}, {Smith},
  {Wu}, {Yang}, {Yanny}, {Zhang}, \& {Zheng}}]{Carlin2015}
{Carlin}, J.~L., {Liu}, C., {Newberg}, H.~J., {et~al.} 2015, \aj, 150, 4

\bibitem[{{Castelli} \& {Kurucz}(2003)}]{Castelli2003}
{Castelli}, F., \& {Kurucz}, R.~L. 2003, in IAU Symposium, Vol. 210, Modelling
  of Stellar Atmospheres, ed. N.~{Piskunov}, W.~W. {Weiss}, \& D.~F. {Gray},
  A20

\bibitem[{{Chen} {et~al.}(2019{\natexlab{a}}){Chen}, {Huang}, {Hou}, {Tian},
  {Li}, {Yuan}, {Wang}, {Wang}, {Tian}, \& {Liu}}]{Chen2019}
{Chen}, B.~Q., {Huang}, Y., {Hou}, L.~G., {et~al.} 2019{\natexlab{a}}, \mnras,
  487, 1400

\bibitem[{{Chen} {et~al.}(2019{\natexlab{b}}){Chen}, {Huang}, {Yuan}, {Wang},
  {Fan}, {Xiang}, {Zhang}, {Tian}, \& {Liu}}]{Chen2019b}
{Chen}, B.~Q., {Huang}, Y., {Yuan}, H.~B., {et~al.} 2019{\natexlab{b}}, \mnras,
  483, 4277

\bibitem[{{Cheng} {et~al.}(2019){Cheng}, {Liu}, {Mao}, \& {Cui}}]{Cheng2019}
{Cheng}, X., {Liu}, C., {Mao}, S., \& {Cui}, W. 2019, \apjl, 872, L1

\bibitem[{{Chieffi} \& {Limongi}(2004)}]{Chieffi2004}
{Chieffi}, A., \& {Limongi}, M. 2004, \apj, 608, 405

\bibitem[{{Choi} {et~al.}(2016){Choi}, {Dotter}, {Conroy}, {Cantiello},
  {Paxton}, \& {Johnson}}]{Choi2016}
{Choi}, J., {Dotter}, A., {Conroy}, C., {et~al.} 2016, \apj, 823, 102

\bibitem[{{Coronado} {et~al.}(2018{\natexlab{a}}){Coronado}, {Rix}, \&
  {Trick}}]{Coronado2018}
{Coronado}, J., {Rix}, H.-W., \& {Trick}, W.~H. 2018{\natexlab{a}}, \mnras,
  481, 2970

\bibitem[{{Coronado} {et~al.}(2018{\natexlab{b}}){Coronado}, {Sep{\'u}lveda},
  {Gould}, \& {Chanam{\'e}}}]{Coronado2018b}
{Coronado}, J., {Sep{\'u}lveda}, M.~P., {Gould}, A., \& {Chanam{\'e}}, J.
  2018{\natexlab{b}}, \mnras, 480, 4302

\bibitem[{{Deng} {et~al.}(2020){Deng}, {Sun}, {Jian}, {Jiang}, \&
  {Yuan}}]{Deng2020}
{Deng}, D., {Sun}, Y., {Jian}, M., {Jiang}, B., \& {Yuan}, H. 2020, \aj, 159,
  208

\bibitem[{{Deng} {et~al.}(2012){Deng}, {Newberg}, {Liu}, {Carlin}, {Beers},
  {Chen}, {Chen}, {Christlieb}, {Grillmair}, {Guhathakurta}, {Han}, {Hou},
  {Lee}, {L{\'e}pine}, {Li}, {Liu}, {Pan}, {Sellwood}, {Wang}, {Wang}, {Yang},
  {Yanny}, {Zhang}, {Zhang}, {Zheng}, \& {Zhu}}]{Deng2012}
{Deng}, L.-C., {Newberg}, H.~J., {Liu}, C., {et~al.} 2012, Research in
  Astronomy and Astrophysics, 12, 735

\bibitem[{{Duch{\^e}ne} \& {Kraus}(2013)}]{Duchene2013}
{Duch{\^e}ne}, G., \& {Kraus}, A. 2013, \araa, 51, 269

\bibitem[{{El-Badry} \& {Quataert}(2020{\natexlab{a}})}]{El-Badry2020a}
{El-Badry}, K., \& {Quataert}, E. 2020{\natexlab{a}}, \mnras, 493, L22

\bibitem[{{El-Badry} \& {Quataert}(2020{\natexlab{b}})}]{El-Badry2020b}
---. 2020{\natexlab{b}}, arXiv e-prints, arXiv:2006.11974

\bibitem[{{El-Badry} \& {Rix}(2018)}]{El-Badry2018c}
{El-Badry}, K., \& {Rix}, H.-W. 2018, \mnras, 480, 4884

\bibitem[{{El-Badry} {et~al.}(2018{\natexlab{a}}){El-Badry}, {Rix}, {Ting},
  {Weisz}, {Bergemann}, {Cargile}, {Conroy}, \& {Eilers}}]{El-Badry2018}
{El-Badry}, K., {Rix}, H.-W., {Ting}, Y.-S., {et~al.} 2018{\natexlab{a}},
  \mnras, 473, 5043

\bibitem[{{El-Badry} {et~al.}(2018{\natexlab{b}}){El-Badry}, {Ting}, {Rix},
  {Quataert}, {Weisz}, {Cargile}, {Conroy}, {Hogg}, {Bergemann}, \&
  {Liu}}]{El-Badry2018b}
{El-Badry}, K., {Ting}, Y.-S., {Rix}, H.-W., {et~al.} 2018{\natexlab{b}},
  \mnras, 476, 528

\bibitem[{{Eldridge} {et~al.}(2020){Eldridge}, {Stanway}, {Breivik}, {Casey},
  {Steeghs}, \& {Stevance}}]{Eldridge2020}
{Eldridge}, J.~J., {Stanway}, E.~R., {Breivik}, K., {et~al.} 2020, \mnras, 495,
  2786

\bibitem[{{Evans} {et~al.}(2018){Evans}, {Riello}, {De Angeli}, {Carrasco},
  {Montegriffo}, {Fabricius}, {Jordi}, {Palaversa}, {Diener}, {Busso},
  {Cacciari}, {van Leeuwen}, {Burgess}, {Davidson}, {Harrison}, {Hodgkin},
  {Pancino}, {Richards}, {Altavilla}, {Balaguer-N{\'u}{\~n}ez}, {Barstow},
  {Bellazzini}, {Brown}, {Castellani}, {Cocozza}, {De Luise}, {Delgado},
  {Ducourant}, {Galleti}, {Gilmore}, {Giuffrida}, {Holl}, {Kewley}, {Koposov},
  {Marinoni}, {Marrese}, {Osborne}, {Piersimoni}, {Portell}, {Pulone},
  {Ragaini}, {Sanna}, {Terrett}, {Walton}, {Wevers}, \&
  {Wyrzykowski}}]{Evans2018}
{Evans}, D.~W., {Riello}, M., {De Angeli}, F., {et~al.} 2018, \aap, 616, A4

\bibitem[{{Fernandez} {et~al.}(2017){Fernandez}, {Covey}, {De Lee},
  {Chojnowski}, {Nidever}, {Ballantyne}, {Cottaar}, {Da Rio}, {Foster},
  {Majewski}, {Meyer}, {Reyna}, {Roberts}, {Skinner}, {Stassun}, {Tan},
  {Troup}, \& {Zasowski}}]{Fernandez2017}
{Fernandez}, M.~A., {Covey}, K.~R., {De Lee}, N., {et~al.} 2017, \pasp, 129,
  084201

\bibitem[{{Fitzpatrick}(1999)}]{Fitzpatrick1999}
{Fitzpatrick}, E.~L. 1999, \pasp, 111, 63

\bibitem[{{Freyer} {et~al.}(2003){Freyer}, {Hensler}, \& {Yorke}}]{Freyer2003}
{Freyer}, T., {Hensler}, G., \& {Yorke}, H.~W. 2003, \apj, 594, 888

\bibitem[{{Freyer} {et~al.}(2006){Freyer}, {Hensler}, \& {Yorke}}]{Freyer2006}
---. 2006, \apj, 638, 262

\bibitem[{{Gaia Collaboration} {et~al.}(2016){Gaia Collaboration}, {Prusti},
  {de Bruijne}, {Brown}, {Vallenari}, {Babusiaux}, {Bailer-Jones}, {Bastian},
  {Biermann}, {Evans}, {Eyer}, {Jansen}, {Jordi}, {Klioner}, {Lammers},
  {Lindegren}, {Luri}, {Mignard}, {Milligan}, {Panem}, {Poinsignon},
  {Pourbaix}, {Randich}, {Sarri}, {Sartoretti}, {Siddiqui}, {Soubiran},
  {Valette}, {van Leeuwen}, {Walton}, {Aerts}, {Arenou}, {Cropper}, {Drimmel},
  {H{\o}g}, {Katz}, {Lattanzi}, {O'Mullane}, {Grebel}, {Holland}, {Huc},
  {Passot}, {Bramante}, {Cacciari}, {Casta{\~n}eda}, {Chaoul}, {Cheek}, {De
  Angeli}, {Fabricius}, {Guerra}, {Hern{\'a}ndez}, {Jean-Antoine-Piccolo},
  {Masana}, {Messineo}, {Mowlavi}, {Nienartowicz}, {Ord{\'o}{\~n}ez-Blanco},
  {Panuzzo}, {Portell}, {Richards}, {Riello}, {Seabroke}, {Tanga},
  {Th{\'e}venin}, {Torra}, {Els}, {Gracia-Abril}, {Comoretto},
  {Garcia-Reinaldos}, {Lock}, {Mercier}, {Altmann}, {Andrae}, {Astraatmadja},
  {Bellas-Velidis}, {Benson}, {Berthier}, {Blomme}, {Busso}, {Carry},
  {Cellino}, {Clementini}, {Cowell}, {Creevey}, {Cuypers}, {Davidson}, {De
  Ridder}, {de Torres}, {Delchambre}, {Dell'Oro}, {Ducourant}, {Fr{\'e}mat},
  {Garc{\'\i}a-Torres}, {Gosset}, {Halbwachs}, {Hambly}, {Harrison}, {Hauser},
  {Hestroffer}, {Hodgkin}, {Huckle}, {Hutton}, {Jasniewicz}, {Jordan},
  {Kontizas}, {Korn}, {Lanzafame}, {Manteiga}, {Moitinho}, {Muinonen},
  {Osinde}, {Pancino}, {Pauwels}, {Petit}, {Recio-Blanco}, {Robin}, {Sarro},
  {Siopis}, {Smith}, {Smith}, {Sozzetti}, {Thuillot}, {van Reeven}, {Viala},
  {Abbas}, {Abreu Aramburu}, {Accart}, {Aguado}, {Allan}, {Allasia},
  {Altavilla}, {{\'A}lvarez}, {Alves}, {Anderson}, {Andrei}, {Anglada Varela},
  {Antiche}, {Antoja}, {Ant{\'o}n}, {Arcay}, {Atzei}, {Ayache}, {Bach},
  {Baker}, {Balaguer-N{\'u}{\~n}ez}, {Barache}, {Barata}, {Barbier}, {Barblan},
  {Baroni}, {Barrado y Navascu{\'e}s}, {Barros}, {Barstow}, {Becciani},
  {Bellazzini}, {Bellei}, {Bello Garc{\'\i}a}, {Belokurov}, {Bendjoya},
  {Berihuete}, {Bianchi}, {Bienaym{\'e}}, {Billebaud}, {Blagorodnova},
  {Blanco-Cuaresma}, {Boch}, {Bombrun}, {Borrachero}, {Bouquillon}, {Bourda},
  {Bouy}, {Bragaglia}, {Breddels}, {Brouillet}, {Br{\"u}semeister},
  {Bucciarelli}, {Budnik}, {Burgess}, {Burgon}, {Burlacu}, {Busonero}, {Buzzi},
  {Caffau}, {Cambras}, {Campbell}, {Cancelliere}, {Cantat-Gaudin}, {Carlucci},
  {Carrasco}, {Castellani}, {Charlot}, {Charnas}, {Charvet}, {Chassat},
  {Chiavassa}, {Clotet}, {Cocozza}, {Collins}, {Collins}, {Costigan}, {Crifo},
  {Cross}, {Crosta}, {Crowley}, {Dafonte}, {Damerdji}, {Dapergolas}, {David},
  {David}, {De Cat}, {de Felice}, {de Laverny}, {De Luise}, {De March}, {de
  Martino}, {de Souza}, {Debosscher}, {del Pozo}, {Delbo}, {Delgado},
  {Delgado}, {di Marco}, {Di Matteo}, {Diakite}, {Distefano}, {Dolding}, {Dos
  Anjos}, {Drazinos}, {Dur{\'a}n}, {Dzigan}, {Ecale}, {Edvardsson}, {Enke},
  {Erdmann}, {Escolar}, {Espina}, {Evans}, {Eynard Bontemps}, {Fabre},
  {Fabrizio}, {Faigler}, {Falc{\~a}o}, {Farr{\`a}s Casas}, {Faye}, {Federici},
  {Fedorets}, {Fern{\'a}ndez-Hern{\'a}ndez}, {Fernique}, {Fienga}, {Figueras},
  {Filippi}, {Findeisen}, {Fonti}, {Fouesneau}, {Fraile}, {Fraser}, {Fuchs},
  {Furnell}, {Gai}, {Galleti}, {Galluccio}, {Garabato}, {Garc{\'\i}a-Sedano},
  {Gar{\'e}}, {Garofalo}, {Garralda}, {Gavras}, {Gerssen}, {Geyer}, {Gilmore},
  {Girona}, {Giuffrida}, {Gomes}, {Gonz{\'a}lez-Marcos},
  {Gonz{\'a}lez-N{\'u}{\~n}ez}, {Gonz{\'a}lez-Vidal}, {Granvik}, {Guerrier},
  {Guillout}, {Guiraud}, {G{\'u}rpide}, {Guti{\'e}rrez-S{\'a}nchez}, {Guy},
  {Haigron}, {Hatzidimitriou}, {Haywood}, {Heiter}, {Helmi}, {Hobbs},
  {Hofmann}, {Holl}, {Holland }, {Hunt}, {Hypki}, {Icardi}, {Irwin}, {Jevardat
  de Fombelle}, {Jofr{\'e}}, {Jonker}, {Jorissen}, {Julbe}, {Karampelas},
  {Kochoska}, {Kohley}, {Kolenberg}, {Kontizas}, {Koposov}, {Kordopatis},
  {Koubsky}, {Kowalczyk}, {Krone-Martins}, {Kudryashova}, {Kull}, {Bachchan},
  {Lacoste-Seris}, {Lanza}, {Lavigne}, {Le Poncin-Lafitte}, {Lebreton},
  {Lebzelter}, {Leccia}, {Leclerc}, {Lecoeur-Taibi}, {Lemaitre}, {Lenhardt},
  {Leroux}, {Liao}, {Licata}, {Lindstr{\o}m}, {Lister}, {Livanou}, {Lobel},
  {L{\"o}ffler}, {L{\'o}pez}, {Lopez-Lozano}, {Lorenz}, {Loureiro},
  {MacDonald}, {Magalh{\~a}es Fernandes}, {Managau}, {Mann}, {Mantelet},
  {Marchal}, {Marchant}, {Marconi}, {Marie}, {Marinoni}, {Marrese},
  {Marschalk{\'o}}, {Marshall}, {Mart{\'\i}n-Fleitas}, {Martino}, {Mary},
  {Matijevi{\v{c}}}, {Mazeh}, {McMillan}, {Messina}, {Mestre}, {Michalik},
  {Millar}, {Miranda}, {Molina}, {Molinaro}, {Molinaro}, {Moln{\'a}r},
  {Moniez}, {Montegriffo}, {Monteiro}, {Mor}, {Mora}, {Morbidelli}, {Morel},
  {Morgenthaler}, {Morley}, {Morris}, {Mulone}, {Muraveva}, {Musella},
  {Narbonne}, {Nelemans}, {Nicastro}, {Noval}, {Ord{\'e}novic},
  {Ordieres-Mer{\'e}}, {Osborne}, {Pagani}, {Pagano}, {Pailler}, {Palacin},
  {Palaversa}, {Parsons}, {Paulsen}, {Pecoraro}, {Pedrosa}, {Pentik{\"a}inen},
  {Pereira}, {Pichon}, {Piersimoni}, {Pineau}, {Plachy}, {Plum}, {Poujoulet},
  {Pr{\v{s}}a}, {Pulone}, {Ragaini}, {Rago}, {Rambaux}, {Ramos-Lerate},
  {Ranalli}, {Rauw}, {Read}, {Regibo}, {Renk}, {Reyl{\'e}}, {Ribeiro},
  {Rimoldini}, {Ripepi}, {Riva}, {Rixon}, {Roelens}, {Romero-G{\'o}mez},
  {Rowell}, {Royer}, {Rudolph}, {Ruiz-Dern}, {Sadowski}, {Sagrist{\`a}
  Sell{\'e}s}, {Sahlmann}, {Salgado}, {Salguero}, {Sarasso}, {Savietto},
  {Schnorhk}, {Schultheis}, {Sciacca}, {Segol}, {Segovia}, {Segransan},
  {Serpell}, {Shih}, {Smareglia}, {Smart}, {Smith}, {Solano}, {Solitro},
  {Sordo}, {Soria Nieto}, {Souchay}, {Spagna}, {Spoto}, {Stampa}, {Steele},
  {Steidelm{\"u}ller}, {Stephenson}, {Stoev}, {Suess}, {S{\"u}veges}, {Surdej},
  {Szabados}, {Szegedi-Elek}, {Tapiador}, {Taris}, {Tauran}, {Taylor},
  {Teixeira}, {Terrett}, {Tingley}, {Trager}, {Turon}, {Ulla}, {Utrilla},
  {Valentini}, {van Elteren}, {Van Hemelryck}, {van Leeuwen}, {Varadi},
  {Vecchiato}, {Veljanoski}, {Via}, {Vicente}, {Vogt}, {Voss}, {Votruba},
  {Voutsinas}, {Walmsley}, {Weiler}, {Weingrill}, {Werner}, {Wevers},
  {Whitehead}, {Wyrzykowski}, {Yoldas}, {{\v{Z}}erjal}, {Zucker}, {Zurbach},
  {Zwitter}, {Alecu}, {Allen}, {Allende Prieto}, {Amorim},
  {Anglada-Escud{\'e}}, {Arsenijevic}, {Azaz}, {Balm}, {Beck}, {Bernstein},
  {Bigot}, {Bijaoui}, {Blasco}, {Bonfigli}, {Bono}, {Boudreault}, {Bressan},
  {Brown}, {Brunet}, {Bunclark}, {Buonanno}, {Butkevich}, {Carret}, {Carrion},
  {Chemin}, {Ch{\'e}reau}, {Corcione}, {Darmigny}, {de Boer}, {de Teodoro}, {de
  Zeeuw}, {Delle Luche}, {Domingues}, {Dubath}, {Fodor}, {Fr{\'e}zouls},
  {Fries}, {Fustes}, {Fyfe}, {Gallardo}, {Gallegos}, {Gardiol}, {Gebran},
  {Gomboc}, {G{\'o}mez}, {Grux}, {Gueguen}, {Heyrovsky}, {Hoar}, {Iannicola},
  {Isasi Parache}, {Janotto}, {Joliet}, {Jonckheere}, {Keil}, {Kim},
  {Klagyivik}, {Klar}, {Knude}, {Kochukhov}, {Kolka}, {Kos}, {Kutka}, {Lainey},
  {LeBouquin}, {Liu}, {Loreggia}, {Makarov}, {Marseille}, {Martayan},
  {Martinez-Rubi}, {Massart}, {Meynadier}, {Mignot}, {Munari}, {Nguyen},
  {Nordlander}, {Ocvirk}, {O'Flaherty}, {Olias Sanz}, {Ortiz}, {Osorio},
  {Oszkiewicz}, {Ouzounis}, {Palmer}, {Park}, {Pasquato}, {Peltzer}, {Peralta},
  {P{\'e}turaud}, {Pieniluoma}, {Pigozzi}, {Poels}, {Prat}, {Prod'homme},
  {Raison}, {Rebordao}, {Risquez}, {Rocca-Volmerange}, {Rosen}, {Ruiz-Fuertes},
  {Russo}, {Sembay}, {Serraller Vizcaino}, {Short}, {Siebert}, {Silva},
  {Sinachopoulos}, {Slezak}, {Soffel}, {Sosnowska}, {Strai{\v{z}}ys}, {ter
  Linden}, {Terrell}, {Theil}, {Tiede}, {Troisi}, {Tsalmantza}, {Tur},
  {Vaccari}, {Vachier}, {Valles}, {Van Hamme}, {Veltz}, {Virtanen}, {Wallut},
  {Wichmann}, {Wilkinson}, {Ziaeepour}, \& {Zschocke}}]{Prusti2016}
{Gaia Collaboration}, {Prusti}, T., {de Bruijne}, J.~H.~J., {et~al.} 2016,
  \aap, 595, A1

\bibitem[{{Gaia Collaboration} {et~al.}(2018{\natexlab{a}}){Gaia
  Collaboration}, {Brown}, {Vallenari}, {Prusti}, {de Bruijne}, {Babusiaux},
  {Bailer-Jones}, {Biermann}, {Evans}, {Eyer}, {Jansen}, {Jordi}, {Klioner},
  {Lammers}, {Lindegren}, {Luri}, {Mignard}, {Panem}, {Pourbaix}, {Randich}, \&
  {Sartoretti}}]{Brown2018}
{Gaia Collaboration}, {Brown}, A.~G.~A., {Vallenari}, A., {et~al.}
  2018{\natexlab{a}}, \aap, 616, A1

\bibitem[{{Gaia Collaboration} {et~al.}(2018{\natexlab{b}}){Gaia
  Collaboration}, {Babusiaux}, {van Leeuwen}, {Barstow}, {Jordi}, {Vallenari},
  {Bossini}, {Bressan}, {Cantat-Gaudin}, {van Leeuwen}, {Brown}, {Prusti}, {de
  Bruijne}, {Bailer-Jones}, {Biermann}, {Evans}, {Eyer}, {Jansen}, {Klioner},
  {Lammers}, {Lindegren}, {Luri}, {Mignard}, {Panem}, {Pourbaix}, {Randich},
  {Sartoretti}, {Siddiqui}, {Soubiran}, {Walton}, {Arenou}, {Bastian},
  {Cropper}, {Drimmel}, {Katz}, {Lattanzi}, {Bakker}, {Cacciari},
  {Casta{\~n}eda}, {Chaoul}, {Cheek}, {De Angeli}, {Fabricius}, {Guerra},
  {Holl}, {Masana}, {Messineo}, {Mowlavi}, {Nienartowicz}, {Panuzzo},
  {Portell}, {Riello}, {Seabroke}, {Tanga}, {Th{\'e}venin}, {Gracia-Abril},
  {Comoretto}, {Garcia-Reinaldos}, {Teyssier}, {Altmann}, {Andrae}, {Audard},
  {Bellas-Velidis}, {Benson}, {Berthier}, {Blomme}, {Burgess}, {Busso},
  {Carry}, {Cellino}, {Clementini}, {Clotet}, {Creevey}, {Davidson}, {De
  Ridder}, {Delchambre}, {Dell'Oro}, {Ducourant},
  {Fern{\'a}ndez-Hern{\'a}ndez}, {Fouesneau}, {Fr{\'e}mat}, {Galluccio},
  {Garc{\'\i}a-Torres}, {Gonz{\'a}lez-N{\'u}{\~n}ez}, {Gonz{\'a}lez-Vidal},
  {Gosset}, {Guy}, {Halbwachs}, {Hambly}, {Harrison}, {Hern{\'a}ndez},
  {Hestroffer}, {Hodgkin}, {Hutton}, {Jasniewicz}, {Jean-Antoine-Piccolo},
  {Jordan}, {Korn}, {Krone-Martins}, {Lanzafame}, {Lebzelter}, {L{\"o}ffler},
  {Manteiga}, {Marrese}, {Mart{\'\i}n-Fleitas}, {Moitinho}, {Mora}, {Muinonen},
  {Osinde}, {Pancino}, {Pauwels}, {Petit}, {Recio-Blanco}, {Richards},
  {Rimoldini}, {Robin}, {Sarro}, {Siopis}, {Smith}, {Sozzetti}, {S{\"u}veges},
  {Torra}, {van Reeven}, {Abbas}, {Abreu Aramburu}, {Accart}, {Aerts},
  {Altavilla}, {{\'A}lvarez}, {Alvarez}, {Alves}, {Anderson}, {Andrei},
  {Anglada Varela}, {Antiche}, {Antoja}, {Arcay}, {Astraatmadja}, {Bach},
  {Baker}, {Balaguer-N{\'u}{\~n}ez}, {Balm}, {Barache}, {Barata}, {Barbato},
  {Barblan}, {Barklem}, {Barrado}, {Barros}, {Bartholom{\'e} Mu{\~n}oz},
  {Bassilana}, {Becciani}, {Bellazzini}, {Berihuete}, {Bertone}, {Bianchi},
  {Bienaym{\'e}}, {Blanco-Cuaresma}, {Boch}, {Boeche}, {Bombrun}, {Borrachero},
  {Bouquillon}, {Bourda}, {Bragaglia}, {Bramante}, {Breddels}, {Brouillet},
  {Br{\"u}semeister}, {Brugaletta}, {Bucciarelli}, {Burlacu}, {Busonero},
  {Butkevich}, {Buzzi}, {Caffau}, {Cancelliere}, {Cannizzaro}, {Carballo},
  {Carlucci}, {Carrasco}, {Casamiquela}, {Castellani}, {Castro-Ginard},
  {Charlot}, {Chemin}, {Chiavassa}, {Cocozza}, {Costigan}, {Cowell}, {Crifo},
  {Crosta}, {Crowley}, {Cuypers}, {Dafonte}, {Damerdji}, {Dapergolas}, {David},
  {David}, {de Laverny}, {De Luise}, {De March}, {de Martino}, {de Souza}, {de
  Torres}, {Debosscher}, {del Pozo}, {Delbo}, {Delgado}, {Delgado}, {Diakite},
  {Diener}, {Distefano}, {Dolding}, {Drazinos}, {Dur{\'a}n}, {Edvardsson},
  {Enke}, {Eriksson}, {Esquej}, {Eynard Bontemps}, {Fabre}, {Fabrizio},
  {Faigler}, {Falc{\~a}o}, {Farr{\`a}s Casas}, {Federici}, {Fedorets},
  {Fernique}, {Figueras}, {Filippi}, {Findeisen}, {Fonti}, {Fraile}, {Fraser},
  {Fr{\'e}zouls}, {Gai}, {Galleti}, {Garabato}, {Garc{\'\i}a-Sedano},
  {Garofalo}, {Garralda}, {Gavel}, {Gavras}, {Gerssen}, {Geyer}, {Giacobbe},
  {Gilmore}, {Girona}, {Giuffrida}, {Glass}, {Gomes}, {Granvik}, {Gueguen},
  {Guerrier}, {Guiraud}, {Guti{\'e}}, {Haigron}, {Hatzidimitriou}, {Hauser},
  {Haywood}, {Heiter}, {Helmi}, {Heu}, {Hilger}, {Hobbs}, {Hofmann}, {Holland},
  {Huckle}, {Hypki}, {Icardi}, {Jan{\ss}en}, {Jevardat de Fombelle}, {Jonker},
  {Juh{\'a}sz}, {Julbe}, {Karampelas}, {Kewley}, {Klar}, {Kochoska}, {Kohley},
  {Kolenberg}, {Kontizas}, {Kontizas}, {Koposov}, {Kordopatis},
  {Kostrzewa-Rutkowska}, {Koubsky}, {Lambert}, {Lanza}, {Lasne}, {Lavigne}, {Le
  Fustec}, {Le Poncin-Lafitte}, {Lebreton}, {Leccia}, {Leclerc},
  {Lecoeur-Taibi}, {Lenhardt}, {Leroux}, {Liao}, {Licata}, {Lindstr{\o}m},
  {Lister}, {Livanou}, {Lobel}, {L{\'o}pez}, {Managau}, {Mann}, {Mantelet},
  {Marchal}, {Marchant}, {Marconi}, {Marinoni}, {Marschalk{\'o}}, {Marshall},
  {Martino}, {Marton}, {Mary}, {Massari}, {Matijevi{\v{c}}}, {Mazeh},
  {McMillan}, {Messina}, {Michalik}, {Millar}, {Molina}, {Molinaro},
  {Moln{\'a}r}, {Montegriffo}, {Mor}, {Morbidelli}, {Morel}, {Morris},
  {Mulone}, {Muraveva}, {Musella}, {Nelemans}, {Nicastro}, {Noval},
  {O'Mullane}, {Ord{\'e}novic}, {Ord{\'o}{\~n}ez-Blanco}, {Osborne}, {Pagani},
  {Pagano}, {Pailler}, {Palacin}, {Palaversa}, {Panahi}, {Pawlak},
  {Piersimoni}, {Pineau}, {Plachy}, {Plum}, {Poggio}, {Poujoulet},
  {Pr{\v{s}}a}, {Pulone}, {Racero}, {Ragaini}, {Rambaux}, {Ramos-Lerate},
  {Regibo}, {Reyl{\'e}}, {Riclet}, {Ripepi}, {Riva}, {Rivard}, {Rixon},
  {Roegiers}, {Roelens}, {Romero-G{\'o}mez}, {Rowell}, {Royer}, {Ruiz-Dern},
  {Sadowski}, {Sagrist{\`a} Sell{\'e}s}, {Sahlmann}, {Salgado}, {Salguero},
  {Sanna}, {Santana-Ros}, {Sarasso}, {Savietto}, {Schultheis}, {Sciacca},
  {Segol}, {Segovia}, {S{\'e}gransan}, {Shih}, {Siltala}, {Silva}, {Smart},
  {Smith}, {Solano}, {Solitro}, {Sordo}, {Soria Nieto}, {Souchay}, {Spagna},
  {Spoto}, {Stampa}, {Steele}, {Steidelm{\"u}ller}, {Stephenson}, {Stoev},
  {Suess}, {Surdej}, {Szabados}, {Szegedi-Elek}, {Tapiador}, {Taris}, {Tauran},
  {Taylor}, {Teixeira}, {Terrett}, {Teyssand ier}, {Thuillot}, {Titarenko},
  {Torra Clotet}, {Turon}, {Ulla}, {Utrilla}, {Uzzi}, {Vaillant}, {Valentini},
  {Valette}, {van Elteren}, {Van Hemelryck}, {Vaschetto}, {Vecchiato},
  {Veljanoski}, {Viala}, {Vicente}, {Vogt}, {von Essen}, {Voss}, {Votruba},
  {Voutsinas}, {Walmsley}, {Weiler}, {Wertz}, {Wevers}, {Wyrzykowski},
  {Yoldas}, {{\v{Z}}erjal}, {Ziaeepour}, {Zorec}, {Zschocke}, {Zucker},
  {Zurbach}, \& {Zwitter}}]{Babusiaux2018}
{Gaia Collaboration}, {Babusiaux}, C., {van Leeuwen}, F., {et~al.}
  2018{\natexlab{b}}, \aap, 616, A10

\bibitem[{{Gao} {et~al.}(2014){Gao}, {Liu}, {Zhang}, {Justham}, {Deng}, \&
  {Yang}}]{Gao2014}
{Gao}, S., {Liu}, C., {Zhang}, X., {et~al.} 2014, \apjl, 788, L37

\bibitem[{{Gao} {et~al.}(2017){Gao}, {Zhao}, {Yang}, \& {Gao}}]{Gao2017}
{Gao}, S., {Zhao}, H., {Yang}, H., \& {Gao}, R. 2017, \mnras, 469, L68

\bibitem[{{Green} {et~al.}(2020){Green}, {Rix}, {Tschesche}, {Finkbeiner},
  {Zucker}, {Schlafly}, {Rybizki}, \& {Speagle}}]{Green2020}
{Green}, G.~M., {Rix}, H.-W., {Tschesche}, L., {et~al.} 2020, arXiv e-prints,
  arXiv:2006.16258

\bibitem[{{Green} {et~al.}(2019){Green}, {Schlafly}, {Zucker}, {Speagle}, \&
  {Finkbeiner}}]{Green2019}
{Green}, G.~M., {Schlafly}, E., {Zucker}, C., {Speagle}, J.~S., \&
  {Finkbeiner}, D. 2019, \apj, 887, 93

\bibitem[{{Henden} {et~al.}(2012){Henden}, {Levine}, {Terrell}, {Smith}, \&
  {Welch}}]{Henden2012}
{Henden}, A.~A., {Levine}, S.~E., {Terrell}, D., {Smith}, T.~C., \& {Welch}, D.
  2012, Journal of the American Association of Variable Star Observers
  (JAAVSO), 40, 430

\bibitem[{{Hogg} {et~al.}(2019){Hogg}, {Eilers}, \& {Rix}}]{Hogg19}
{Hogg}, D.~W., {Eilers}, A.-C., \& {Rix}, H.-W. 2019, \aj, 158, 147

\bibitem[{{Hollands} {et~al.}(2018){Hollands}, {Tremblay}, {G{\"a}nsicke},
  {Gentile-Fusillo}, \& {Toonen}}]{Hollands2018}
{Hollands}, M.~A., {Tremblay}, P.~E., {G{\"a}nsicke}, B.~T., {Gentile-Fusillo},
  N.~P., \& {Toonen}, S. 2018, \mnras, 480, 3942

\bibitem[{{Hopkins} {et~al.}(2014){Hopkins}, {Kere{\v{s}}}, {O{\~n}orbe},
  {Faucher-Gigu{\`e}re}, {Quataert}, {Murray}, \& {Bullock}}]{Hopkins2014}
{Hopkins}, P.~F., {Kere{\v{s}}}, D., {O{\~n}orbe}, J., {et~al.} 2014, \mnras,
  445, 581

\bibitem[{{Humphreys} \& {McElroy}(1984)}]{Humphreys1984}
{Humphreys}, R.~M., \& {McElroy}, D.~B. 1984, \apj, 284, 565

\bibitem[{{Hurley} \& {Tout}(1998)}]{Hurley1998}
{Hurley}, J., \& {Tout}, C.~A. 1998, \mnras, 300, 977

\bibitem[{{Irrgang} {et~al.}(2020){Irrgang}, {Geier}, {Kreuzer}, {Pelisoli}, \&
  {Heber}}]{Irrgang2020}
{Irrgang}, A., {Geier}, S., {Kreuzer}, S., {Pelisoli}, I., \& {Heber}, U. 2020,
  \aap, 633, L5

\bibitem[{{Ivezi{\'c}} {et~al.}(2008){Ivezi{\'c}}, {Sesar}, {Juri{\'c}},
  {Bond}, {Dalcanton}, {Rockosi}, {Yanny}, {Newberg}, {Beers}, {Allende
  Prieto}, {Wilhelm}, {Lee}, {Sivarani}, {Norris}, {Bailer-Jones}, {Re
  Fiorentin}, {Schlegel}, {Uomoto}, {Lupton}, {Knapp}, {Gunn}, {Covey}, {Allyn
  Smith}, {Miknaitis}, {Doi}, {Tanaka}, {Fukugita}, {Kent}, {Finkbeiner},
  {Munn}, {Pier}, {Quinn}, {Hawley}, {Anderson}, {Kiuchi}, {Chen}, {Bushong},
  {Sohi}, {Haggard}, {Kimball}, {Barentine}, {Brewington}, {Harvanek},
  {Kleinman}, {Krzesinski}, {Long}, {Nitta}, {Snedden}, {Lee}, {Harris},
  {Brinkmann}, {Schneider}, \& {York}}]{Ivezic2008}
{Ivezi{\'c}}, {\v{Z}}., {Sesar}, B., {Juri{\'c}}, M., {et~al.} 2008, \apj, 684,
  287

\bibitem[{{Jofr{\'e}} {et~al.}(2015){Jofr{\'e}}, {M{\"a}dler}, {Gilmore},
  {Casey}, {Soubiran}, \& {Worley}}]{Jofre2015}
{Jofr{\'e}}, P., {M{\"a}dler}, T., {Gilmore}, G., {et~al.} 2015, \mnras, 453,
  1428

\bibitem[{{Juri{\'c}} {et~al.}(2008){Juri{\'c}}, {Ivezi{\'c}}, {Brooks},
  {Lupton}, {Schlegel}, {Finkbeiner}, {Padmanabhan}, {Bond}, {Sesar},
  {Rockosi}, {Knapp}, {Gunn}, {Sumi}, {Schneider}, {Barentine}, {Brewington},
  {Brinkmann}, {Fukugita}, {Harvanek}, {Kleinman}, {Krzesinski}, {Long},
  {Neilsen}, {Nitta}, {Snedden}, \& {York}}]{Juric2008}
{Juri{\'c}}, M., {Ivezi{\'c}}, {\v{Z}}., {Brooks}, A., {et~al.} 2008, \apj,
  673, 864

\bibitem[{{Kervella} {et~al.}(2019){Kervella}, {Arenou}, {Mignard}, \&
  {Th{\'e}venin}}]{Kervella2019}
{Kervella}, P., {Arenou}, F., {Mignard}, F., \& {Th{\'e}venin}, F. 2019, \aap,
  623, A72

\bibitem[{{Kouwenhoven} {et~al.}(2005){Kouwenhoven}, {Brown}, {Zinnecker},
  {Kaper}, \& {Portegies Zwart}}]{Kouwenhoven2005}
{Kouwenhoven}, M.~B.~N., {Brown}, A.~G.~A., {Zinnecker}, H., {Kaper}, L., \&
  {Portegies Zwart}, S.~F. 2005, \aap, 430, 137

\bibitem[{{Kudritzki} {et~al.}(2003){Kudritzki}, {Bresolin}, \&
  {Przybilla}}]{Kudritzki03}
{Kudritzki}, R.~P., {Bresolin}, F., \& {Przybilla}, N. 2003, \apjl, 582, L83

\bibitem[{{Kudritzki} {et~al.}(2020){Kudritzki}, {Urbaneja}, \&
  {Rix}}]{Kudritzki2020}
{Kudritzki}, R.-P., {Urbaneja}, M.~A., \& {Rix}, H.-W. 2020, \apj, 890, 28

\bibitem[{{Kurucz}(1970)}]{Kurucz1970}
{Kurucz}, R.~L. 1970, SAO Special Report, 309

\bibitem[{{Kurucz}(1993)}]{Kurucz1993}
---. 1993, {SYNTHE spectrum synthesis programs and line data}

\bibitem[{{Kurucz}(2005)}]{Kurucz2005}
---. 2005, Memorie della Societa Astronomica Italiana Supplementi, 8, 14

\bibitem[{{Lequeux}(1979)}]{Lequeux1979}
{Lequeux}, J. 1979, \aap, 80, 35

\bibitem[{{Leung} \& {Bovy}(2019)}]{Leung2019}
{Leung}, H.~W., \& {Bovy}, J. 2019, \mnras, 489, 2079

\bibitem[{{Li} {et~al.}(2013){Li}, {de Grijs}, \& {Deng}}]{Li2013}
{Li}, C., {de Grijs}, R., \& {Deng}, L. 2013, \mnras, 436, 1497

\bibitem[{{Li} {et~al.}(2019){Li}, {Zhao}, {Jia}, {Liao}, {Yang}, \&
  {Wang}}]{Li2019}
{Li}, C., {Zhao}, G., {Jia}, Y., {et~al.} 2019, \apj, 871, 208

\bibitem[{{Lindegren} {et~al.}(2018){Lindegren}, {Hern{\'a}ndez}, {Bombrun},
  {Klioner}, {Bastian}, {Ramos-Lerate}, {de Torres}, {Steidelm{\"u}ller},
  {Stephenson}, {Hobbs}, {Lammers}, {Biermann}, {Geyer}, {Hilger}, {Michalik},
  {Stampa}, {McMillan}, {Casta{\~n}eda}, {Clotet}, {Comoretto}, {Davidson},
  {Fabricius}, {Gracia}, {Hambly}, {Hutton}, {Mora}, {Portell}, {van Leeuwen},
  {Abbas}, {Abreu}, {Altmann}, {Andrei}, {Anglada}, {Balaguer-N{\'u}{\~n}ez},
  {Barache}, {Becciani}, {Bertone}, {Bianchi}, {Bouquillon}, {Bourda},
  {Br{\"u}semeister}, {Bucciarelli}, {Busonero}, {Buzzi}, {Cancelliere},
  {Carlucci}, {Charlot}, {Cheek}, {Crosta}, {Crowley}, {de Bruijne}, {de
  Felice}, {Drimmel}, {Esquej}, {Fienga}, {Fraile}, {Gai}, {Garralda},
  {Gonz{\'a}lez-Vidal}, {Guerra}, {Hauser}, {Hofmann}, {Holl}, {Jordan},
  {Lattanzi}, {Lenhardt}, {Liao}, {Licata}, {Lister}, {L{\"o}ffler},
  {Marchant}, {Martin-Fleitas}, {Messineo}, {Mignard}, {Morbidelli}, {Poggio},
  {Riva}, {Rowell}, {Salguero}, {Sarasso}, {Sciacca}, {Siddiqui}, {Smart},
  {Spagna}, {Steele}, {Taris}, {Torra}, {van Elteren}, {van Reeven}, \&
  {Vecchiato}}]{Lindegren2018}
{Lindegren}, L., {Hern{\'a}ndez}, J., {Bombrun}, A., {et~al.} 2018, \aap, 616,
  A2

\bibitem[{{Liu}(2019)}]{LiuC2019}
{Liu}, C. 2019, \mnras, 490, 550

\bibitem[{{Liu} {et~al.}(2019{\natexlab{a}}){Liu}, {Zhang}, {Howard}, {Bai},
  {Lu}, {Soria}, {Justham}, {Li}, {Zheng}, {Wang}, {Belczynski}, {Casares},
  {Zhang}, {Yuan}, {Dong}, {Lei}, {Isaacson}, {Wang}, {Bai}, {Shao}, {Gao},
  {Wang}, {Niu}, {Cui}, {Zheng}, {Mu}, {Zhang}, {Wang}, {Heger}, {Qi}, {Liao},
  {Lattanzi}, {Gu}, {Wang}, {Wu}, {Shao}, {Shen}, {Wang}, {Bregman}, {Di
  Stefano}, {Liu}, {Han}, {Zhang}, {Wang}, {Ren}, {Zhang}, {Zhang}, {Wang},
  {Cabrera-Lavers}, {Corradi}, {Rebolo}, {Zhao}, {Zhao}, {Chu}, \&
  {Cui}}]{LiuJ2019}
{Liu}, J., {Zhang}, H., {Howard}, A.~W., {et~al.} 2019{\natexlab{a}}, \nat,
  575, 618

\bibitem[{{Liu} {et~al.}(2020){Liu}, {Zheng}, {Soria}, {Aceituno}, {Zhang},
  {Lu}, {Wang}, {Hamann}, {Oskinova}, {Ramachandran}, {Yuan}, {Bai}, {Wang},
  {McKee}, {Wu}, {Wang}, {Lattanzi}, {Belczynski}, {Casares}, {Simon-Diaz},
  {Gonz{\'a}lez Hern{\'a}ndez}, \& {Rebolo}}]{LiuJ2020}
{Liu}, J., {Zheng}, Z., {Soria}, R., {et~al.} 2020, arXiv e-prints,
  arXiv:2005.12595

\bibitem[{{Liu} {et~al.}(2018){Liu}, {Fu}, {Zong}, {Wang}, {Uddin}, {Zhang},
  {Zhang}, {Cang}, {Li}, {Yang}, {Yang}, {Mould}, \& {Morrell}}]{LiuN2018}
{Liu}, N., {Fu}, J.~N., {Zong}, W., {et~al.} 2018, \aj, 155, 168

\bibitem[{{Liu} {et~al.}(2014){Liu}, {Yuan}, {Huo}, {Deng}, {Hou}, {Zhao},
  {Zhao}, {Shi}, {Luo}, {Xiang}, {Zhang}, {Huang}, \& {Zhang}}]{Liu2014}
{Liu}, X.~W., {Yuan}, H.~B., {Huo}, Z.~Y., {et~al.} 2014, in IAU Symposium,
  Vol. 298, Setting the scene for Gaia and LAMOST, ed. S.~{Feltzing},
  G.~{Zhao}, N.~A. {Walton}, \& P.~{Whitelock}, 310--321

\bibitem[{{Liu} {et~al.}(2019{\natexlab{b}}){Liu}, {Cui}, {Liu}, {Huang},
  {Zhao}, \& {Zhang}}]{LiuZ2019}
{Liu}, Z., {Cui}, W., {Liu}, C., {et~al.} 2019{\natexlab{b}}, \apjs, 241, 32

\bibitem[{{Mackey} {et~al.}(2015){Mackey}, {Gvaramadze}, {Mohamed}, \&
  {Langer}}]{Mackey2015}
{Mackey}, J., {Gvaramadze}, V.~V., {Mohamed}, S., \& {Langer}, N. 2015, \aap,
  573, A10

\bibitem[{{Majewski} {et~al.}(2011){Majewski}, {Zasowski}, \&
  {Nidever}}]{Majewski2011}
{Majewski}, S.~R., {Zasowski}, G., \& {Nidever}, D.~L. 2011, \apj, 739, 25

\bibitem[{{Majewski} {et~al.}(2017){Majewski}, {Schiavon}, {Frinchaboy},
  {Allende Prieto}, {Barkhouser}, {Bizyaev}, {Blank}, {Brunner}, {Burton},
  {Carrera}, {Chojnowski}, {Cunha}, {Epstein}, {Fitzgerald}, {Garc{\'{\i}}a
  P{\'e}rez}, {Hearty}, {Henderson}, {Holtzman}, {Johnson}, {Lam}, {Lawler},
  {Maseman}, {M{\'e}sz{\'a}ros}, {Nelson}, {Nguyen}, {Nidever}, {Pinsonneault},
  {Shetrone}, {Smee}, {Smith}, {Stolberg}, {Skrutskie}, {Walker}, {Wilson},
  {Zasowski}, {Anders}, {Basu}, {Beland}, {Blanton}, {Bovy}, {Brownstein},
  {Carlberg}, {Chaplin}, {Chiappini}, {Eisenstein}, {Elsworth}, {Feuillet},
  {Fleming}, {Galbraith-Frew}, {Garc{\'{\i}}a}, {Garc{\'{\i}}a-Hern{\'a}ndez},
  {Gillespie}, {Girardi}, {Gunn}, {Hasselquist}, {Hayden}, {Hekker}, {Ivans},
  {Kinemuchi}, {Klaene}, {Mahadevan}, {Mathur}, {Mosser}, {Muna}, {Munn},
  {Nichol}, {O'Connell}, {Parejko}, {Robin}, {Rocha-Pinto}, {Schultheis},
  {Serenelli}, {Shane}, {Silva Aguirre}, {Sobeck}, {Thompson}, {Troup},
  {Weinberg}, \& {Zamora}}]{Majewski2017}
{Majewski}, S.~R., {Schiavon}, R.~P., {Frinchaboy}, P.~M., {et~al.} 2017, \aj,
  154, 94

\bibitem[{{Matijevi{\v{c}}} {et~al.}(2011){Matijevi{\v{c}}}, {Zwitter},
  {Bienaym{\'e}}, {Bland -Hawthorn}, {Freeman}, {Gilmore}, {Grebel}, {Helmi},
  {Munari}, {Navarro}, {Parker}, {Reid}, {Seabroke}, {Siebert}, {Siviero},
  {Steinmetz}, {Watson}, {Williams}, \& {Wyse}}]{Matijevic2011}
{Matijevi{\v{c}}}, G., {Zwitter}, T., {Bienaym{\'e}}, O., {et~al.} 2011, \aj,
  141, 200

\bibitem[{{Merle} {et~al.}(2017){Merle}, {Van Eck}, {Jorissen}, {Van der
  Swaelmen}, {Masseron}, {Zwitter}, {Hatzidimitriou}, {Klutsch}, {Pourbaix},
  {Blomme}, {Worley}, {Sacco}, {Lewis}, {Abia}, {Traven}, {Sordo}, {Bragaglia},
  {Smiljanic}, {Pancino}, {Damiani}, {Hourihane}, {Gilmore}, {Randich},
  {Koposov}, {Casey}, {Morbidelli}, {Franciosini}, {Magrini}, {Jofre},
  {Costado}, {Jeffries}, {Bergemann}, {Lanzafame}, {Bayo}, {Carraro},
  {Flaccomio}, {Monaco}, \& {Zaggia}}]{Merle2017}
{Merle}, T., {Van Eck}, S., {Jorissen}, A., {et~al.} 2017, \aap, 608, A95

\bibitem[{{Moe} \& {Di Stefano}(2017)}]{Moe2017}
{Moe}, M., \& {Di Stefano}, R. 2017, \apjs, 230, 15

\bibitem[{{Munari} {et~al.}(2014){Munari}, {Henden}, {Frigo}, {Zwitter},
  {Bienaym{\'e}}, {Bland-Hawthorn}, {Boeche}, {Freeman}, {Gibson}, {Gilmore},
  {Grebel}, {Helmi}, {Kordopatis}, {Levine}, {Navarro}, {Parker}, {Reid},
  {Seabroke}, {Siebert}, {Siviero}, {Smith}, {Steinmetz}, {Templeton},
  {Terrell}, {Welch}, {Williams}, \& {Wyse}}]{Munari2014}
{Munari}, U., {Henden}, A., {Frigo}, A., {et~al.} 2014, \aj, 148, 81

\bibitem[{{Nomoto} {et~al.}(2006){Nomoto}, {Tominaga}, {Umeda}, {Kobayashi}, \&
  {Maeda}}]{Nomoto2006}
{Nomoto}, K., {Tominaga}, N., {Umeda}, H., {Kobayashi}, C., \& {Maeda}, K.
  2006, \nphysa, 777, 424

\bibitem[{{Oh} {et~al.}(2017){Oh}, {Price-Whelan}, {Hogg}, {Morton}, \&
  {Spergel}}]{Oh2017}
{Oh}, S., {Price-Whelan}, A.~M., {Hogg}, D.~W., {Morton}, T.~D., \& {Spergel},
  D.~N. 2017, \aj, 153, 257

\bibitem[{{Penoyre} {et~al.}(2020){Penoyre}, {Belokurov}, {Wyn Evans},
  {Everall}, \& {Koposov}}]{Penoyre2020}
{Penoyre}, Z., {Belokurov}, V., {Wyn Evans}, N., {Everall}, A., \& {Koposov},
  S.~E. 2020, \mnras, 495, 321

\bibitem[{{Perryman} {et~al.}(1997){Perryman}, {Lindegren}, {Kovalevsky},
  {Hog}, {Bastian}, {Bernacca}, {Creze}, {Donati}, {Grenon}, {Grewing}, {van
  Leeuwen}, {van der Marel}, {Mignard}, {Murray}, {Le Poole}, {Schrijver},
  {Turon}, {Arenou}, {Froeschle}, \& {Petersen}}]{Perryman1997}
{Perryman}, M.~A.~C., {Lindegren}, L., {Kovalevsky}, J., {et~al.} 1997, \aap,
  500, 501

\bibitem[{{Pourbaix} {et~al.}(2004){Pourbaix}, {Ivezi{\'c}}, {Knapp}, {Gunn},
  \& {Lupton}}]{Pourbaix2004}
{Pourbaix}, D., {Ivezi{\'c}}, {\v{Z}}., {Knapp}, G.~R., {Gunn}, J.~E., \&
  {Lupton}, R.~H. 2004, \aap, 423, 755

\bibitem[{{Price-Whelan} {et~al.}(2017){Price-Whelan}, {Hogg},
  {Foreman-Mackey}, \& {Rix}}]{Price-Whelan2017}
{Price-Whelan}, A.~M., {Hogg}, D.~W., {Foreman-Mackey}, D., \& {Rix}, H.-W.
  2017, \apj, 837, 20

\bibitem[{{Qian} {et~al.}(2017){Qian}, {He}, {Zhang}, {Zhu}, {Shi}, {Zhao}, \&
  {Zhou}}]{Qian2017}
{Qian}, S.-B., {He}, J.-J., {Zhang}, J., {et~al.} 2017, Research in Astronomy
  and Astrophysics, 17, 087

\bibitem[{{Queiroz} {et~al.}(2020){Queiroz}, {Anders}, {Chiappini},
  {Khalatyan}, {Santiago}, {Steinmetz}, {Valentini}, {Miglio}, {Bossini},
  {Barbuy}, {Minchev}, {Minniti}, {Garc{\'\i}a Hern{\'a}ndez}, {Schultheis},
  {Beaton}, {Beers}, {Bizyaev}, {Brownstein}, {Cunha},
  {Fern{\'a}ndez-Trincado}, {Frinchaboy}, {Lane}, {Majewski}, {Nataf},
  {Nitschelm}, {Pan}, {Roman-Lopes}, {Sobeck}, {Stringfellow}, \&
  {Zamora}}]{Queiroz2020}
{Queiroz}, A.~B.~A., {Anders}, F., {Chiappini}, C., {et~al.} 2020, \aap, 638,
  A76

\bibitem[{{Raghavan} {et~al.}(2010){Raghavan}, {McAlister}, {Henry}, {Latham},
  {Marcy}, {Mason}, {Gies}, {White}, \& {ten Brummelaar}}]{Raghavan2010}
{Raghavan}, D., {McAlister}, H.~A., {Henry}, T.~J., {et~al.} 2010, \apjs, 190,
  1

\bibitem[{{Reed}(2005)}]{Reed2005}
{Reed}, B.~C. 2005, \aj, 130, 1652

\bibitem[{{Rivinius} {et~al.}(2020){Rivinius}, {Baade}, {Hadrava}, {Heida}, \&
  {Klement}}]{Rivinius2020}
{Rivinius}, T., {Baade}, D., {Hadrava}, P., {Heida}, M., \& {Klement}, R. 2020,
  \aap, 637, L3

\bibitem[{{Sana} {et~al.}(2012){Sana}, {de Mink}, {de Koter}, {Langer},
  {Evans}, {Gieles}, {Gosset}, {Izzard}, {Le Bouquin}, \&
  {Schneider}}]{Sana2012}
{Sana}, H., {de Mink}, S.~E., {de Koter}, A., {et~al.} 2012, Science, 337, 444

\bibitem[{{Sana} {et~al.}(2013){Sana}, {de Koter}, {de Mink}, {Dunstall},
  {Evans}, {H{\'e}nault-Brunet}, {Ma{\'\i}z Apell{\'a}niz},
  {Ram{\'\i}rez-Agudelo}, {Taylor}, {Walborn}, {Clark}, {Crowther}, {Herrero},
  {Gieles}, {Langer}, {Lennon}, \& {Vink}}]{Sana2013}
{Sana}, H., {de Koter}, A., {de Mink}, S.~E., {et~al.} 2013, \aap, 550, A107

\bibitem[{{Shenar} {et~al.}(2020){Shenar}, {Bodensteiner}, {Abdul-Masih},
  {Fabry}, {Mahy}, {Marchant}, {Banyard}, {Bowman}, {Dsilva}, {Hawcroft},
  {Reggiani}, \& {Sana}}]{Shenar2020}
{Shenar}, T., {Bodensteiner}, J., {Abdul-Masih}, M., {et~al.} 2020, arXiv
  e-prints, arXiv:2004.12882

\bibitem[{{Shu}(2016)}]{Shu2016}
{Shu}, F.~H. 2016, \araa, 54, 667

\bibitem[{{Shull} \& {Danforth}(2019)}]{Shull2019}
{Shull}, J.~M., \& {Danforth}, C.~W. 2019, \apj, 882, 180

\bibitem[{{Sim{\'o}n-D{\'\i}az} {et~al.}(2020){Sim{\'o}n-D{\'\i}az}, {Ma{\'\i}z
  Apell{\'a}niz}, {Lennon}, {Gonz{\'a}lez Hern{\'a}ndez}, {Allende Prieto},
  {Castro}, {de Burgos}, {Dufton}, {Herrero}, {Toledo-Padr{\'o}n}, \&
  {Smartt}}]{SimonDaz2020}
{Sim{\'o}n-D{\'\i}az}, S., {Ma{\'\i}z Apell{\'a}niz}, J., {Lennon}, D.~J.,
  {et~al.} 2020, \aap, 634, L7

\bibitem[{{Skinner} {et~al.}(2018){Skinner}, {Covey}, {Bender}, {Rivera}, {De
  Lee}, {Souto}, {Chojnowski}, {Troup}, {Badenes}, {Bizyaev}, {Blake},
  {Burgasser}, {Ca{\~n}as}, {Carlberg}, {G{\'o}mez Maqueo Chew}, {Deshpande},
  {Fleming}, {Fern{\'a}ndez-Trincado}, {Garc{\'\i}a-Hern{\'a}ndez}, {Hearty},
  {Kounkel}, {Longa-Pe{\~n}e}, {Mahadevan}, {Majewski}, {Minniti}, {Nidever},
  {Oravetz}, {Pan}, {Stassun}, {Terrien}, \& {Zamora}}]{Skinner2018}
{Skinner}, J., {Covey}, K.~R., {Bender}, C.~F., {et~al.} 2018, \aj, 156, 45

\bibitem[{{Skrutskie} {et~al.}(2006){Skrutskie}, {Cutri}, {Stiening},
  {Weinberg}, {Schneider}, {Carpenter}, {Beichman}, {Capps}, {Chester},
  {Elias}, {Huchra}, {Liebert}, {Lonsdale}, {Monet}, {Price}, {Seitzer},
  {Jarrett}, {Kirkpatrick}, {Gizis}, {Howard}, {Evans}, {Fowler}, {Fullmer},
  {Hurt}, {Light}, {Kopan}, {Marsh}, {McCallon}, {Tam}, {Van Dyk}, \&
  {Wheelock}}]{Skrutskie2006}
{Skrutskie}, M.~F., {Cutri}, R.~M., {Stiening}, R., {et~al.} 2006, \aj, 131,
  1163

\bibitem[{Srivastava {et~al.}(2014)Srivastava, Hinton, Krizhevsky, Sutskever,
  \& Salakhutdinov}]{srivastava14a}
Srivastava, N., Hinton, G., Krizhevsky, A., Sutskever, I., \& Salakhutdinov, R.
  2014, Journal of Machine Learning Research, 15, 1929.
\newblock \url{http://jmlr.org/papers/v15/srivastava14a.html}

\bibitem[{{Struck}(2020)}]{Struck2020}
{Struck}, C. 2020, \mnras, 494, 1838

\bibitem[{{Thielemann} \& {Arnett}(1985)}]{Thielemann1985}
{Thielemann}, F.~K., \& {Arnett}, W.~D. 1985, \apj, 295, 604

\bibitem[{{Tian} {et~al.}(2020){Tian}, {Liu}, {Yuan}, {Fang}, {Chen}, {Xiang},
  {Huang}, {Bi}, {Yang}, {Wu}, {Wang}, {Zhang}, {Huo}, {Yang}, {Liu}, {Guo}, \&
  {Zhang}}]{Tian2020}
{Tian}, Z., {Liu}, X., {Yuan}, H., {et~al.} 2020, arXiv e-prints,
  arXiv:2006.01394

\bibitem[{{Tian} {et~al.}(2018){Tian}, {Liu}, {Yuan}, {Chen}, {Xiang}, {Huang},
  {Wang}, {Zhang}, {Guo}, {Ren}, {Huo}, {Yang}, {Zhang}, {Bi}, {Yang}, {Liu},
  {Zhang}, {Li}, {Wu}, \& {Zhang}}]{Tian2018}
{Tian}, Z.-J., {Liu}, X.-W., {Yuan}, H.-B., {et~al.} 2018, Research in
  Astronomy and Astrophysics, 18, 052

\bibitem[{{Timmes} {et~al.}(1995){Timmes}, {Woosley}, \& {Weaver}}]{Timmes1995}
{Timmes}, F.~X., {Woosley}, S.~E., \& {Weaver}, T.~A. 1995, \apjs, 98, 617

\bibitem[{{Ting} {et~al.}(2019){Ting}, {Conroy}, {Rix}, \&
  {Cargile}}]{Ting2019}
{Ting}, Y.-S., {Conroy}, C., {Rix}, H.-W., \& {Cargile}, P. 2019, \apj, 879, 69

\bibitem[{{Torra} {et~al.}(2000){Torra}, {Fern{\'a}ndez}, \&
  {Figueras}}]{Torra2000}
{Torra}, J., {Fern{\'a}ndez}, D., \& {Figueras}, F. 2000, \aap, 359, 82

\bibitem[{{Traven} {et~al.}(2017){Traven}, {Matijevi{\v{c}}}, {Zwitter},
  {{\v{Z}}erjal}, {Kos}, {Asplund}, {Bland -Hawthorn}, {Casey}, {De Silva},
  {Freeman}, {Lin}, {Martell}, {Schlesinger}, {Sharma}, {Simpson}, {Zucker},
  {Anguiano}, {Da Costa}, {Duong}, {Horner}, {Hyde}, {Kafle}, {Munari},
  {Nataf}, {Navin}, {Reid}, \& {Ting}}]{Traven2017}
{Traven}, G., {Matijevi{\v{c}}}, G., {Zwitter}, T., {et~al.} 2017, \apjs, 228,
  24

\bibitem[{{Traven} {et~al.}(2020){Traven}, {Feltzing}, {Merle}, {Van der
  Swaelmen}, {{\v{C}}otar}, {Church}, {Zwitter}, {Ting}, {Sahlholdt},
  {Asplund}, {Bland-Hawthorn}, {De Silva}, {Freeman}, {Martell}, {Sharma},
  {Zucker}, {Buder}, {Casey}, {D'Orazi}, {Kos}, {Lewis}, {Lin}, {Lind},
  {Simpson}, {Stello}, {Munari}, \& {Wittenmyer}}]{Traven2020}
{Traven}, G., {Feltzing}, S., {Merle}, T., {et~al.} 2020, \aap, 638, A145

\bibitem[{{Wang} {et~al.}(2020){Wang}, {Huang}, {Zhang},
  {L{\'o}pez-Corredoira}, {Chen}, {Guo}, \& {Chang}}]{Wang2020}
{Wang}, H.~F., {Huang}, Y., {Zhang}, H.~W., {et~al.} 2020, arXiv e-prints,
  arXiv:2005.14362

\bibitem[{{Wang} {et~al.}(2016){Wang}, {Shi}, {Zhao}, {Zhang}, {Huo}, {Zhang},
  {Chen}, {Wu}, {Zhang}, \& {Hou}}]{Wang2016}
{Wang}, J., {Shi}, J., {Zhao}, Y., {et~al.} 2016, \mnras, 456, 672

\bibitem[{{Wang} \& {Chen}(2019)}]{Wang2019}
{Wang}, S., \& {Chen}, X. 2019, \apj, 877, 116

\bibitem[{{Woosley} \& {Weaver}(1995)}]{Woosley1995}
{Woosley}, S.~E., \& {Weaver}, T.~A. 1995, \apjs, 101, 181

\bibitem[{{Xiang} {et~al.}(2019){Xiang}, {Ting}, {Rix}, {Sand ford}, {Buder},
  {Lind}, {Liu}, {Shi}, \& {Zhang}}]{Xiang2019}
{Xiang}, M., {Ting}, Y.-S., {Rix}, H.-W., {et~al.} 2019, arXiv e-prints,
  arXiv:1908.09727

\bibitem[{{Xiang} {et~al.}(2015){Xiang}, {Liu}, {Yuan}, {Huo}, {Huang},
  {Zheng}, {Zhang}, {Chen}, {Zhang}, {Sun}, {Wang}, {Zhao}, {Shi}, {Luo}, {Li},
  {Bai}, {Zhang}, {Hou}, {Yuan}, \& {Li}}]{Xiang2015}
{Xiang}, M.~S., {Liu}, X.~W., {Yuan}, H.~B., {et~al.} 2015, \mnras, 448, 90

\bibitem[{{Xiang} {et~al.}(2017{\natexlab{a}}){Xiang}, {Liu}, {Yuan}, {Huo},
  {Huang}, {Wang}, {Chen}, {Ren}, {Zhang}, {Tian}, {Yang}, {Shi}, {Zhao}, {Li},
  {Zhao}, {Cui}, {Li}, {Hou}, {Zhang}, {Zhang}, {Wang}, {Wu}, {Cao}, {Yan},
  {Yan}, {Luo}, {Zhang}, {Bai}, {Yuan}, {Dong}, {Lei}, \& {Li}}]{Xiang2017a}
---. 2017{\natexlab{a}}, \mnras, 467, 1890

\bibitem[{{Xiang} {et~al.}(2017{\natexlab{b}}){Xiang}, {Liu}, {Shi}, {Yuan},
  {Huang}, {Luo}, {Zhang}, {Zhao}, {Zhang}, {Ren}, {Chen}, {Wang}, {Li}, {Huo},
  {Zhang}, {Wang}, {Zhang}, {Hou}, \& {Wang}}]{Xiang2017b}
{Xiang}, M.~S., {Liu}, X.~W., {Shi}, J.~R., {et~al.} 2017{\natexlab{b}},
  \mnras, 464, 3657

\bibitem[{{Xu} {et~al.}(2006){Xu}, {Reid}, {Zheng}, \& {Menten}}]{Xu2006}
{Xu}, Y., {Reid}, M.~J., {Zheng}, X.~W., \& {Menten}, K.~M. 2006, Science, 311,
  54

\bibitem[{{Xu} {et~al.}(2018){Xu}, {Bian}, {Reid}, {Li}, {Zhang}, {Yan},
  {Dame}, {Menten}, {He}, {Liao}, \& {Tang}}]{Xu2018}
{Xu}, Y., {Bian}, S.~B., {Reid}, M.~J., {et~al.} 2018, \aap, 616, L15

\bibitem[{{Yang} {et~al.}(2020){Yang}, {Long}, {Shan}, {Zhang}, {Guo}, {Bai},
  {Bai}, {Cui}, {Wang}, \& {Liu}}]{Yang2020}
{Yang}, F., {Long}, R.~J., {Shan}, S.-S., {et~al.} 2020, arXiv e-prints,
  arXiv:2006.04430

\bibitem[{{Yuan} {et~al.}(2015{\natexlab{a}}){Yuan}, {Liu}, {Xiang}, {Huang},
  {Chen}, {Wu}, {Hou}, \& {Zhang}}]{Yuan2015b}
{Yuan}, H., {Liu}, X., {Xiang}, M., {et~al.} 2015{\natexlab{a}}, \apj, 799, 135

\bibitem[{{Yuan} {et~al.}(2013){Yuan}, {Liu}, \& {Xiang}}]{Yuan2013}
{Yuan}, H.-B., {Liu}, X.-W., \& {Xiang}, M.-S. 2013, \mnras, 430, 2188

\bibitem[{{Yuan} {et~al.}(2015{\natexlab{b}}){Yuan}, {Liu}, {Huo}, {Xiang},
  {Huang}, {Chen}, {Zhang}, {Sun}, {Wang}, {Zhang}, {Zhao}, {Luo}, {Shi}, {Li},
  {Yuan}, {Dong}, {Li}, {Hou}, \& {Zhang}}]{Yuan2015}
{Yuan}, H.~B., {Liu}, X.~W., {Huo}, Z.~Y., {et~al.} 2015{\natexlab{b}}, \mnras,
  448, 855

\bibitem[{{Yungelson} {et~al.}(2020){Yungelson}, {Kuranov}, {Postnov}, \&
  {Kolesnikov}}]{Yungelson2020}
{Yungelson}, L.~R., {Kuranov}, A.~G., {Postnov}, K.~A., \& {Kolesnikov}, D.~A.
  2020, \mnras, 496, L6

\bibitem[{{Zhang} {et~al.}(2014){Zhang}, {Liu}, {Yuan}, {Zhao}, {Yao}, {Zhang},
  {Xiang}, \& {Huang}}]{Zhang2014}
{Zhang}, H.-H., {Liu}, X.-W., {Yuan}, H.-B., {et~al.} 2014, Research in
  Astronomy and Astrophysics, 14, 456

\bibitem[{{Zhang} {et~al.}(2017){Zhang}, {Qian}, \& {He}}]{Zhang2017}
{Zhang}, J., {Qian}, S.-B., \& {He}, J.-D. 2017, Research in Astronomy and
  Astrophysics, 17, 22

\bibitem[{{Zhao} {et~al.}(2012){Zhao}, {Zhao}, {Chu}, {Jing}, \&
  {Deng}}]{Zhao2012}
{Zhao}, G., {Zhao}, Y.-H., {Chu}, Y.-Q., {Jing}, Y.-P., \& {Deng}, L.-C. 2012,
  Research in Astronomy and Astrophysics, 12, 723

\end{thebibliography}

\begin{appendix}
\section{Examples of problematic LAMOST spectra}
\begin{figure*}
\centering
\includegraphics[width=180mm]{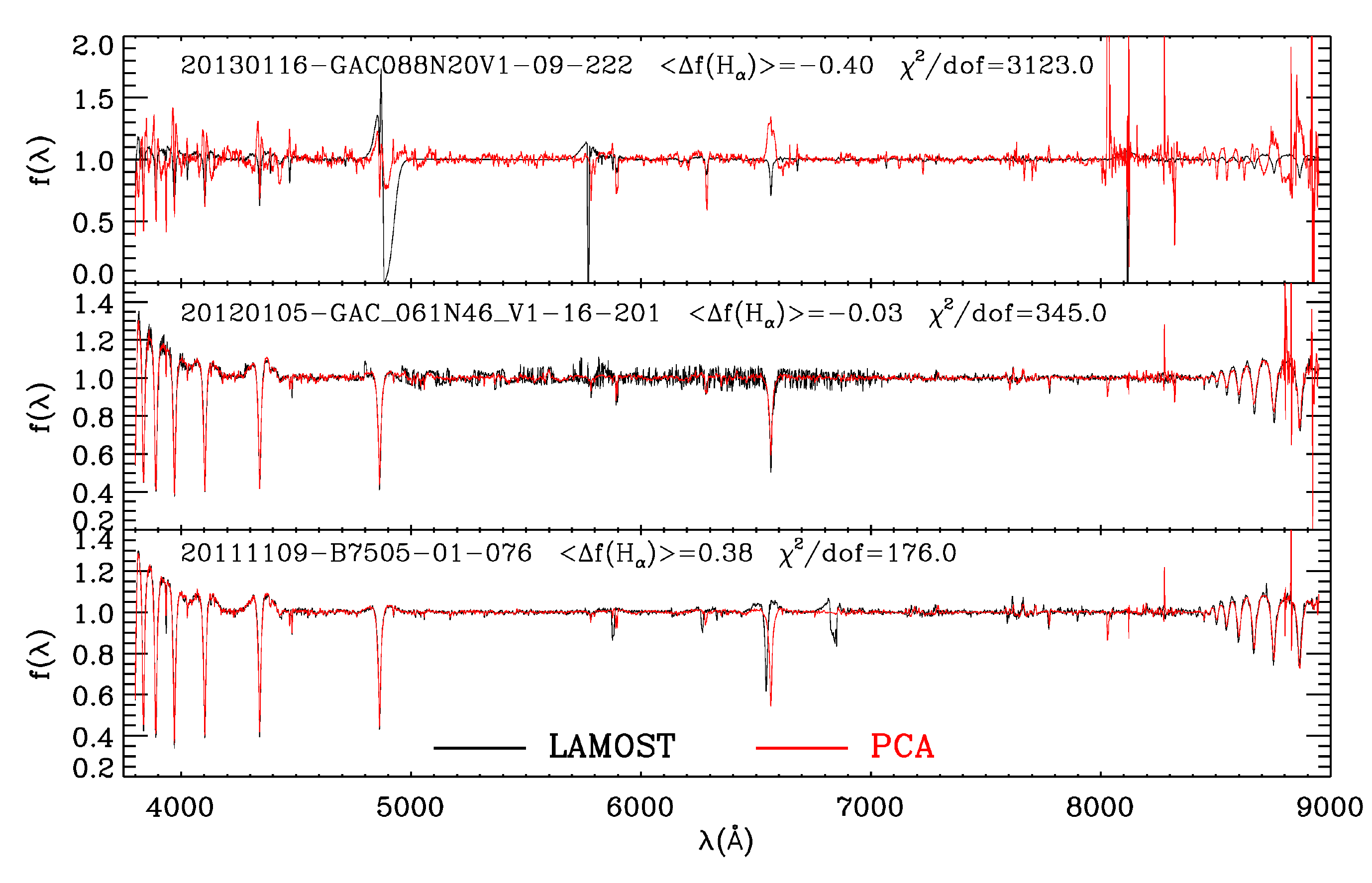}
\caption{A few typical examples that exhibit abnormal residuals between the LAMOST spectrum and the PCA reconstruction. The top panel shows a spectrum with problematic fluxes over the wavelength range of $\lambda$4700--5000\AA, leading to poor PCA reconstruction. The middle panel shows a spectrum with artifacts in the LAMOST spectra across $\lambda$5000--7000\AA. The PCA reconstruction is visually reasonable, but the $\chi^2$ between the LAMOST and PCA-reconstructed spectra is suboptimal. The bottom panel shows a spectrum that has problematic wavelength calibration from the LAMOST pipeline in the wavelength range of $\lambda$5800--7000\AA. }
\label{fig:FigA1}
\end{figure*}

As shown in Fig.\,\ref{fig:Fig2}, there are some stars with large $\chi^2$ in the residuals between the LAMOST spectra and the PCA reconstruction. We find that, for these stars, the LAMOST spectra are often problematic due to various reasons, including instrument problems, erroneous wavelength calibration, and other reasons. Fig.\,\ref{fig:FigA1} shows a few examples of the erroneous LAMOST spectra.  

\section{Examples of mock binary spectra}
\begin{figure*}
\centering
\includegraphics[width=150mm]{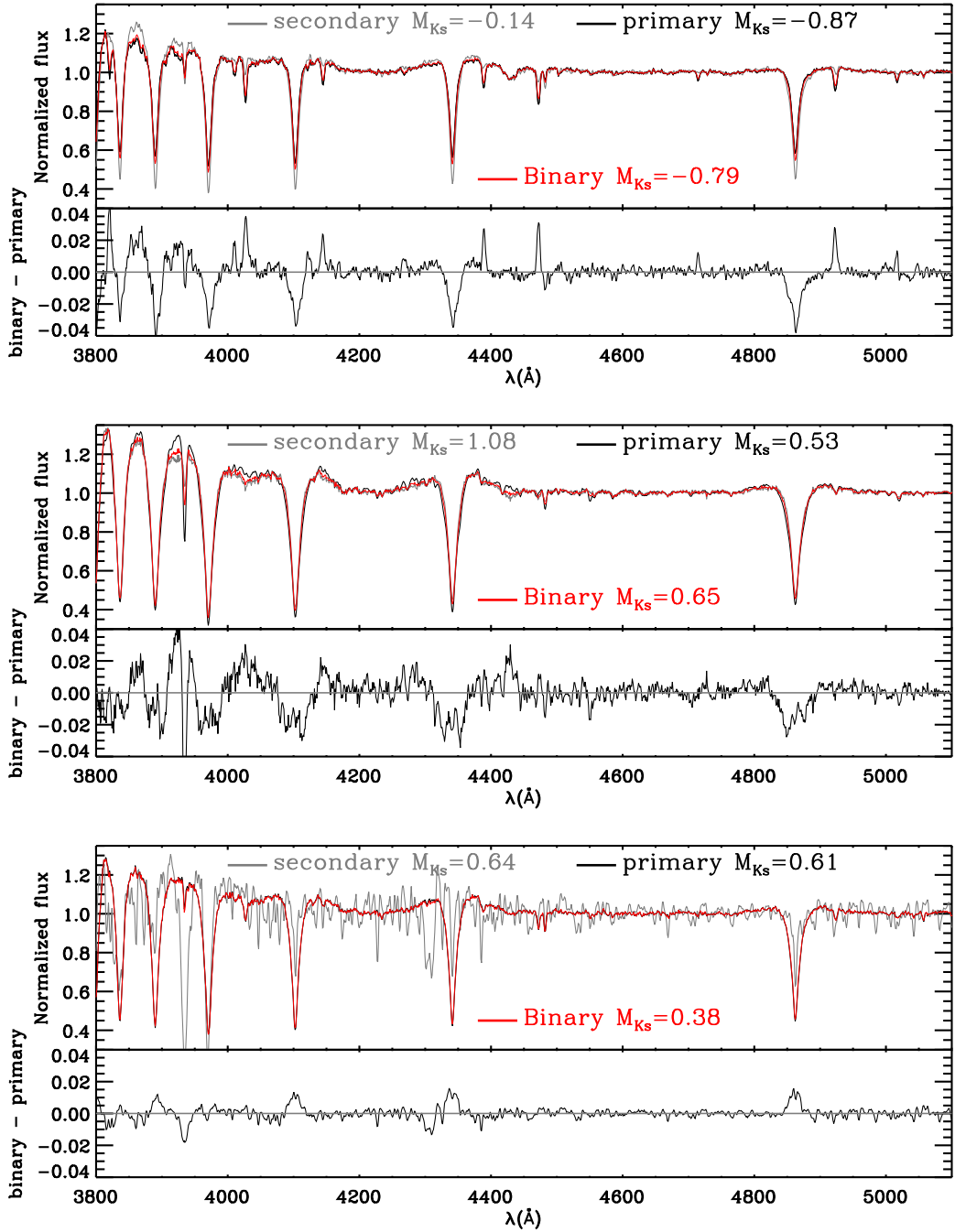}
\caption{A few typical examples of mock composite spectra. We focus on the wavelength range of $\lambda$3800--5100{\AA}. The spectra of the primary, secondary, and binary are shown in black, grey, and red, respectively. For all these cases, the primary is a B-type star. The secondary are B-, A- and G-type stars from the top to bottom, respectively. The spectroscopic \mk estimates applied to the primary, secondary, and composite binary spectra are also shown in the figure. The spectroscopic \mk estimate of the binary spectra typically agrees with the one from the primary spectra, with a difference $\lesssim$0.2 mag.}
\label{fig:FigA2}
\end{figure*}

As discussed in Section\,\ref{binaryeffect}, in order to study the spectroscopic \mk derived from composite spectra, we build an empirical library of binary spectra using the LAMOST spectra of single stars. Fig.\,\ref{fig:FigA2} shows a few examples of our mock composite spectra as well as their inferred spectroscopic \mk.  The spectroscopic \mk estimates of the binary spectra typically agree with those from the spectra of the primary stars, with a difference $\lesssim$0.2\,mag.

\end{appendix}

\end{document}